\documentclass[12pt,a4paper,dvips]{article}
\usepackage{amsmath}
\usepackage{graphicx}
\newcommand{\plaatje}[2]{\centerline{\mbox{\includegraphics[width=#2\textwidth]{#1}}}}
\newcommand{\os}{\underline}
\newcommand{\mc}{\multicolumn}
\newcommand{\pr}{\rightarrow}

\newcommand{\ba}{\begin{array}}
\newcommand{\ea}{\end{array}}

\newcommand{\varp}{\varphi}
\newcommand{\eps}{\varepsilon}

\newcommand{\il}{\int\limits}
\newenvironment{inspring}[1]%
{\begin{list}{}{\setlength{\rightmargin}{0cm}
                \setlength{\listparindent}{0cm}
                \settowidth{\labelwidth}{\mbox{#1}}
                \setlength{\leftmargin}{1.1\labelwidth}
                \setlength{\labelsep}{.1\labelwidth}}}%
{\end{list}}
\newcommand{\ITEM}[1]{\item[#1\hfill]}
\newcommand{\bi}[1]{\begin{inspring}{#1}}
\newcommand{\ei}{\end{inspring}}

\newcommand{\beq}{\begin{equation}}
\newcommand{\eq}{\end{equation}}
\catcode`\@=11

\font\tenmsa=msam10 \font\sevenmsa=msam7 \font\fivemsa=msam5
\font\tenmsb=msbm10 \font\sevenmsb=msbm7 \font\fivemsb=msbm5
\newfam\msafam
\newfam\msbfam
\textfont\msafam=\tenmsa  \scriptfont\msafam=\sevenmsa
  \scriptscriptfont\msafam=\fivemsa
\textfont\msbfam=\tenmsb  \scriptfont\msbfam=\sevenmsb
  \scriptscriptfont\msbfam=\fivemsb

\def\Bbb{\ifmmode\let\next\Bbb@\else
 \def\next{\errmessage{Use \string\Bbb\space only in math mode}}\fi\next}
\def\Bbb@#1{{\Bbb@@{#1}}}
\def\Bbb@@#1{\fam\msbfam#1}

\title{Truncation strategy for the series expressions in the advanced ENZ-theory of diffraction integrals}
\author{S.\ van Haver \\
S[\&]T Experts Pool (STEP) \\
P.O.\ Box 608, 2600 AP Delft, The Netherlands, and \\
Optics Research Group, Faculty of Applied Sciences, \\
Technical University Delft, \\
Van der Waalsweg 8, \\
2618 CH Delft, The Netherlands. \\
E-mail svenvanhaver@gmail.com \\
\mbox{} \\
A.J.E.M.\ Janssen \\
Department of Mathematics and Computer Science, \\
Eindhoven University of Technology, \\
P.O.\ Box 513, 5600 MB Eindhoven, The Netherlands. \\
E-mail a.j.e.m.janssen@tue.nl}
\date{}

\begin{document}
\maketitle
\mbox{} \\
\noindent
{\bf Abstract.} \\
The advanced ENZ-theory of diffraction integrals, as published recently in J.\ Europ.\ Opt.\ Soc.\ Rap.\ Public.\ {\bf 8}, 13044 (2013), presents the diffraction integrals per Zernike term in the form of doubly infinite series. These double series involve, aside from an overall azimuthal factor, the products of Jinc functions Jinc$_h$ for the radial dependence and structural quantities $c_t$ that depend on the optical parameters of the optical system (such as NA and refractive indices) and the defocus value. The products in the double series have coefficients that are related to Clebsch-Gordan coefficients and that depend on the order $h$ of the Jinc function and the index $t$ of the structural quantity, as well as on the azimuthal order $m$ and degree $n$ of the involved Zernike term $Z_n^m$. The structural quantities themselves are also given in the form of doubly infinite series, the terms of which are products of Zernike coefficients $a_l$, pertaining to an algebraic function containing the optical parameters, and Zernike coefficients $b_k$, pertaining to a focal factor, and these products have coefficients that are again related to Clebsch-Gordan coefficients. Finally, the $a_l$, are also given in the form of an infinite series. In this paper, we give truncation rules for the various infinite series depending on required accuracy. In particular, we make precise the following rule-of-thumb for truncation of the double series per Zernike term: For a given value of the radial variable $r$ and the defocus parameter $f$, it is enough to include in the double series \\
-- all Jinc functions of order $h$ less than $H$, \\
-- all structural quantities with index $t$ less than $T$, \\
where $H$ is somewhat larger than $2\pi r$ and $T$ is somewhat larger than $\frac12\,|f|$. We present of this rule both a global version, which can be used for all Zernike terms at the same time, and a dedicated version, in which the $H$ and $T$ take into account order and degree of the involved Zernike term. 
\newpage
\noindent
\section{Introduction and overview} \label{sec1}
\mbox{} \\[-9mm]

The advanced ENZ-theory of diffraction integrals, as presented in \cite{ref1}, aims at the computation of the Debye approximation of the Rayleigh integral for the optical point-spread functions of radially symmetric optical systems that range from as basic as having low NA and small defocus value to advanced high-NA systems, with vector fields and polarization, that are meant for imaging of extended objects into a multilayer structure. As in the classical Nijboer-Zernike theory, the generalized pupil function is developed into a series of Zernike terms. This gives rise to diffraction integrals per Zernike term that are expressed in \cite{ref1} as doubly infinite series
\beq \label{e1}
I=I_n^m=\sum_{h,t}\,A_{2t,n,h}^{0mm}({-}1)^{\frac{h-m}{2}}\,c_t\,\frac{J_{h+1}(2\pi r)}{2\pi r}~.
\eq
In Eq.~(\ref{e1}), $m$ and $n$ are the azimuthal order and degree of the involved Zernike term $Z_n^m$, the $c_t=c_t(OS,f)$ are the Zernike coefficients of the radially symmetric front factor composed of an algebraic factor comprising the parameters of the optical system and a factor comprising the defocus parameter $f$, the $J_{h+1}(2\pi r)/2\pi r$ are Jinc functions whose order $h$ has the same parity as $m$ with argument $2\pi r$ where $r$ is the value of the radial parameter, and the $A$ are to Clebsch-Gordan coefficients related numbers. In \cite{ref1}, Eq.~(59), there occurs a slightly more general expression, in which the vectorial nature and polarization conditions are accounted for, leading to 5 series expressions involving an integer $j$, $|j|=0,1,2\,$, of which Eq.~(\ref{e1}) is the case $j=0$. We shall not consider this generalization, since for truncation matters all these 5 cases behave the same. Furthermore, in the low-NA, small-defocus case, where a scalar treatment is allowed, the only required diffraction integral is the one with $j=0$.

The $A$-coefficients in the double series in Eq.~(\ref{e1}) have attractive properties with respect to their size and the set of $h,t$ for which they are non-vanishing. The main effort in getting truncation rules goes therefore into bounding Jinc functions Jinc$_h$ and structural quantities $c_t$. The Jinc functions are directly given in terms of Bessel functions while the structural quantities involve products of spherical Bessel and Hankel functions evaluated at $f/2$ and $f/2v_0$, respectively, where $v_0$, $0<v_0<1$, is a quantity determined by the optical system. Now it is a fact that (spherical) Bessel functions, considered as a function of the order, are of constant magnitude as long as the order is less than the value of the argument. Beyond this point a super exponential decay as a function of order takes place. The situation for the structural quantities is somewhat complicated by the occurrence of the Hankel functions (causing decay to slow down to exponential for $t$ beyond $|f|/2v_0$). These observations are basic to the approach taken in this paper and lead to the general rule-of-thumb that it suffices to include in Eq.~(\ref{e1}) all terms $h$, $t$ with $0\leq h\leq H$, $0\leq t\leq T$ in which $H$ is slightly larger than $2\pi r$ and $T$ is slightly larger than $|f|/2$. It is the aim of this paper to give a more precise meaning to this rule-of-thumb, in which the required absolute accuracy is included. Furthermore, by taking advantage of the $(m,n)$-dependent support properties of the $A$-coefficients, it is possible to formulate a truncation rule per Zernike term $Z_n^m$ that achieves a particular accuracy with substantially less terms than when the general rule were used.

We shall do this in all detail for the diffraction integral $I=I_{VM}$ of \cite{ref1}, Sec.~8, which is meant for systems with high NA, vector fields and magnification. Explicitly, $I$ assumes the form
\beq \label{e2}
I=I_{VM}=I_{n,VM}^m=\il_0^1\,a(\rho)\,f(\rho)\,p(\rho)\,b(\rho)\,\rho\,d\rho~,
\eq
where
\beq \label{e3}
a(\rho)=\frac{(1-s_0^2\rho^2)^{1/2}+(1-s_{0,M}^2\rho^2)^{1/2}}{(1-s_0^2\rho^2)^{1/4}\,(1-s_{0,M}^2\rho^2)^{3/4}}~,
\eq
\beq \label{e4}
f(\rho)=\exp\,\Bigl[\frac{if}{u_0}\,(1-\sqrt{1-s_0^2\rho^2})\Bigr]~,
\eq
\beq \label{e5}
p(\rho)=R_n^{|m|}(\rho)~,~~~~~~b(\rho)=J_m(2\pi r\rho)~,
\eq
are the algebraic, focal, polynomial and Bessel function factor, respectively. Here $s_0$ is the NA in image space, $s_{0,M}$ is built from the refractive indices in image and object space and the magnification factor in object space according to \cite{ref1}, Eq.~(31), and $u_0=1-\sqrt{1-s_0^2}$.

The $I_{VM}$-case is with respect to truncation issues quite representative for all diffraction integrals considered in \cite{ref1}, except for the case of $I_{VMML}$ in \cite{ref1}, Sec.~9, with backward propagating waves in a layer of the multilayer structure in image space. The $I_{VM}$-case is also general enough to illustrate the various intricacies that come with the computation of the Zernike coefficients $c_t$, the structural quantities, of the front factor $a(\rho)\,f(\rho)$, see \cite{ref1}, Sec.~4, requiring truncation rules as well.

In Sec.~\ref{sec2} we consider rules for the truncation of the double series in Eq.~(\ref{e1}) for the $I_{VM}$-case for which we use bounds on the Jinc functions and on the structural quantities that follow from Debye's asymptotics for Bessel functions. In Sec.~\ref{sec3} we consider the truncation issues associated with the computation of the structural quantities. In Sec.~\ref{sec4} the whole computation scheme and the truncation rules are summarized. In Sec.~\ref{sec5} we illustrate the performance of the truncation rules by plotting actually achieved accuracy and computation times against required accuracy. In Sec.~\ref{sec6} we present our conclusions. In Appendix~A we present basic properties of $\varp$-functions that arise in bounding the (spherical) Bessel and Hankel functions using Debye's asymptotics. The results of Appendix~A are used in Appendix~B and C where we develop bounds on Jinc functions and structural quantities. In Appendix~D we present some proofs concerning the validity of the truncation rules. In Appendix~E we present a number of results containing the computation and asymptotics for the Zernike coefficients of the algebraic factors that occur in the $I_{VM}$-case.

\section{Truncation rules for the double series for $I_{VM}$} \label{sec2}

\subsection{Double series for $I_{VM}$ and truncation strategy} \label{subsec2.1}
\mbox{} \\[-9mm]

We have
\beq \label{e6}
I_{VM}=\sum_{h,t}\,A_{2t,n,h}^{0mm}({-}1)^{\frac{h-m}{2}}\,c_t\,\frac{J_{h+1}(2\pi r)}{2\pi r}
\eq
as in Eq.~(\ref{e1}), where $c_t$ are the Zernike coefficients of the front factor $a(\rho)\,f(\rho)$, with $a(\rho)$ and $f(\rho)$ as in Eqs.~(\ref{e3}--\ref{e4}) so that
\begin{eqnarray} \label{e7}
& \mbox{} & \frac{(1-s_0^2\rho^2)^{1/2}+(1-s_{0,M}^2\rho^2)^{1/2}}{(1-s_0^2\rho^2)^{1/4}\,(1-s_{0,M}^2\rho^2)^{3/4}} \exp\,\Bigl[\frac{if}{u_0}\,(1-\sqrt{1-s_0^2\rho^2})\Bigr] \nonumber \\[3mm]
& & =~\sum_{t=0}^{\infty}\,c_t\,R_{2t}^0(\rho)~.
\end{eqnarray}

Our approach to get truncation rules for the double series uses the following observations. The coefficients $A$ are all non-negative and bounded by 1 and satisfy other boundedness properties such as
\beq \label{e8}
\sum_h\,A_{2t,n,h}^{0mm}=1=\sum_t\,\frac{2t+1}{h+1}\,A_{2t,n,h}^{0mm}~.
\eq
In Subsec.~\ref{subsec2.2} we give bounds on the Jinc functions $J_{h+1}(2\pi r)/2\pi r$ and the coefficients $c_t$ that show rapid decay after $h=2\pi r$ and $t=\frac12\,|f|$, respectively. For values of absolute accuracy $\eps$ that are relevant in the optical practice, the double series in Eq.~(\ref{e6}) is truncated at values $h=H$ and $t=T$ where both the Jinc functions and the coefficients have reached their plunge ranges. Accordingly, the absolute truncation error in approximating $I_{VM}$ in Eq.~(\ref{e6}) by
\beq \label{e9}
\sum_{h+1\leq H,t\leq T}\,A_{2t,n,h}^{0mm}({-}1)^{\frac{h-m}{2}}\,c_t\,\frac{J_{h+1}(2\pi r)}{2\pi r}
\eq
is safely bounded by
\beq \label{e10}
\max_{(h,2t)\in S_n^m\,;\,h+1>H\,{\rm or}\,t>T}\,\Bigl|c_t\, \frac{J_{h+1}(2\pi r)}{2\pi r}\Bigr|~,
\eq
where $S_n^m$ is the set of all $h$, $t$ such that $A_{2t,n,h}^{0mm}\neq0$.

In the general truncation rule, the dependence on $n$ and $m$ of the supporting set $S_n^m$ is totally ignored and the functions bounding Jinc$_{h+1}$ and $c_t$ are replaced by simple functions allowing convenient determination of set points $H$ and $T$ for which
\beq \label{e11}
\max_{h+1>H\,{\rm or}\,t>T}\,\Bigl|c_t\,\frac{J_{h+1}(2\pi r)}{2\pi r}\Bigr|
\eq
is below a specified $\eps>0$.

In the dedicated rule, we use a more careful approximation of the bounding functions, and we include explicitly the supporting set $S_n^m$. It thus appears that an inspection of the product of the approximated bounding functions along the boundary $\partial\,S_n^m$ of the supporting set in the $(h,2t)$-plane produces numbers $H=H_n^m$ and $T=T_n^m$ such that the quantity in Eq.~(\ref{e10}) is below a specified $\eps>0$.

\subsection{Bounding Jinc functions and structural quantities} \label{subsec2.2}
\mbox{} \\[-9mm]

We let for $c>0$ and $x\geq0$
\beq \label{e12}
\varp(x\,;\,c)=\left\{\ba{llll}
0 & \!\!, & ~~~0\leq x\leq c & \!\!, \\[3mm]
x\,{\rm arccosh}(x/c)-c\,\sqrt{(x/c)^2-1} & \!\!, & ~~~x\geq c & \!\!,
\ea\right.
\eq
where ${\rm arccosh}(y)={\rm ln}(y+\sqrt{y^2-1})$. In Appendix~B, the following is shown. Let $r>0$, and set
\beq \label{e13}
R=\max\Bigl(\frac{1}{2\pi}\,,r\Bigr)~.
\eq
Then
\beq \label{e14}
\Bigl|\frac{J_{h+1}(2\pi r)}{2\pi r}\Bigr|\leq\frac{1}{2\pi^2\,R\,\sqrt{R}}\,\exp({-}\varp(h+1\,;\,2\pi R))~.
\eq
The bound in Eq.~(\ref{e14}) is valid for all $h\geq0$, except for a small range of $h$'s near $2\pi r$ with $r\pr\infty$. In fact, Eq.~(\ref{e14}) is valid for all $r\geq0$ and $h\leq 2$, it is valid within a factor of 2 for all $r\geq0$ and all $h\leq 175$, it is valid within a factor of 4 for all $r\geq0$ and all $h\leq 11194$, and so on. Of course, we also have the general bound $|J_{h+1}(2\pi r)/2\pi r|\leq \tfrac12$.

\begin{figure}
\begin{center}
\plaatje{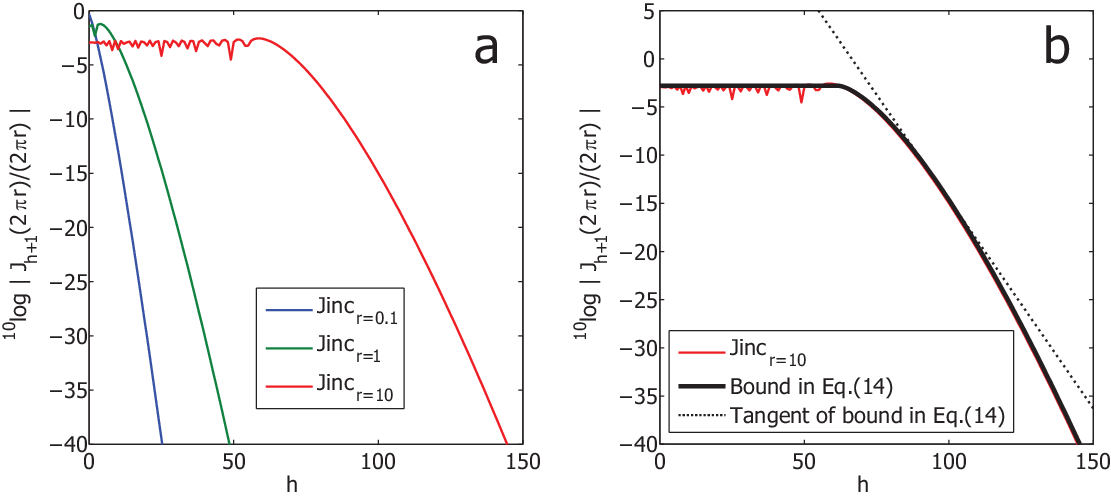}{0.9}
\caption{(a) Plot of $\log_{10}|J_{h+1}(2\pi r)/2\pi r|$ as a function of $h=0,\;1,\cdots,\; 150$ for the case $r=0.1$ (blue), $1$ (green), $10$ (red). (b) Plot of $\log_{10}|J_{h+1}(2\pi r)/2\pi r|$ as a function of $h=0,\;1,\cdots,\; 150$ case $r=10$ (red), together with the $\log_{10}$ of the bound at the right-hand side of Eq.~(\ref{e14}) (solid black) and the tangent line (dashed) corresponding to the right-hand side of Eq.~(\ref{e20}). }
\label{fig1}
\end{center}
\end{figure}

In Figure \ref{fig1}a, we show $\log_{10}|J_{h+1}(2\pi r)/2\pi r|$ as a function of $h,\;0\leq h \leq 150$, for $r=0.1,\;1$ and $10$, respectively. It can be seen that there is rapid decay from $h+1=2\pi r=0.63,\;6.28$ and $62.83$, respectively onwards. For the case that $r=R=10$, we have plotted in Figure \ref{fig1}b both $\log_{10}|J_{h+1}(2\pi r)/2\pi r|$ and the bound $\log_{10}[\,\exp{\{-\varphi(h+1;2\pi R)\}}/2\pi^2 R\sqrt{R}\,]$, see Eq.~(\ref{e14}). The (asymptotic) maximum of $\log_{10}|J_{h+1}(2\pi r)/2\pi r|$ can be found from Appendix \ref{appB} and equals $-2.5609$, assumed at $h = 58.67$ when $r=10$. At this point $h$, the upper bound $\log_{10}[1/2\pi^2 R\sqrt{R}]=-2.7953$ is slightly lower than the asymptotic maximum. We have also shown in Fig.~\ref{fig1}b the linear function $\log_{10}[\,\exp{\{-(h+1-2\pi R \sinh{(1)})\}}/(2\pi^2 R\sqrt{R})\,]=28.8387-0.4343h$ which is a tangent line of the bounding function, see Subsec.~\ref{subsec2.3}.

For the structural quantities $c_t$ a similar result holds. In Appendix~C the following is shown. let $f$ be a real number, and set
\beq \label{e15}
g=\max(1,|f|)~.
\eq
Then
\beq \label{e16}
|c_t|\leq 4w_0\,a_0\,\exp({-}\varp(t\,;\,g/2)+\varp(t\,;\,g/2v_0))~,
\eq
where
\beq \label{e17}
a_0=2\il_0^1\,a(\rho)\,\sqrt{1-s_0^2\rho^2}\,\rho\,d\rho
\eq
is the $R_0^0$-coefficient of $A(\rho)=a(\rho)\,\sqrt{1-s_0^2\rho^2}$, and
\beq \label{e18}
w_0=\frac{1}{1+\sqrt{1-s_0^2}}~,~~~~~~v_0=\frac{1-\sqrt{1-s_0^2}}{1+\sqrt{1-s_0^2}}~.
\eq
Here it has been assumed that $s_0\geq s_{0,M}$. In the case that $s_{0,M}>s_0$, we should replace $s_0$ in Eqs.~(\ref{e17}-\ref{e18}) by $s_{0,M}$ and change the right-hand side of Eq.~(\ref{e16}) accordingly.
The value of $a_0$ is in almost all cases well approximated by
\beq \label{e19}
A(\tfrac12\,\sqrt{2})~~{\rm or}~~\tfrac16\,A(0)+\tfrac23\,A(\tfrac12\,\sqrt{2})+\tfrac16\,A(1)
\eq
(midpoint rule or Simpson rule for integration over $x=\rho^2$). The bound in Eq.~(\ref{e16}) is shown in Appendix~C using a somewhat heuristic approach so as to arrive at manageable expressions. As with the bound in Eq.~(\ref{e14}) there are small exceptional ranges of $t$ near $\tfrac12\,g$ and $g\pr\infty$, where Eq.~(\ref{e16}) holds safe for a factor that grows to infinity very slowly as $g\pr\infty$.

\begin{figure}
\begin{center}
\plaatje{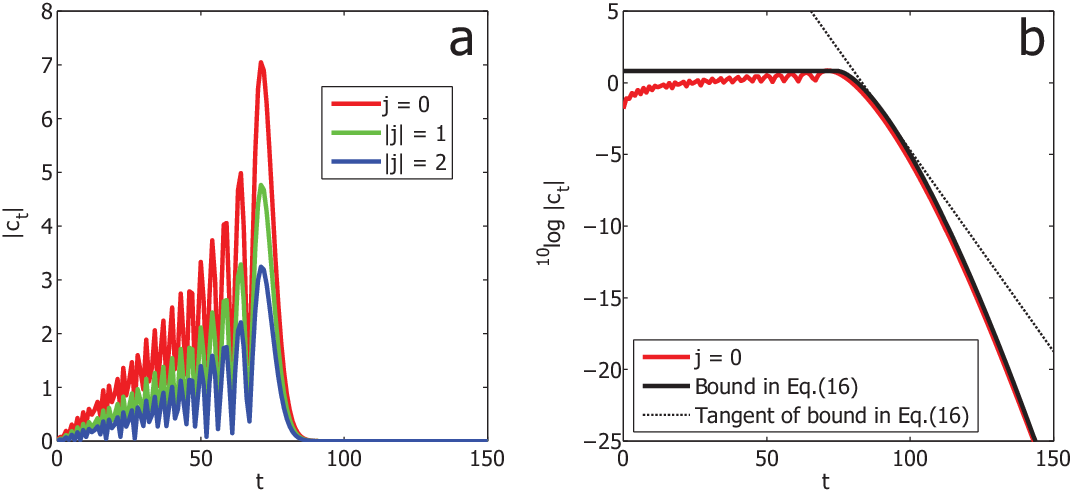}{0.9}
\caption{(a) Plot of $\log_{10}|c_t|$ as a function of $t=0,\;1,\cdots,\;150$, for $f=150$, $s_0=0.95$, $s_{0,M}=0.50$, where $c_t$ are the Zernike coefficients of the front factors that occur in accordance with \cite{ref1}, Eq.~(30) for $|j|=0$ (red), $1$ (green), $2$ (blue) and of which $c_t$ in Eq.~(\ref{e7}) gives the case $|j|=0$. (b) Plot of $\log_{10}|c_t|$ as in (a) for the case $|j|=0$ (red), together with the $\log_{10}$ of the bound at the right-hand side of Eq.~(\ref{e16}) (solid black) and the tangent line (dashed) corresponding to the right-hand side of Eq.(\ref{e21}).}     
\label{fig2}
\end{center}
\end{figure}

In Figure \ref{fig2}a, we show $|c_t|$ as a function of $t,\;0\leq t\leq 150$, for $f=150$, $s_0=0.95$ and $s_{0,M}=0.50$, with $j=0,\;1,\;2$ determining the precise form of the algebraic function in the vectorial setting according to \cite{ref1}, Eq.~(30). It can be seen that the graphs for these three cases are qualitatively the same, except for an overall amplitude factor that is related to the $R^0_0$-coefficient $a_0$ of $a(\rho)\sqrt{1-s_0^2\rho^2}$. There is rapid decay from $t=\tfrac{1}{2}f=75$ onwards. For the case $j=0$, we have plotted in Figure \ref{fig2}b both $\log_{10}|c_t|$ and the bound $\log_{10}[4w_0a_0\exp{(-\varphi(t;g/2)+\varphi(t;g/2v_0))}]$, see Eq.~(\ref{e16}). The (asymptotic) maximum of $\log_{10}|c_t|$ occurs somewhat before $t=75$ and exceeds the value $\log_{10}[4w_0a_0]$ obtained from the bounding function somewhat. We also show in Figure \ref{fig2}b the linear function $\log_{10}[4w_0a_0\exp{(\tfrac12g\sinh(\gamma_0)-\gamma_0t)}]=23.1718-0.2806t$, where $\gamma_0=\ln{(1/v_0)}=0.6461$, which is a tangent line of the bounding function, see Subsec.~\ref{subsec2.4}.

\begin{figure}
\begin{center}
\plaatje{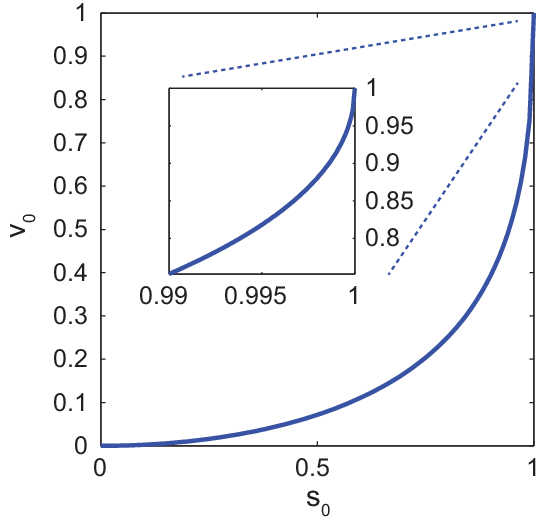}{0.5}
\caption{Graph of $v_0=(1-\sqrt{1-s_0^2})/(1+\sqrt{1-s_0^2})$ as a function of $s_0,\;0\leq s_0 \leq 1$.}
\label{fig3}
\end{center}
\end{figure}

In Figure \ref{fig3}, we show the graph of $v_0$, as given in Eq.~(\ref{e18}), against $s_0,\;0\leq s_0\leq 1$. The asymptotic decay of $c_t$ is $Cv_0^t$, and so there is rapid decay of $c_t$ for all $s_0$ until $s_0=0.95$ (with $v_0=0.5241$), and even cases like $s_0=0.99$ are still practicable.


\subsection{General truncation rule} \label{subsec2.3}
\mbox{} \\[-9mm]

In Appendix~A the functions $\varp(h+1\,;\,2\pi R)$ and $\varp(t\,;\,g/2)-\varp(t\,;\,g/2v_0)$ are bounded from below by piecewise linear functions according to
\beq \label{e20}
\varp(h+1\,;\,2\pi R)\geq\max(0,h+1-2\pi R\sinh(1))~,
\eq
and
\beq \label{e21}
\varp(t\,;\,g/2)-\varp(t\,;\,g/2v_0)\geq\max(0,\gamma t-\tfrac12\,g\sinh(\gamma))~,
\eq
where
\beq \label{e22}
\gamma=\min(1,{\rm ln}(1/v_0))~,
\eq
respectively. This leads to the following general truncation rule: Let $0<\eps<1$, and let
\beq \label{e23}
B=\max\Bigl(0,{\rm ln}\bigl(\frac{2w_0a_0}{\pi^2\,\eps\,R\,\sqrt{R}} \Bigr)\Bigr)~.
\eq
Then the quantity in Eq.~(\ref{e11}) is less than $\eps$ when
\beq \label{e24}
T=T^{\textrm{gen}}=\frac{1}{\gamma}\,B+\tfrac12\,g\,\frac{\sinh(\gamma)}{\gamma}~,~~~~~~H=H^{\textrm{gen}}=B+2\pi R\sinh(1)~.
\eq
See Appendix \ref{appD} for a proof.

By observing that we can write $T$ and $H$ in Eq.~(\ref{e24}) as
\beq \label{e25}
T=\tfrac12\,g+\frac{1}{\gamma}\,B+\tfrac12\,g\,\frac{\sinh(\gamma)-\gamma}{\gamma} ~,~~~~~~H=2\pi R+B+2\pi R(\sinh(1)-1)~,
\eq
where for $0<\gamma\leq1$
\beq \label{e26}
0<\frac{\sinh(\gamma)-\gamma}{\gamma}\leq\sinh(1)-1=0.1752~,
\eq
we have given precision to the rule-of-thumb that the truncation points should be chosen somewhat larger than $\tfrac{1}{2}|f|$ and $2\pi r$, respectively.

\subsection{Dedicated truncation rule} \label{subsec2.4}
\mbox{} \\[-9mm]

We now present a truncation rule that takes into account the $(n,m)$-dependence of the supporting set $S_n^m$ of the $A$'s in Eq.~(\ref{e6}). We also use better approximations for the functions $\varp(h+1\,;\,2\pi R)$ and $\varp(t\,;\,g/2)-\varp(t\,;\,g/2v_0)$ on the left-hand sides of Eqs.~(\ref{e20}--\ref{e21}). Thus we consider
\beq \label{e27}
F(h,t)=\varp(h+1\,;\,2\pi R)+\varp(t\,;\,g/2,g/2v_0)~,
\eq
where
\beq \label{e28}
\varp(t\,;\,g/2,g/2v_0)=\left\{\ba{llll}
\varp(t\,;\,g/2) & \!\!, & ~~~0\leq t\leq\tfrac12\,g\cosh(\gamma_0) & \!\!, \\[3mm]
\gamma_0t-\tfrac12\,g\sinh(\gamma_0) & \!\!, & ~~~t\geq\tfrac12\,g\cosh(\gamma_0) & \!\!,
\ea\right.
\eq
with $\gamma_0={\rm ln}(1/v_0)$. The function $\varp(t\,;\,g/2,g/2v_0)$ is the largest convex function bounding $\varp(t\,;\,g/2)-\varp(t\,;\,g/2v_0)$, which is convex in $t\leq g/2$ but concave in $t\geq g/2v_0$, from below. The function $\varp(h+1\,;\,2\pi R)$ is convex in $h\geq0$. See Appendix \ref{appA}.

\begin{figure}
\begin{center}
\plaatje{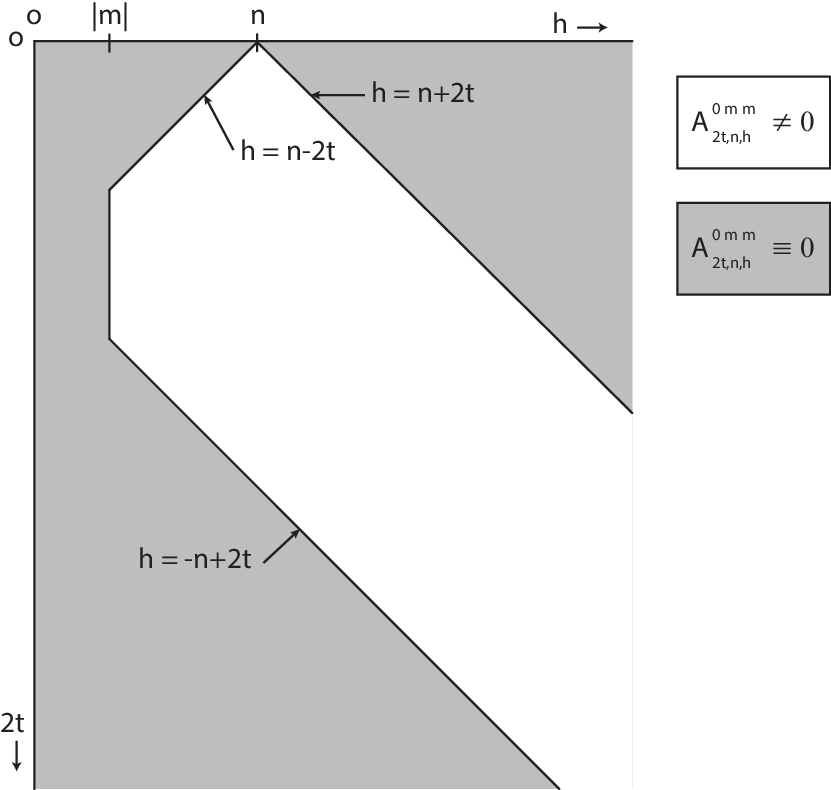}{0.6}
\caption{For given integers $n$ and $m$ with $n-|m|$ even and non-negative, the unshaded set $h\geq|m|$, $|h-n|\leq2t\leq h+n$ contains all points $(h,2t)$ with non-negative integer $h$ and $t$ such that $A^{0mm}_{2t,n,h}\neq 0$.}
\label{fig4}
\end{center}
\end{figure}

In Figure \ref{fig4} we depict, for given $n$ and $m$ such that $n-|m|$ is even an non-negative, the set $S_n^m$ in the $(h,2t)$-plane that contains all non-zero coefficients $A_{2t,n,h}^{0mm}$ ($S_n^m$ is the convex hull of those points $(h,2t)$). The boundary $\partial\,S_n^m$ of $S_n^m$ consists of 4 line segments I, II, III, IV in accordance with the conditions, see \cite{ref1}, Sec.~5,
\beq \label{e29}
h\geq|m|~,~~~~~~|h-n|\leq 2t \leq h+n~.
\eq
We consider the function $F(h,t)$ of Eq.~(\ref{e27}) along $\partial\,S_n^m$ with continuous $t\geq0$, $h\geq0$. We have that $F(h,t)$ is non-negative and increasing and convex in both $h$ and $t$, and
\beq \label{e30}
\Bigl|c_t\,\frac{J_{h+1}(2\pi r)}{2\pi r}\Bigr|\leq\frac{2w_0a_0}{\pi^2\,R\,\sqrt{R}}\,\exp({-}F(h,t))~.
\eq
We let $B$ as in Subsec.~\ref{subsec2.3}, and we let
\beq \label{e31}
M=\min\,\{F(h,t)\,|\,(h,2t)\in\partial\,S_n^m,~h+1\leq H^{\textrm{gen}},~t\leq T^{\textrm{gen}}\}
\eq
with $H^{\textrm{gen}}$ and $T^{\textrm{gen}}$ from Subsec~\ref{subsec2.3}. From the monotonicity and convexity properties of $F$, we then get, see Appendix~\ref{appD},
\bi{--0}
\ITEM{--} when $M>B$, we have that
\beq \label{e32}
\max_{(h,2t)\in S_n^m}\,\Bigl|c_t\,\frac{J_{h+1}(2\pi r)}{2\pi r}\Bigr|<\eps~,
\eq
\ITEM{--} when $M\leq B$, there are two points $(h_1,2t_1)$ and $(h_2,2t_2)\in\partial\,S_n^m$ such that for any $(h,2t)\in S_n^m$
\beq \label{e33}
h\geq\max(h_1,h_2)~~{\rm or}~~t\geq\max(t_1,t_2)\Rightarrow F(h,t)\geq B~.
\eq
\ei
The dedicated truncation rule becomes then as follows. Determine $M$ in Eq.~(\ref{e31}). When $M>B$, we set $H=H_n^m=1$, $T=T_n^m=0$. When $M\leq B$, we search the boundary $\partial\,S_n^m$, as long as contained in the box $h+1\leq H^{\textrm{gen}}\; \&\; t\leq T^{\textrm{gen}}$, for the two points $(h_1,2t_1)$ and $(h_2,2t_2)$ satisfying Eq.~(\ref{e33}), and we set $H=H_n^m=\max(h_1,h_2)+1$, $T=T_n^m=\max(t_1,t_2)$. With $H$ and $T$ defined this way, we have that the quantity in Eq.~(\ref{e10}) is less than $\eps$.

By the monotonicity and convexity properties of $F$, the minimum $M$ of $F$ along $\partial\,S_n^m$ is assumed on edge II. Hence, it is sufficient to inspect $F$ along this edge to find $M$.

The actual variables $h$, $t$ are non-negative integer, and this should be accounted for. We intersect $\partial\,S_n^m$ with the box $(h,2t)$, $h\leq \hat{H}-1$ or $t\leq\hat{T}$, where $\hat{H}-1$ is the smallest integer of same parity as $n$ with $\hat{H}\geq H^{\textrm{gen}}$ and $\hat{T}$ is the smallest integer with $\hat{T}\geq T^{\textrm{gen}}$. In case that we find 0 or 1 point $(h,2t)$ in the intersection, the inspection is a trivial matter. In the case that we find two intersection points, we let the inspection start at the point with largest value of $h$ and lowest values of $2t$, and we end the inspection at or before the point with lowest value of $h$ and largest value of $2t$, following the boundary curve counterclockwise with points $(h,2t)$, integer $h$ and $t$ and $h$ same parity as $n$.

\section{Computation of structural quantities and truncation issues} \label{sec3}

\subsection{Series expressions for structural quantities} \label{subsec3.1}
\mbox{} \\[-9mm]

We consider in this section computation of the Zernike coefficients of the front factor $a(\rho)\,f(\rho)$, with $a(\rho)$ and $f(\rho)$ given in Eqs.~(\ref{e3}--\ref{e4}). We make a slight variation of the approach in \cite{ref1}, Sec.~4 and 8, in that we write
\beq \label{e34}
a(\rho)\,\sqrt{1-s_0^2\rho^2}=\sum_{l=0}^{\infty}\,a_l\,R_{2l}^0(\rho)~,
\eq
\beq \label{e35}
f(\rho)/\sqrt{1-s_0^2\rho^2}=\sum_{k=0}^{\infty}\,b_k\,R_{2k}^0(\rho)~,
\eq
and we use linearization coefficients $A$ to write
\beq \label{e36}
a(\rho)\,f(\rho)=\sum_{t=0}^{\infty}\,c_t\,R_{2t}^0(\rho)~,
\eq
where
\beq \label{e37}
c_t=\sum_{l,k=0}^{\infty}\,A_{2l,2k,2t}^{000}\,a_l\,b_k~.
\eq

The reason for moving a factor $\sqrt{1-s_0^2\rho^2}$ from the focal factor $f(\rho)$ to the algebraic factor $a(\rho)$ is the fact that this yields the most convenient expression for the expansion coefficients $b_k$, viz.
\beq \label{e38}
b_k=\frac{1}{iu_0}\,\exp\,[if/u_0]\,(2k+1)\,j_k(f/2)\,h_k^{(2)}(f/2v_0)~.
\eq
Here $j_k$ and $h_k^{(2)}$ are the spherical Bessel and Hankel functions of order $k$, given as
\begin{eqnarray} \label{e39}
j_k(z)&=&\sqrt{\frac{\pi}{2z}}\,J_{k+1/2}(z)~,\\[3mm]
h_k(z) & = & j_k(z)-i\,y_k(z) \nonumber \\[3mm]
& = & \sqrt{\frac{\pi}{2z}}\,(J_{k+1/2}(z)-i\,Y_{k+1/2}(z)) \nonumber \\[3mm]
& = & \sqrt{\frac{\pi}{2z}}\,H_{k+1/2}^{(2)}(z)~,\label{e40}
\end{eqnarray}
with $J_{\nu}$, $Y_{\nu}$ and $H_{\nu}^{(2)}$ the Bessel function of first, second and third kind (Hankel function) and of order $\nu$, see \cite{ref2}, Ch.~10. The quantities $b_k$ can be computed, via Eqs.~(\ref{e39}--\ref{e40}) using MatLab routines, efficiently at any desired accuracy.

As to the coefficients $a_l$, we first write, see Eq.~(\ref{e3}),
\begin{eqnarray} \label{e41}
a(\rho)\,\sqrt{1-s_0^2\rho^2} & = & (1-s_0^2\rho^2)^{3/4}\,(1-s_{0,M}^2\rho^2)^{-3/4} \nonumber \\[3mm]
& & +~(1-s_0^2\rho^2)^{1/4}\,(1-s_{0,M}^2\rho^2)^{-1/4}~.
\end{eqnarray}
Next, either term on the right-hand side of Eq.~(\ref{e41}) is developed into a power series
\beq \label{e42}
a_{\alpha\beta}(\rho)=(1-s_{\alpha}^2\rho^2)^{\alpha}\,(1-s_{\beta}^2\rho^2)^{\beta}=\sum_{N=0}^{\infty} \,r_N\rho^{2N}~,
\eq
where the coefficients $r_N$ are computed recursively according to \cite{ref1}, Eqs.~(37--39) and \cite{ref1}, Eq.~(106). Finally, the Zernike coefficients $a_{l,\alpha\beta}$ are computed from $r_N$ according to
\beq \label{e43}
a_{l,\alpha\beta}=\sum_{N=l}^{\infty}\,b_N(l)\,r_N~,~~~~~~l=0,1,...~,
\eq
with $b_N(l)$ given explicitly and computed recursively in \cite{ref1}, Eqs.~(41--44).

\subsection{Truncation and accuracy issues}\label{subsec3.2}
\mbox{} \\[-9mm]

The accuracy by which the $c_t$ must be computed is dictated by the absolute accuracy $\eps$ in the truncation analysis of Sec.~\ref{sec2} that involves the products of $c_t$'s and Jinc functions $J_{h+1}(2\pi r)/2\pi r$ as in Eqs.~(\ref{e10}--\ref{e11}). Now $|J_{h+1}(z)/z|\leq 1/2$ for $z\geq0$. Hence, when $c_t$ is computed with absolute accuracy $\eps$, and the truncation rules of Subsecs.~\ref{subsec2.3}--\ref{subsec2.4} are used with $\eps/2$ instead of $\eps$, a final absolute accuracy better than $\eps$ results.

Next, given integers $L,K>0$, the absolute error due to approximating $c_t$ of Eq.~(\ref{e37}) by
\beq \label{e44}
c_{t,LK}=\sum_{l=0}^L\,\sum_{k=0}^K\,A_{2l,2k,2t}^{000}\,a_l\,b_k
\eq
is, as in Eqs.~(\ref{e9}--\ref{e10}), safely bounded by
\beq \label{e45}
\max_{l>L\,{\rm or}\,k>K}\,|a_lb_k|~.
\eq
Now there are the bounds
\beq \label{e46}
|a_l|\leq\tfrac{16}{3}\,,~~|b_k|\leq 4~,~~~~~~l,k=0,1,...~.
\eq
The second bound in Eq.~(\ref{e46}) follows from Appendix~\ref{appC}, Eq.~(\ref{c18}), while the first bound is obtained by considering in Appendix~\ref{appE}, Eq.~(\ref{E1}) the worst case $l=0$ with $s_0=0$ and $s_{0,M}$ close to $1$.
Hence, when $\eps\in(0,1)$, we have that the quantity in Eq.~(\ref{e45}) is less than $\eps$ when $L$ and $K$ are such that
\beq \label{e47}
l>L\Rightarrow|a_l|<\tfrac14\,\eps~~~\&~~~k>K\Rightarrow|b_k|<\tfrac{3}{16}\,\eps~.
\eq

According to Appendix~C we have
\beq \label{e48}
|b_k|\leq 4\,\exp({-}\varp(k\,;\,g/2)+\varp(k\,;\,g/2v_0))~,
\eq
and this is less than $\frac{3}{16}\,\eps$ when
\beq \label{e49}
k>\frac{1}{\gamma}\,\max\Bigl(0,{\rm ln}\Bigl(\frac{64}{3\eps}\Bigr)\Bigr)+\tfrac12\,g\,\frac{\sinh(\gamma)}{\gamma}~,
\eq
with $\gamma$ as in Eq.~(\ref{e22}).

The quantities $b_k$ are computed using Eq.~(\ref{e38}), involving the spherical Bessel and Hankel functions $j_k$ and $h_k^{(2)}$ that can be computed using Matlab routines. From Appendix~\ref{appC} we have that
\beq\label{e49i}
|j_k(f/2)|\leq \tfrac{2}{g}\;,\;\; |h_k(f/2v_0)|\leq \frac{2^{7/4}v_0}{g}\exp(\varphi(k;g/2v_0))\,,
\eq
where the first inequality holds for all $f$ and the second inequality only holds when $|f/v_0|\geq 1$. In the case that $|f/v_0|<1$, the $b_k$ of Eq.~(\ref{e38}) is best evaluated using the power series representations of $j_k$ and $h^{92)}_k$ that follow from \cite{ref2}, 10.53.
Thus it follows that $b_k$ is computed with absolute accuracy $3\eps/16$ for $k=0,\;1,\;\cdots,\;K$ when $j_k(f/2)$ and $h_k^{(2)}(f/2v_0)$ are computed with absolute accuracy
\beq\label{e49ii}
\frac{3\eps}{32}\cdot\frac{u_0\exp(-\varphi(K;g/2v_0))}{2^{7/4}(2K+1)v_0} \;\;\;\textrm{and}\;\;\;\frac{3\eps}{32}\cdot\frac{u_0}{2(2K+1)}\;
\eq
respectively.

As to the first condition in Eq.~(\ref{e47}), we consider the decomposition of $a(\rho)\,\sqrt{1-s_0^2\rho^2}$ in terms $a_{\alpha\beta}(\rho)$ as in Eq.~(\ref{e42}) with $\alpha+\beta=0$ and Zernike coefficients $a_{l,\alpha\beta}$ as in Eq.~(\ref{e43}). In Appendix~E the following is shown. Let $\delta=|\alpha|=|\beta|$, and let $S=\max(s_{\alpha},s_{\beta})$. Denoting ``the $R_{2l}^0$-coefficient of $A(\rho)$'' by $Z\,C_l[A(\rho)]$, we have
\beq \label{e50}
|a_{l,\alpha\beta}|\leq Z\,C_l[(1-S^2\rho^2)^{-\delta}]\sim\frac{E\,V^l}{(l+1)^{-\delta+1/2}}~,
\eq
where
\beq \label{e51}
E=\frac{2\sqrt{\pi}}{\Gamma(\delta)}~\frac{(1-S^2)^{-\tfrac12\delta+\tfrac14}}{1+\sqrt{1-S^2}}~,~~~~~~ V=\frac{1-\sqrt{1-S^2}}{1+\sqrt{1-S^2}}~.
\eq
Furthermore, the right-hand side of Eq.~(\ref{e50}) is less than $\eta:=\eps/8$ when
\beq \label{e52}
l\geq\frac{{\rm ln}(E\eta^{-1})-({-}\delta+1/2)\,{\rm ln}(1+{\rm ln}(E\eta^{-1})/{\rm ln}(1/V))} {{\rm ln}(1/V)}~.
\eq
Therefore, the first condition in Eq.~(\ref{e47}) is satisfied when $L$ is the maximum of the two numbers that occur at the right-hand side of Eq.~(\ref{e52}) for the choices $\delta=3/4,1/4$ (where evidently $\delta=3/4$ yields the largest value of the two).

We finally address the issue of truncating the series in Eq.~(\ref{e43}). It is shown in Appendix~E that for a given $\eps>0$ and an integer $L>0$ such that all $|a_{l,\alpha\beta}|<\tfrac18\,\eps$ when $l>L$, we have that all numbers $a_{l,\alpha\beta}$, $l=0,1,...,L\,$, are computed with absolute accuracy $\eps/16$ when the infinite series in Eq.~(\ref{e43}) is truncated at $N=2L/\sqrt{1-S^2}$.

\begin{figure}
\begin{center}
\plaatje{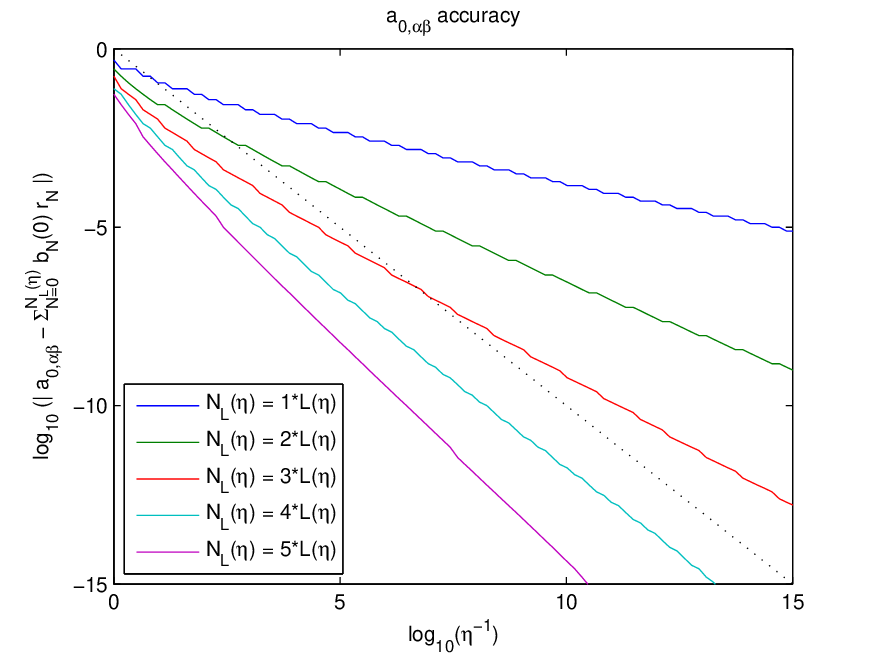}{0.8}
\caption{Plot of $\log_{10}|a_{0,\alpha\beta}-\sum_{N=0}^{N_L(\eta)} b_N(0)r_N|$ as a function of $\log_{10}\eta^{-1} \in [0,15]$, for the case that $a_{0,\alpha\beta}$ is the $R_0^0$-coefficient of $a_{\alpha\beta}(\rho)=(1-s_0^2\rho^2)^\alpha (1-s_{0,M}^2\rho^2)^\beta$ with $\alpha=-\beta=3/4$ and $s_0=0.50,\;s_{0,M}=0.90$. The colored solid lines represent different summation limits $N_L(\eta)=L(\eta),\;2L(\eta),\;4L(\eta),\;5L(\eta)$, respectively, with $L(\eta)$ given by the right-hand side of Eq.~(\ref{e52}). The black (dotted) curve indicates those positions at which the observed accuracy of $a_{0,\alpha\beta}$ is equal to $\eta$.}
\label{fig5}
\end{center}
\end{figure}

In Figure \ref{fig5}, we show $\log_{10}|a_{0,\alpha\beta}-\sum_{N=0}^{N_L(\eta)} b_N(0)r_N|$ as a function of $\eta$ with $\log_{10}\eta^{-1} \in [0,15]$, for the case that $a_{0,\alpha\beta}$ is the $R_0^0$-coefficient of $a_{\alpha\beta}(\rho)=(1-s_0^2\rho^2)^\alpha (1-s_{0,M}^2\rho^2)^\beta$ with $\alpha=-\beta=3/4$ and $s_0=0.50,\;s_{0,M}=0.90$ and upper summation limit $N_L(\eta)=L(\eta),\;2L(\eta),\;4L(\eta),\;5L(\eta)$, respectively, with $L(\eta)$ the right-hand side of Eq.~(\ref{e52}).

To summarize, for $\eps \in (0,1)$ we replace $c_t$ by $c_{t,LK}$ given in Eq.~(\ref{e44}) in which
\begin{itemize}
\item[-] $L$ and $K$ are given by the right-hand sides of Eq.~(\ref{e52}) and Eq.~(\ref{e47}), respectively,
\item[-] $b_k$ is as in Eq.~(\ref{e38}) for which $j_k(f/2)$ and $h_k^{(2)}(f/2v_0)$ are computed with absolute accuracy as specified in Eq.~(\ref{e49ii}),
\item[-] $a_l=a_{3/4,-3/4,l}+a_{1/4,-1/4,l}$ and the two $a_{\alpha,\beta,l}$ are computed by summing the series in Eq.~(\ref{e43}) until $N=2L/\sqrt{1-S^2}$ with $S=\max{(s_0,s_{0,M})}$.
\end{itemize}
This results into an absolute error in $c_t$ bounded by $\eps+\tfrac{1}{2}\eps+\tfrac14\eps=\tfrac74 \eps$, due to respectively, truncating the double series over $l$ and $k$, approximating $b_k$ by computing $j_k$ and $h_k^{(2)}$ using the Matlab-code, and approximating $a_l$ by truncating the series for the two $a_{\alpha,\beta,l}$.

\section{Summary of the computation scheme and truncation rules} \label{sec4}
\mbox{} \\[-9mm]

For integer $n$ and $m$ such that $n-|m|$ is even and non-negative, consider
\beq \label{e53}
I=I_{n,VM}^m=\il_0^1\,a(\rho)\,f(\rho)\,p(\rho)\,b(\rho)\,\rho\,d\rho~,
\eq
where
\beq \label{e54}
a(\rho)=\frac{(1-s_0^2\rho^2)^{1/2}+(1-s_{0,M}^2\rho^2)^{1/2}} {(1-s_0^2\rho^2)^{1/4}\,(1-s_{0,M}^2\rho^2)^{3/4}}~,
\eq
\beq \label{e55}
f(\rho)=\exp\,\Bigl[\frac{if}{u_0}\,(1-\sqrt{1-s_0^2\rho^2})\Bigr]~,
\eq
\beq \label{e56}
p(\rho)=R_n^{|m|}(\rho)~,~~~~~~b(\rho)=J_m(2\pi r\rho)
\eq
with given real $f$, $r>0$ and $s_0,s_{0,M}\in[0,1)$, and where $u_0=1-\sqrt{1-s_0^2}$. There is the double series representation
\beq \label{e57}
I=\sum_{h,t}\,A_{2t,n,h}^{0mm}({-}1)^{\frac{h-m}{2}}\,c_t\,\frac{J_{h+1}(2\pi r)}{2\pi r}
\eq
with summation over $h$, $t=0,1,...$ and $h$ same parity as $n$ and $m$. In Eq.~(\ref{e57}), we have
\beq \label{e58}
A_{2t,n,h}^{0mm}=(h+1)\,\left|\Bigl(\ba{rlc}
t & \tfrac12\,n & ~\tfrac12\,h \\[2mm]
0 & \tfrac12\,m & -\tfrac12\,m
\ea\Bigr)\right|^2
\eq
in terms of the Clebsch-Gordan coefficients in $|~|^2$ of \cite{ref2}, Chap.~34; the $A$'s are considered in detail in \cite{ref1}, Sec.~5 and Appendix~C. Furthermore, the $c_t$ are the Zernike coefficients of the front factor $a(\rho)\,f(\rho)$, so that
\beq \label{e59}
a(\rho)\,f(\rho)=\sum_{t=0}^{\infty}\,c_t\,R_{2t}^0(\rho)~.
\eq
The $c_t$ have a double series representation
\beq \label{e60}
c_t=\sum_{l,k=0}^{\infty}\,A_{2l,2k,2t}^{000}\,a_l\,b_k~,
\eq
where the $a_l$ are the Zernike coefficients of $A(\rho)=a(\rho)\,\sqrt{1-s_0^2\rho^2}$, so that
\beq \label{e61}
A(\rho)=a(\rho)\,\sqrt{1-s_0^2\rho^2}=\sum_{l=0}^{\infty}\,a_l\,R_{2l}^0(\rho)~,
\eq
the $b_k$ are the Zernike coefficients of $f(\rho)/\sqrt{1-s_0^2\rho^2}$, so that
\beq \label{e62}
f(\rho)/\sqrt{1-s_0^2\rho^2}=\sum_{k=0}^{\infty}\,b_k\,R_{2k}^0(\rho)~,
\eq
and the $A_{2l,2k,2t}^{000}$ are related to Clebsch-Gordan coefficients as in Eq.~(\ref{e58}). The $b_k$ are given as
\beq \label{e63}
b_k=\frac{1}{iu_0}\,\exp\,[if/u_0]\,(2k+1)\,f\,j_k(f/2)\,h_k^{(2)}(f/2v_0)~,
\eq
with $j_k$ and $h_k^{(2)}$ spherical Bessel and Hankel functions, see \cite{ref2}, Chap.~10, Sec.~10.4.7 and
\beq \label{e64}
v_0=\frac{1-\sqrt{1-s_0^2}}{1+\sqrt{1-s_0^2}}~.
\eq
The $a_l$ are computed by first writing
\begin{eqnarray} \label{e65}
a(\rho)(1-s_0^2\rho^2)^{1/2} & = & (1-s_0^2\rho^2)^{3/4}\,(1-s_{0,M}^2\rho^2)^{-3/4} \nonumber \\[3mm]
& & +~(1-s_0^2\rho^2)^{1/4}\,(1-s_{0,M}^2\rho^2)^{-1/4}~,
\end{eqnarray}
and then expanding both terms $a_{\alpha\beta}(\rho)=(1-s_{\alpha}^2\rho^2)^{\alpha}(1-s_{\beta}^2\rho^2)^{\beta}$ at the right-hand side of Eq.~(\ref{e65}) into a power series and subsequently into a Zernike series according to
\beq \label{e66}
a_{\alpha\beta}(\rho)=(1-s_{\alpha}^2\rho^2)^{\alpha}\, (1-s_{\beta}^2\rho^2)^{\beta}=\sum_{N=0}^{\infty}\, r_{N,\alpha\beta}\,\rho^{2N}=\sum_{l=0}^{\infty}\,a_{l,\alpha\beta}\, R_{2l}^0(\rho)~.
\eq
The $r_{N,\alpha\beta}$ in Eq.~(\ref{e66}) are computed recursively according to
\begin{eqnarray} \label{e67}
r_{-1}=0\,,~~r_0=1~;~~~~~~r_{N+1} & = & \frac{1}{N+1}\,[((N-\alpha)\,s_{\alpha}^2+(N-\beta)\,s_{\beta}^2)\,r_N \nonumber \\[3mm]
& & -~(N-1-\alpha-\beta)\,s_{\alpha}^2\,s_{\beta}^2\,r_{N-1}]
\end{eqnarray}
for $N=0,1,...\,$. The $a_{l,\alpha\beta}$ are computed from the $r_{N,\alpha\beta}$ according to
\beq \label{e68}
a_{l,\alpha\beta}=\sum_{N=l}^{\infty}\,b_N(l)\,r_{N,\alpha\beta}~,~~~~~~l=0,1,...~,
\eq
where the $b_N(l)$ are given by
\beq \label{e69}
b_N(l)=\frac{2l+1}{l+1}\,\Bigl(\!\!\ba{c} N \\ l \ea\!\!\Bigr) \Bigl/\Bigl(\!\!\ba{c} N+l+1 \\ N \ea\!\!\Bigr)~,
\eq
and can be computed recursively according to \cite{ref1}, Eqs.~(42--44).

\subsection{Truncating the double series for $I$} \label{subsec4.1}
\mbox{} \\[-9mm]

We consider replacing the double series for $I$ in Eq.~(\ref{e57}) by
\beq \label{e70}
\sum_{h+1\leq H,\,t\leq T}\,A_{2t,n,h}^{0mm}({-}1)^{\frac{h-m}{2}}\,c_t\, \frac{J_{h+1}(2\pi r)}{2\pi r}~,
\eq
where $H$ and $T$ are to be chosen such that the absolute approximation error is less than $\eps\in(0,1)$. Let $R=\max(1/2\pi,r)$, and let $g=\max(1,|f|)$. Furthermore, let
\beq \label{e71}
B=\max\Bigl(0,{\rm ln}\Bigl(\frac{2w_0a_0}{\pi^2\,\eps\,R\,\sqrt{R}}\Bigr)\Bigr)~,
\eq
where $w_0=(1+\sqrt{1-s_0^2})^{-1}$ and $a_0$ is the $R_0^0$-coefficient in Eq.~(\ref{e61}) so that
\beq \label{e72}
a_0=2\il_0^1\,a(\rho)\,\sqrt{1-s_0^2\rho^2}\,\rho\,d\rho~.
\eq
In Eq.~(\ref{e71}) and in the definitions of $v_0$ in Eq.~(\ref{e64}) and of $w_0$ above, we need to replace $s_0$ by $s_{0,M}$ when $s_{0,M}>s_0$.

\subsubsection{General truncation rule} \label{subsubsec4.1.1}
\mbox{} \\[-9mm]

The absolute approximation error is less than $\eps$, simultaneously for all $n$ and $m$, when
\beq \label{e73}
H=H^\textrm{gen}=B+2\pi\,R\,\sinh(1)~,~~~~~~T=T^\textrm{gen}=\frac{1}{\gamma}\,B+\tfrac12\,g\,\frac{\sinh(\gamma)}{\gamma}~,
\eq
where $\gamma=\min(1,{\rm ln}(1/v_0))$.

\subsubsection{Dedicated truncation rule} \label{subsubsec4.1.2}
\mbox{} \\[-9mm]

For $c>0$ and $x\geq0$, define
\beq \label{e74}
\varp(x\,;\,c)=\left\{\ba{llll}
0 & \!\!, & ~~~0\leq x\leq c & \!\!, \\[3mm]
x\,{\rm arccosh}(x/c)-c\,\sqrt{(x/c)^2-1} & \!\!, & ~~~x\geq c & \!\!,
\ea\right.
\eq
and let for $h\geq0$ and $t\geq0$
\beq \label{e75}
F(h,t)=\varp(h+1\,;\,2\pi R)+\varp(t\,;\,g/2,g/2v_0)~,
\eq
where for $t\geq0$
\beq \label{e76}
\varp(t\,;\,g/2,g/2v_0)=\left\{\ba{llll}
\varp(t\,;\,g/2) & \!\!, & ~~~0\leq t\leq\tfrac12\,g\cosh(\gamma_0) & \!\!, \\[3mm]
\gamma_0t-\tfrac12\,g\sinh(\gamma_0) & \!\!, & ~~~t\geq\tfrac12\,g\cosh(\gamma_0) & \!\!,
\ea\right.
\eq
with $\gamma_0={\rm ln}(1/v_0)$ and $v_0$ given in Eq.~(\ref{e64}).

Let $n$ and $m$ be integers such that $n-|m|$ is even and non-negative. The set $S_n^m$ in the $(h,2t)$-plane containing all non-zero coefficients $A_{2t,n,h}^{0mm}$ in the double series in Eq.~(\ref{e57}) is given by the constraints
\beq \label{e76a}
h\geq|m|~,~~~~~~|h-n|\leq 2t \leq h+n~,~~~~~~h-n~{\rm even}~.
\eq
The convex hull of this set $S_n^m$ has a boundary $\partial\,S_n^m$ which is a curve consisting of 4, possibly degenerate, line segments, listed in counterclockwise order as
\bi{III.0}
\ITEM{I.} $h=n+2t$, $t\geq0$,
\ITEM{II.} $h=n-2t$, $0\leq t\leq\tfrac12\,(n-|m|)$,
\ITEM{III.} $h=|m|$, $\tfrac12\,(n-|m|)\leq t\leq\tfrac12\,(n+|m|)$,
\ITEM{IV.} $h={-}n+2t$, $t\geq\tfrac12\,(n+|m|)$.
\ei
Let
\beq \label{e77}
M=\min\,\{F(h,t)\,|\,(h,2t)\in\partial\,S_n^m,~0\leq h\leq H^\textrm{gen},~0\leq t\leq T^\textrm{gen}\}~,
\eq
with $H^\textrm{gen}$ and $T^\textrm{gen}$ as in Eq.~(\ref{e73}).

The absolute approximation error is less than $\eps$ when $H=H_n^m$ and $T=T_n^m$ in Eq.~(\ref{e70}) are chosen as follows. \\ \\
\os{Case $M>B$}.~~Set
\beq \label{e78}
H=H_n^m=1~,~~~~~~T=T_n^m=0~.
\eq
\mbox{} \\
\os{Case $M\leq B$}.~~Follow the boundary curve counterclockwise through points $(h,2t)$ with integer $t$ and integer $h$ such that $h-n$ is even, starting at the point $(h,2t)$ on edge I or II with lowest value of $h$ such that $h+1\geq H^\textrm{gen}$ and ending at the point $(h,2t)$ on edge II, III or IV with lowest value of $t$ such that $t\geq T^\textrm{gen}$. Let $(h_1,t_1)$ be the first point found in this process for which $F(h_1,t_1)\leq B$, and let $(h_2,t_2)$ be the last point for which $F(h_2,t_2)\leq B$. Set
\beq \label{e79}
H=H_n^m=h_1+1~,~~~~~~T=T_n^m=t_2~.
\eq

\subsection{Truncation issues in computing $c_t$} \label{subsec4.2}
\mbox{} \\[-9mm]

%
For $t=0,\;1,\;\cdots$ and $0<\eps<1$, the quantity
\beq \label{e80}
c_{t,LK}=\sum_{l=0}^L~\sum_{k=0}^K\,A_{2l,2k,2t}^{000}\,a_l\,b_k~.
\eq
approximates $c_t$ with absolute error less than $\eps$ when $L$ and $K$ are such that
\beq \label{e81}
l>L\Rightarrow |a_l|<\tfrac14\,\eps~~~\&~~~k>K\Rightarrow|b_k|<\tfrac{3}{16}\,\eps~.
\eq
With $\gamma=\min(1,{\rm ln}(1/v_0))$, the second item in Eq.~(\ref{e81}) holds when
\beq \label{e82}
K=\frac{1}{\gamma}\,\max\Bigl(0,{\rm ln}\,\frac{64}{3\eps}\Bigr)+\tfrac12\,g\,\frac{\sinh(\gamma)}{\gamma}~.
\eq
Subsequently, let $S=\max(s_0,s_{0,M})$, and set
\beq \label{e83}
E=\frac{2\sqrt{\pi}}{\Gamma(3/4)}~\frac{(1-S^2)^{-1/8}}{1+\sqrt{1-S^2}}~,~~~~~~V=\frac{1-\sqrt{1-S^2}} {1+\sqrt{1-S^2}}~.
\eq
Then the first item in Eq.~(\ref{e81}) is valid when
\beq \label{e84}
L=\frac{{\rm ln}(8E/\eps)+\tfrac14\,{\rm ln}(1+{\rm ln}(8E/\eps)/{\rm ln}(1/V))}{{\rm ln}(1/V)}~.
\eq
Furthermore, when the $a_l$ and $b_k$ required in Eq.~(\ref{e80}) are available with absolute accuracy $\tfrac14\,\eps$ and $\tfrac{3}{16}\,\eps$, respectively, while the $K$ and $L$ of Eqs.~(\ref{e82}, \ref{e84}) are used in Eq.~(\ref{e80}), all $c_t$ are approximated with absolute accuracy $2\eps$.

As to the availability of $a_l$ and $b_k$ for $l=0,...,L$ and $k=0,...,K$ with a required accuracy we give the following comments. The $a_l$ have the form
\beq \label{e85}
a_l=a_{l,3/4,{-}3/4}+a_{l,1/4,{-}1/4}~,
\eq
and either term at the right-hand side of Eq.~(\ref{e85}) is computed using the infinite series expression in Eq.~(\ref{e68}). When this infinite series is truncated at $N=2L/\sqrt{1-S^2}$, with $S=\max(s_0,s_{0,M})$, the absolute error is for all $l=0,...,L$ and either term at the right-hand side of Eq.~(\ref{e85}) less than $\eps/16$, and then all $a_l$, $l=0,...,L\,$, are computed with absolute error less than $\eps/8$. Finally, the $b_k$ are given by Eq.~(\ref{e63}) in terms of spherical Bessel and Hankel functions, and can therefore be computed to any desired accuracy using MatLab routines (employing the expressions for spherical Bessel and Hankel functions in terms of ordinary Bessel and Hankel functions, see \cite{ref2}, Sec.~10.47). 
When this is done with absolute accuracy $\tfrac{3}{32}2^{-7/4}\eps u_0\exp{(-\varphi(K;g/2v_0))}/(2K+1)v_0$ and $3\eps u_0/64(2K+1)$ for $j_k$ and $h_k^{(2)}$, respectively, the $b_k$ are computed for $k=0,\;1,\;\cdots,\;K$ with absolute accuracy $3\eps/16$. Using these approximations of $a_l$ and $b_k$ in Eq.~(\ref{e80}) with $K$ and $L$ as in Eqs.~(\ref{e82}, \ref{e84}) yields an approximation of $c_t$ with absolute error less than $\tfrac{7}{4}\eps$.

\subsection{Accuracy of assembled scheme}\label{subsec4.3}
Let $\eps>0$, and use either one of the truncation rules in Subsec.~\ref{subsec4.1}. Furthermore, compute $c_t$ as in Subsec.~\ref{subsec4.2} with absolute accuracy $\tfrac74\eps$. Finally, compute the Bessel function $J_{h+1}(2\pi r)$ with absolute accuracy $2 \pi r \eps/4w_0a_0$, with $w_0$ and $a_0$ given in Subsec.~\ref{subsec4.1}, using Matlab-codes. Then the quantity $I$ in Eq.~(\ref{e57}) is approximated with an absolute error that can be bounded by $\eps+\tfrac12\tfrac74\eps+\eps=\tfrac{23}8\eps$, due to, respectively, truncation of the double series in Eq.~(\ref{e57}), approximating $c_t$ as in Subsec.~\ref{subsec4.2}, and approximating the Jinc function $J_{h+1}(2\pi r)/2\pi r$ by computing $J_{h+1}$ using the Matlab-code.

\section{Illustration of the truncation rules} \label{sec5}
\mbox{} \\[-9mm]

In this section, we show the absolute truncation error and the computation time, using the general truncation rule of Subsec.~\ref{subsec2.3} and the dedicated truncation rule of Subsec.~\ref{subsec2.4} for approximation of the diffraction integral $I$ in Eqs.~(\ref{e1}-\ref{e2}) as a function of $\eps\in(0,1)$ for a variety of radial values $r$, maximum defocus values $f$, numerical aperture values $s_0$ and $s_{0,M}$, and Zernike circle polynomial degrees and orders $n$ and $m$. The truncation rules are used with $\eps/2$ instead of $\eps$. The structural quantities $c_t$ and Jinc functions $J_{h+1}(2\pi r)/2\pi r$ are computed with absolute accuracies $\eps/2$ and $\eps/16w_0a_0$, respectively, so that the absolute error due to using these computed quantities is bounded by $\eps/2$ for all $n$ and $m$ simultaneously. The total absolute error using the truncated series with the computed quantities is then expected to be less than $\tfrac12\eps+\tfrac12\eps=\eps$.

\begin{figure}[p]
\begin{center}
\plaatje{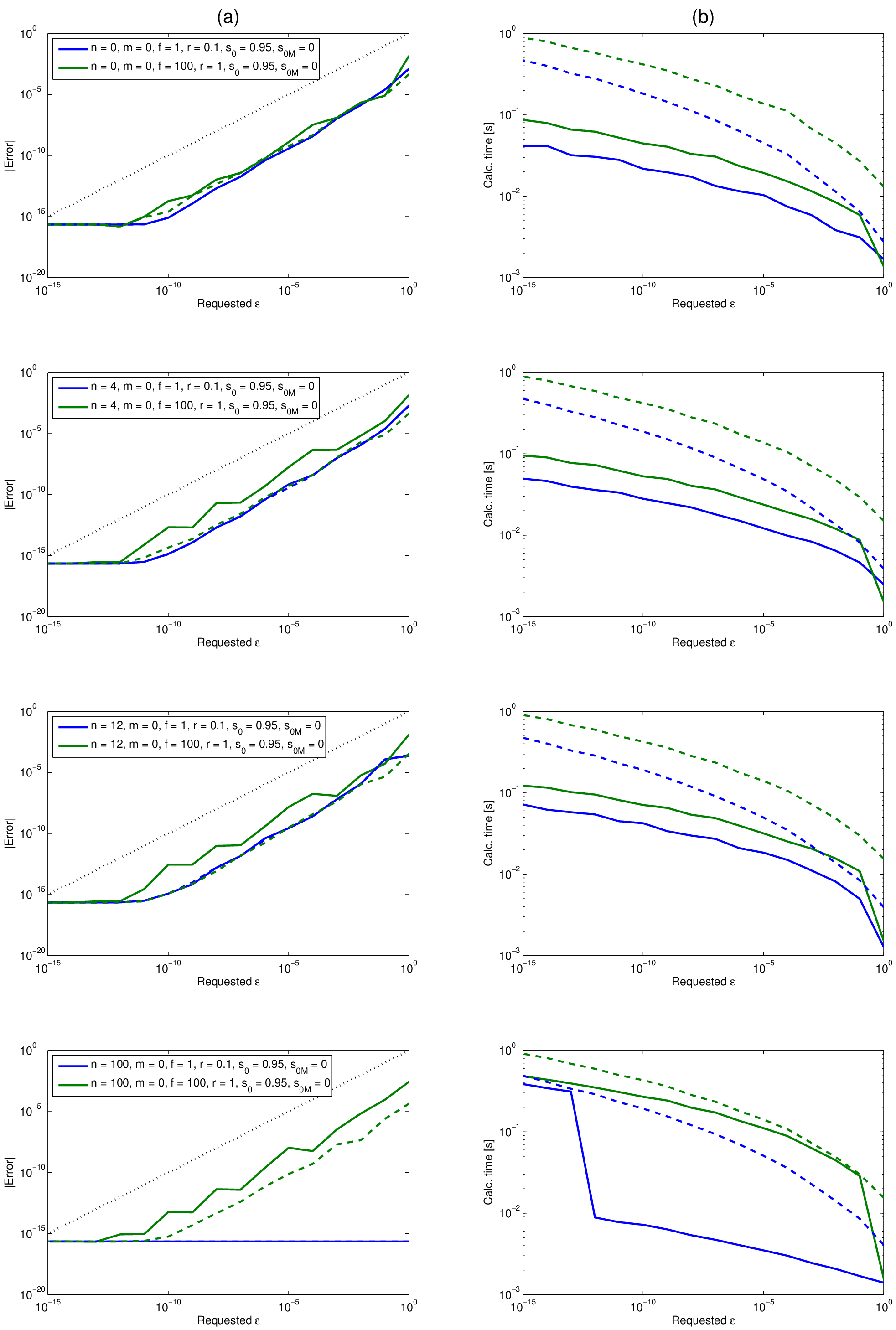}{0.8}
\caption{Absolute accuracy (a) and computation time (b) as a function of requested absolute accuracy $\eps$ using the general truncation rule (dashed lines) and the dedicated truncation rule (solid lines) when varying the degree $n$ and azimuthal order $m$ of the radial polynomial from top to bottom according to $(n,m) = (0,0),\;(4,0),\;(12,0),\;(100,0)$. Setting of aperture variables: $s_0=0.95,\;s_{0,M}=0$, setting of focal and radial variable: $f=1,\;r=0.1$ and $f=100,\; r = 1$. }
\label{fig6}
\end{center}
\end{figure}

\begin{figure}[p]
\begin{center}
\plaatje{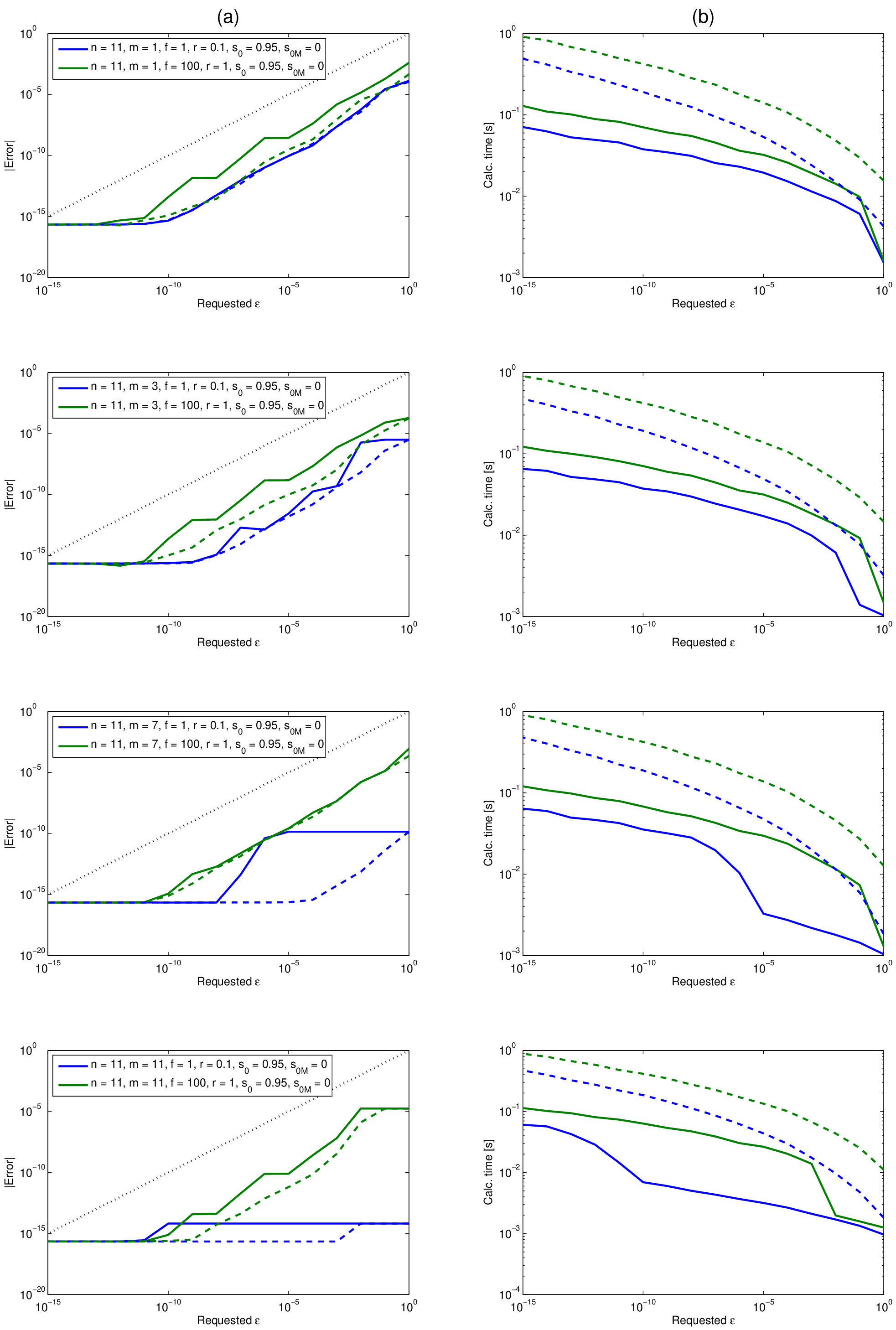}{0.8}
\caption{Absolute accuracy (a) and computation time (b) as a function of requested absolute accuracy $\eps$ using the general truncation rule (dashed lines) and the dedicated truncation rule (solid lines) when varying the degree $n$ and azimuthal order $m$ of the radial polynomial from top to bottom according to $(n,m) = (11,1),\;(11,3),\;(11,7),\;(11,11)$. Setting of aperture variables: $s_0=0.95,\;s_{0,M}=0$, setting of focal and radial variable: $f=1,\;r=0.1$ and $f=100,\; r = 1$. }
\label{fig7}
\end{center}
\end{figure}

\begin{figure}[p]
\begin{center}
\plaatje{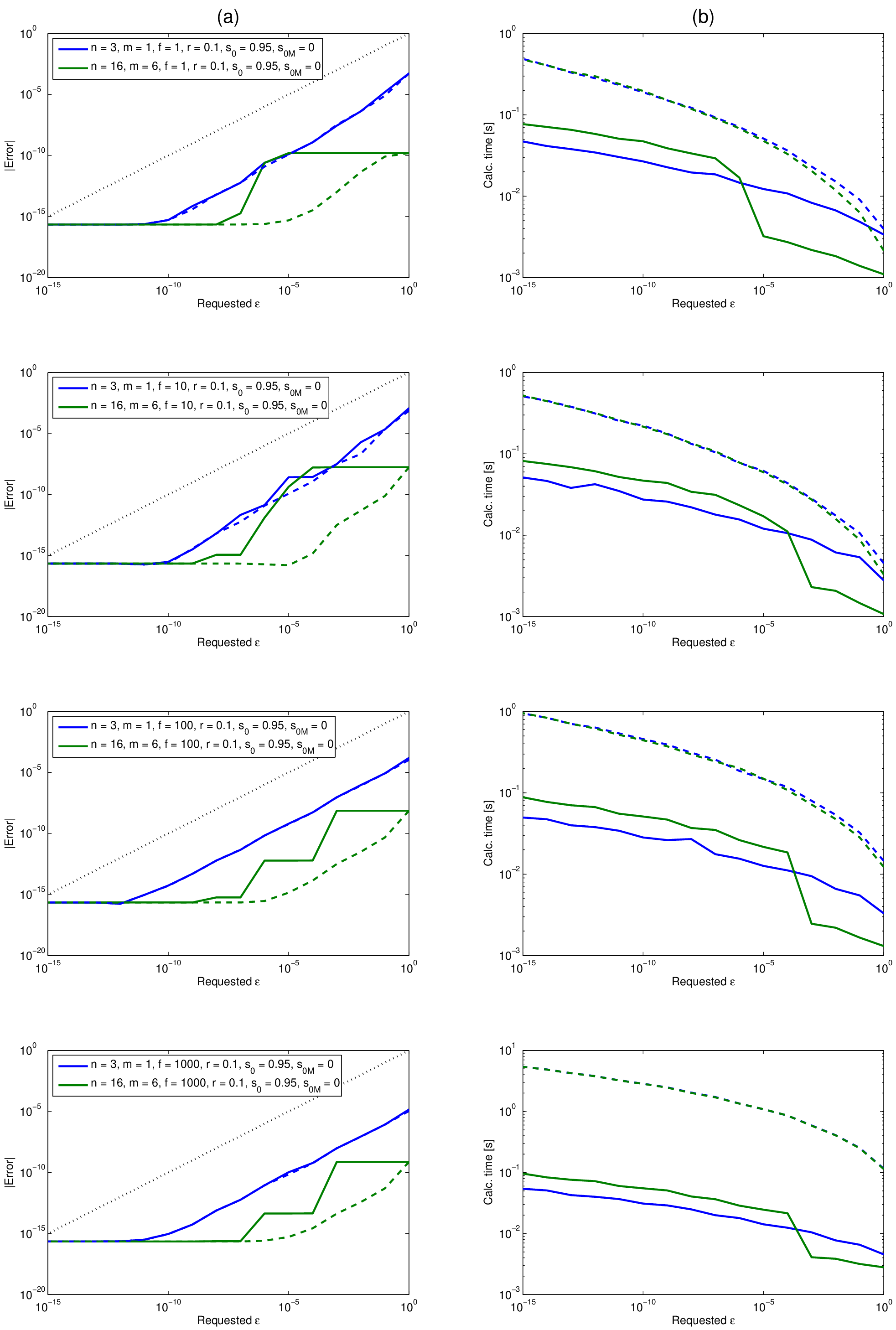}{0.8}
\caption{Absolute accuracy (a) and computation time (b) as a function of requested absolute accuracy $\eps$ using the general truncation rule (dashed lines) and the dedicated truncation rule (solid lines) when varying the focal variable $f$ from top to bottom according to $f = 1,\;10,\;100,\;1000$. Setting of aperture variables: $s_0=0.95,\;s_{0,M}=0$, setting radial variable: $r=0.1$, setting of the degree and azimuthal order of the radial polynomial: $(n,m) = (3,1)$ and $(16,6)$. }
\label{fig8}
\end{center}
\end{figure}

\begin{figure}[p]
\begin{center}
\plaatje{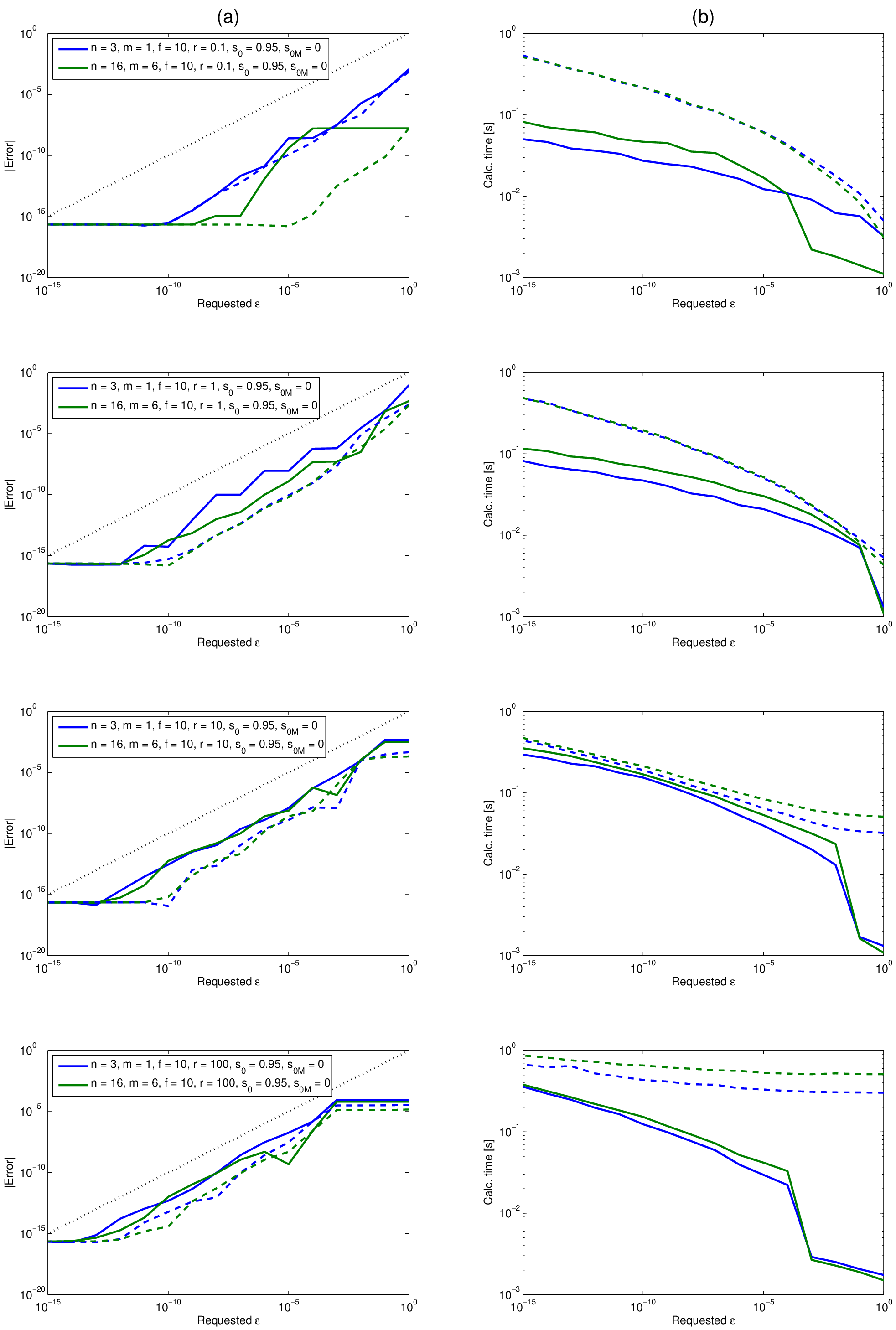}{0.8}
\caption{Absolute accuracy (a) and computation time (b) as a function of requested absolute accuracy $\eps$ using the general truncation rule (dashed lines) and the dedicated truncation rule (solid lines) when varying the radial variable $r$ from top to bottom according to $r = 0.1,\;1,\;10,\;100$. Setting of aperture variables: $s_0=0.95,\;s_{0,M}=0$, setting focal variable: $f=10$, setting of the degree and azimuthal order of the radial polynomial to $(n,m) = (3,1)$ and $(16,6)$. }
\label{fig9}
\end{center}
\end{figure}

\begin{figure}[p]
\begin{center}
\plaatje{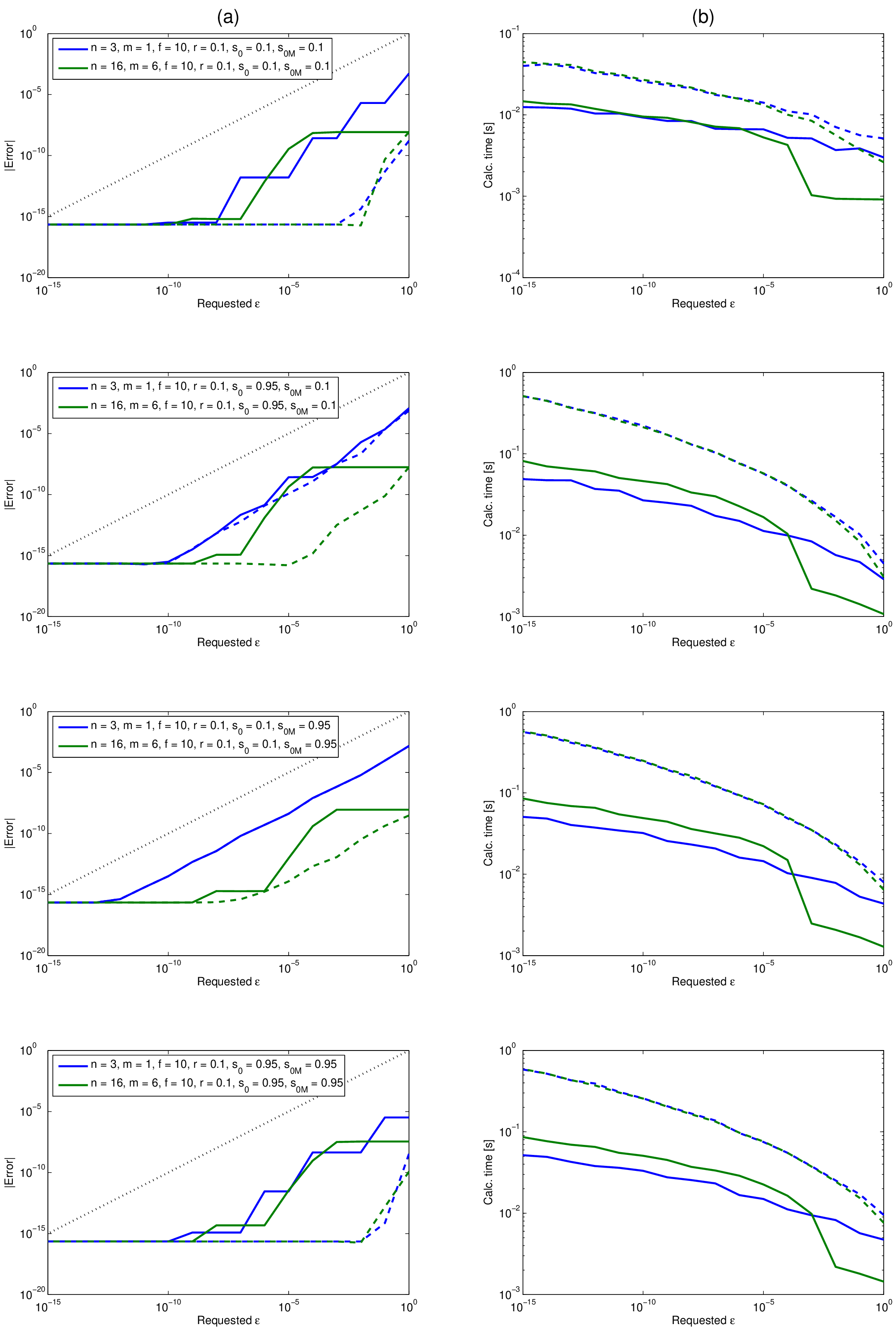}{0.8}
\caption{Absolute accuracy (a) and computation time (b) as a function of requested absolute accuracy $\eps$ using the general truncation rule (dashed lines) and the dedicated truncation rule (solid lines) when varying the aperture variables $s_0$ and $s_{0,M}$ from top to bottom according to $(s_0,s_{0,M}) = (0.1,0.1),\;(0.95,0.1),\;(0.1,0.95),\;(0.95,0.95)$. Setting of degree and azimuthal order of the radial polynomial: $(n,m) = (3,1)$ and $(16,6)$, setting of focal and radial variable: $f=10,\;r=0.1$. }
\label{fig10}
\end{center}
\end{figure}

\begin{figure}[p]
\begin{center}
\plaatje{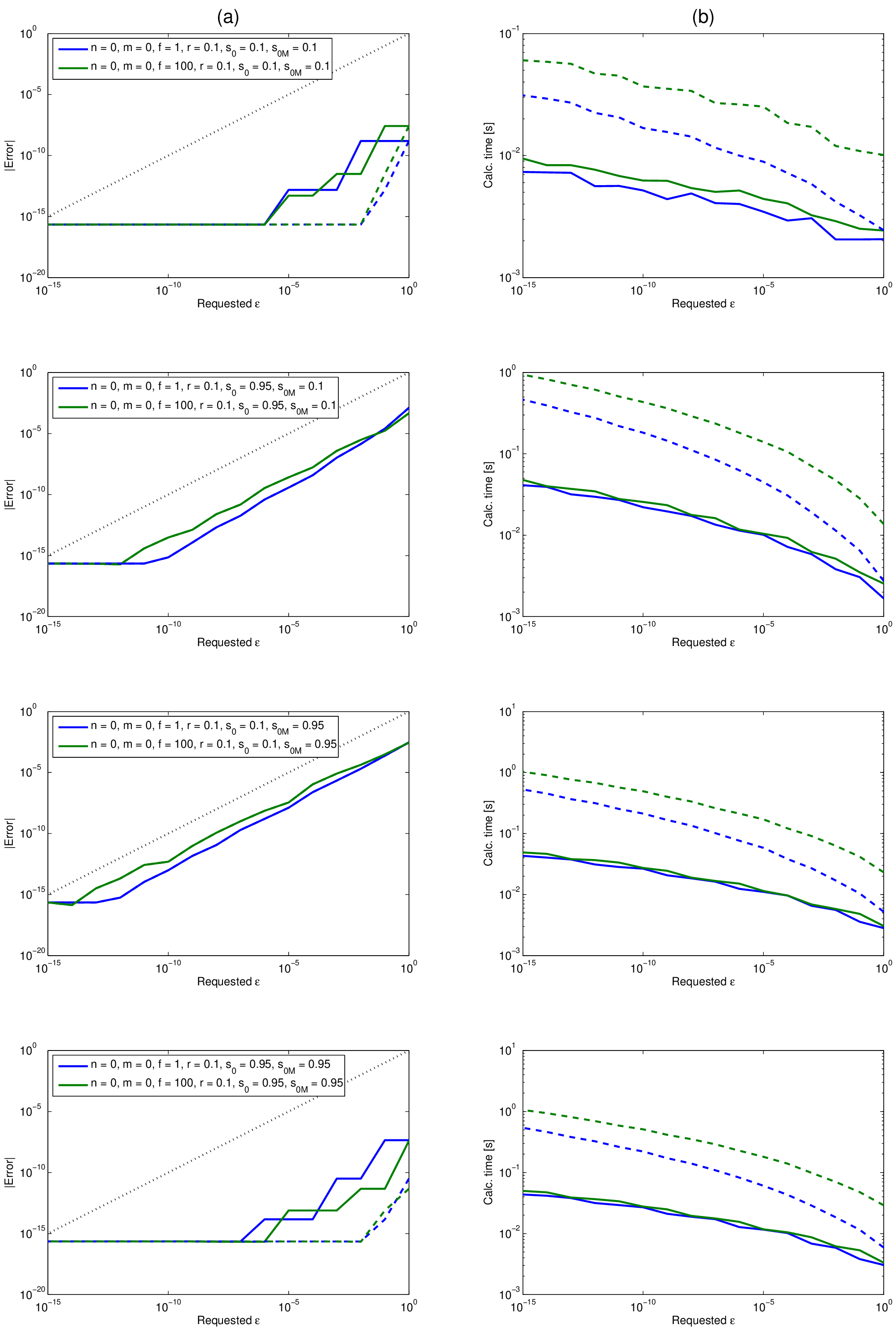}{0.8}
\caption{Absolute accuracy (a) and computation time (b) as a function of requested absolute accuracy $\eps$ using the general truncation rule (dashed lines) and the dedicated truncation rule (solid lines) when varying the aperture variables $s_0$ and $s_{0,M}$ from top to bottom according to $(s_0,s_{0,M}) = (0.1,0.1),\;(0.95,0.1),\;(0.1,0.95),\;(0.95,0.95)$. Setting of degree and azimuthal order of the radial polynomial: $(n,m) = (0,0)$, setting of focal and radial variable: $f=1,\;r=0.1$ and $f=100,\;r=0.1$. }
\label{fig11}
\end{center}
\end{figure}

\begin{figure}[p]
\begin{center}
\plaatje{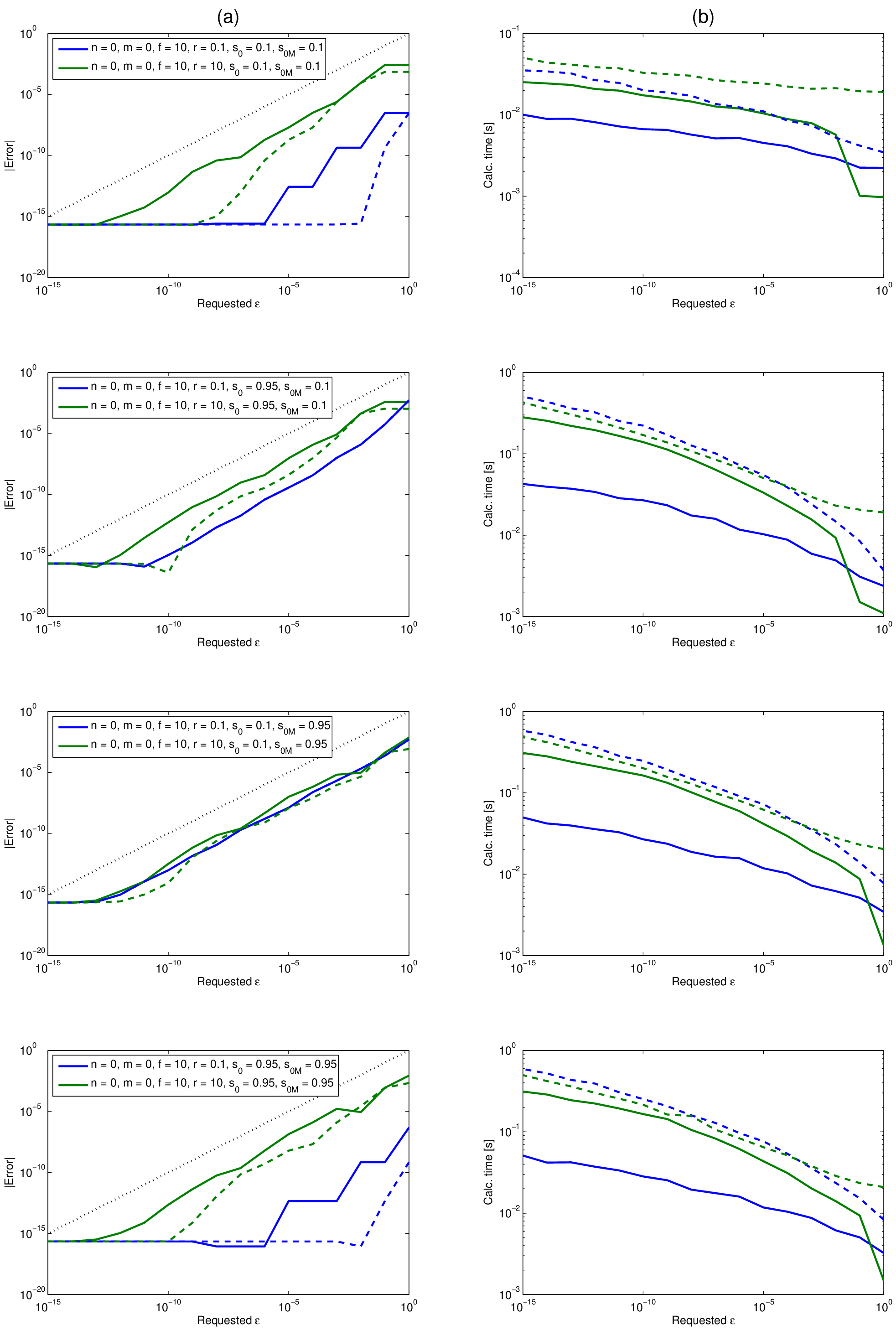}{0.8}
\caption{Absolute accuracy (a) and computation time (b) as a function of requested absolute accuracy $\eps$ using the general truncation rule (dashed lines) and the dedicated truncation rule (solid lines) when varying the aperture variables $s_0$ and $s_{0,M}$ from top to bottom according to $(s_0,s_{0,M}) = (0.1,0.1),\;(0.95,0.1),\;(0.1,0.95),\;(0.95,0.95)$. Setting of degree and azimuthal order of the radial polynomial: $(n,m) = (0,0)$, setting of focal and radial variable: $f=10,\;r=0.1$ and $f=10,\;r=10$. }
\label{fig12}
\end{center}
\end{figure}

\begin{figure}[p]
\begin{center}
\plaatje{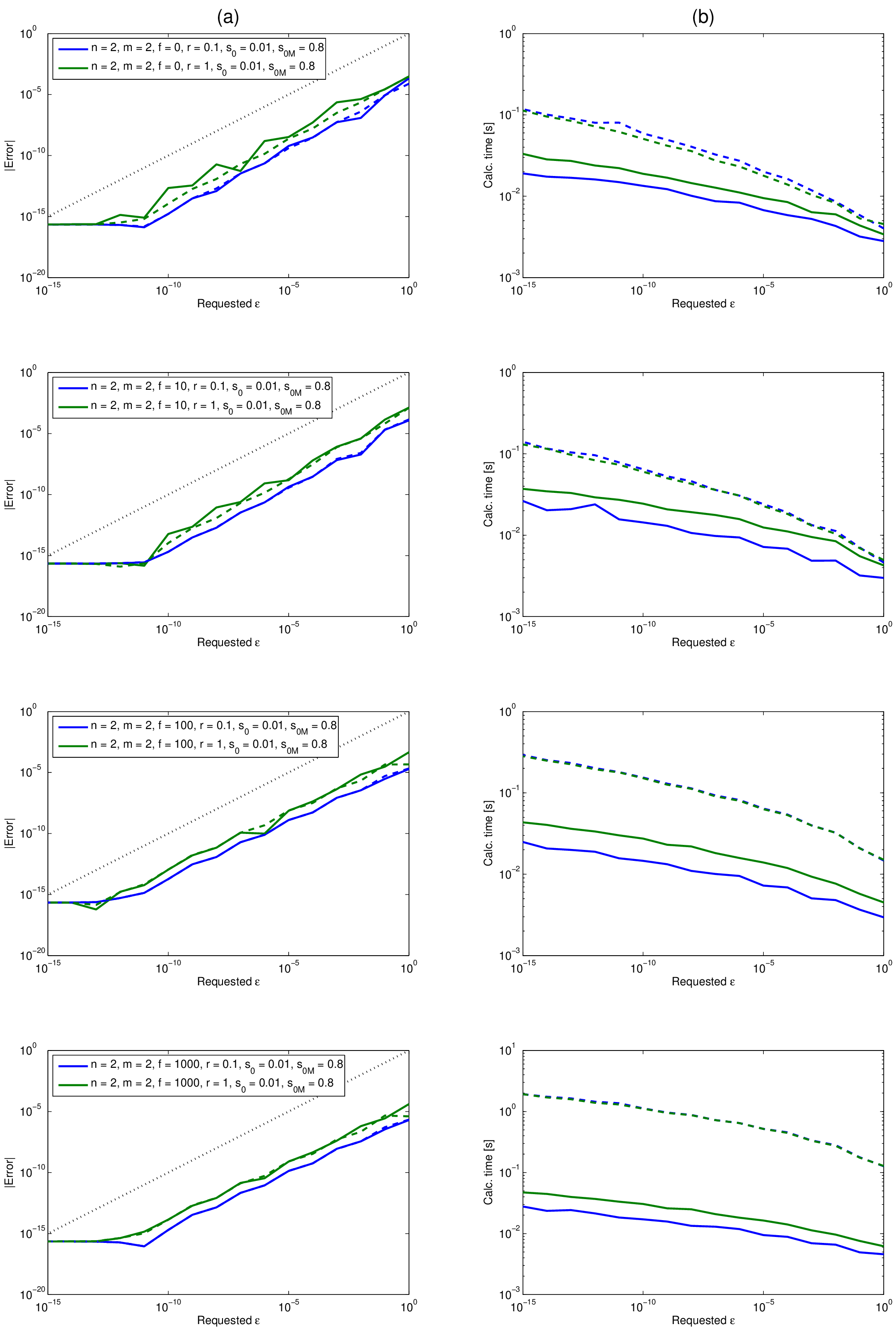}{0.8}
\caption{Absolute accuracy (a) and computation time (b) as a function of requested absolute accuracy $\eps$ using the general truncation rule (dashed lines) and the dedicated truncation rule (solid lines) when varying the focal variable $f$ from top to bottom according to $f = 0,\;10,\;100,\;1000$. Setting of aperture variables: $s_0=0.01,\;s_{0,M}=0.8$, setting radial variable: $r=0.1$ and $r=1$, setting the degree and azimuthal order of the radial polynomial: $(n,m) = (2,2)$. }
\label{fig13}
\end{center}
\end{figure}

\begin{figure}[p]
\begin{center}
\plaatje{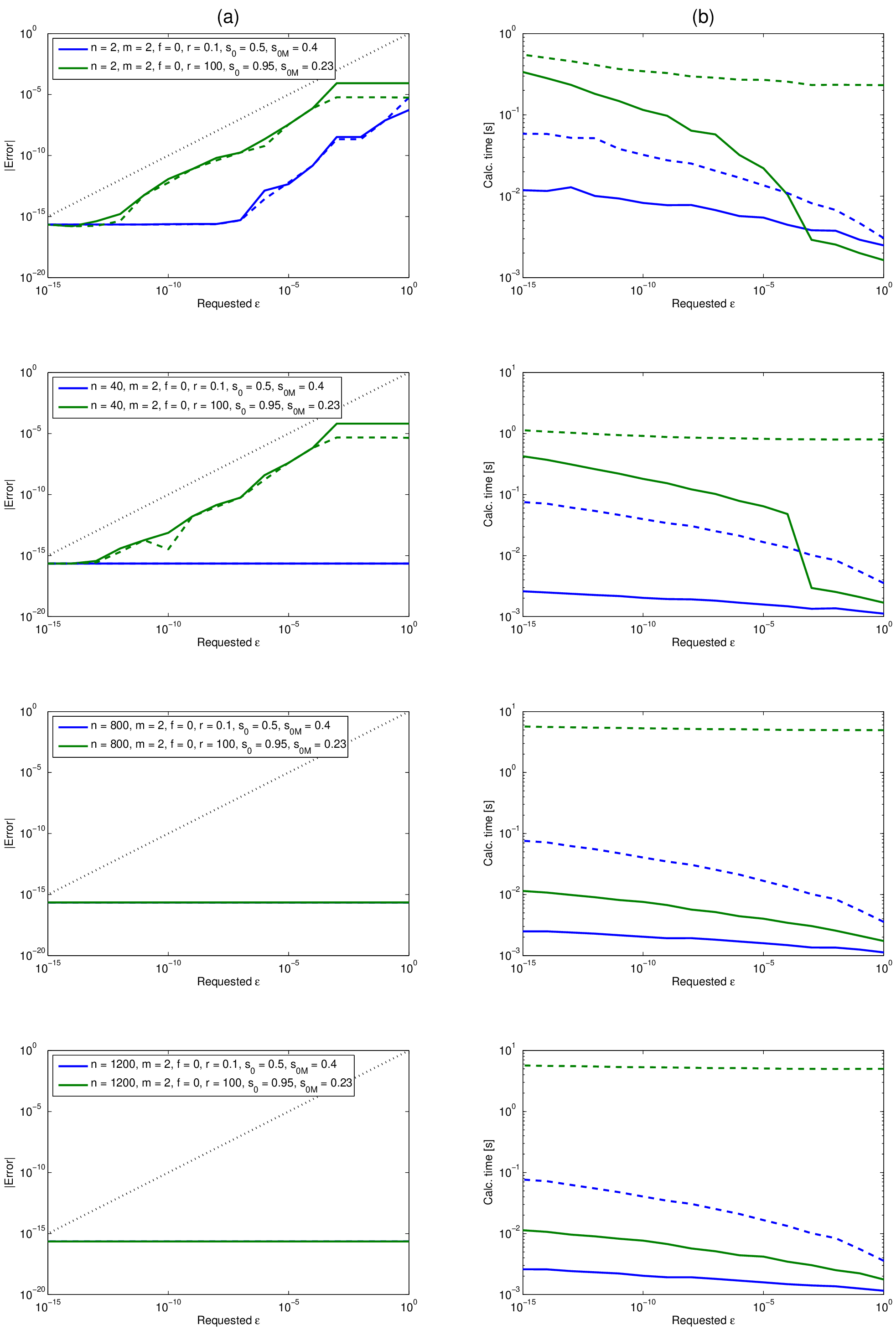}{0.8}
\caption{Absolute accuracy (a) and computation time (b) as a function of requested absolute accuracy $\eps$ using the general truncation rule (dashed lines) and the dedicated truncation rule (solid lines) when varying the degree $n$ and azimuthal order $m$ of the radial polynomial from top to bottom according to $(n,m) = (2,2),\;(40,2),\;(800,2),\;(1200,2)$. Setting of aperture variables: $s_0=0.5,\;s_{0,M}=0.4$ and $s_0=0.95,\;s_{0,M}=0.23$, setting of focal and radial variable: $f=0,\;r=0.1$ and $f=0,\; r = 100$. }
\label{fig14}
\end{center}
\end{figure}

\begin{figure}[p]
\begin{center}
\plaatje{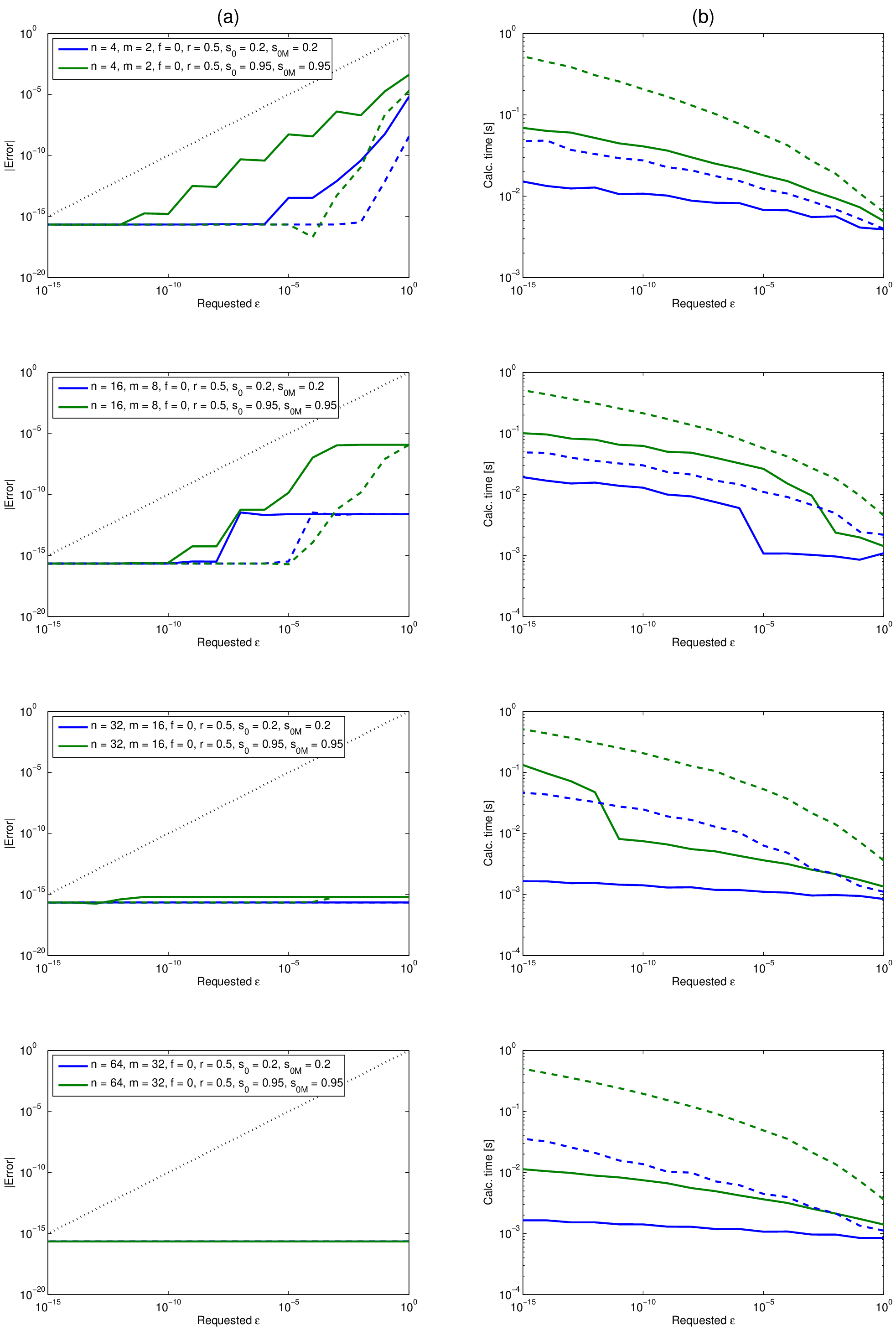}{0.8}
\caption{Absolute accuracy (a) and computation time (b) as a function of requested absolute accuracy $\eps$ using the general truncation rule (dashed lines) and the dedicated truncation rule (solid lines) when varying the degree $n$ and azimuthal order $m$ of the radial polynomial from top to bottom according to $(n,m) = (4,2),\;(16,8),\;(32,16),\;(64,32)$. Setting of aperture variables: $s_0=0.2,\;s_{0,M}=0.2$ and $s_0=0.95,\;s_{0,M}=0.95$, setting of focal and radial variable: $f=0,\;r=0.5$. }
\label{fig15}
\end{center}
\end{figure}

In all figures, we show achieved accuracy (a) and computation time (b) against requested accuracy $\eps$ in the range $10^{-15}-10^{0}$, using the general truncation rule (dashed lines) and the dedicated truncation rule (solid lines). The graphs result from specification of
\begin{description}
\item[A.] the values of the aperture parameters $s_0$, $s_{0,M}$,
\item[B1.] the value of the focal parameter $f$, 
\item[B2.] the value of the radial parameter $r$,
\item[C.] the degree $n$ and order $m$ of the radial polynomial $R^m_n$.
\end{description}
In the presented figures, the item(s) in $3$ of the groups \textbf{A}, \textbf{B1}, \textbf{B2}, \textbf{C} are varied over at most two cases, while the item(s) in the remaining set is varied over several cases. Thus. schematically, we have in Figs. \ref{fig6}-\ref{fig15} the cases as defined in Table \ref{table1}.
\begin{table}[!h]
\begin{center}
\begin{tabular}{c|c|c|c|c}
Figure & $s_0,\;s_{0,M}$ & $f$ & $r$ & $R_n^m$ \\ \hline
6, 7 & fixed & \multicolumn{2}{c|}{2 cases} & varied \\ 
8 & fixed & varied & fixed & 2 cases \\ 
9 & fixed & fixed & varied & 2 cases \\ 
10 & varied & fixed & fixed & 2 cases \\ 
11 & varied & 2 cases & fixed & fixed \\ 
12 & varied & fixed & 2 cases & fixed \\ 
13 & fixed & varied & 2 cases & fixed \\ 
14 & 2 cases & fixed & 2 cases & varied \\ 
15 & 2 cases & fixed & fixed & varied
\end{tabular} 
\caption{Schematic overview indicating the item(s) in the which groups \textbf{A}, \textbf{B1}, \textbf{B2}, \textbf{C} are varied in Figs. \ref{fig6}-\ref{fig15}.}\label{table1}
\end{center}
\end{table}
In general, it can be said that the requested accuracy is achieved amply: the graphs in (a) stay well below and parallel to the graph $(\eps,\eps)$ (dotted lines). The performance of the dedicated rule in terms of accuracy is most of the time slightly worse but comparable to that of the general rule, while the performance in terms of computation time can be significantly better. The latter situation occurs especially when the degree and order of the radial polynomial are large compared to $f/2$ and $2\pi r$.

\section{Conclusions} \label{sec6}
\mbox{} \\[-9mm]

We have formulated and verified truncation rules for the double series expressions that emerge from the advanced ENZ-theory for the computation of the optical diffraction integrals pertaining to optical systems with high NA, vector fields, polarization, and meant for imaging of extended objects. These rules have been devised for the central case $j=0$ in the vectorial framework, which can be considered to be representative for all occurring diffraction integrals. Two versions of the truncation rule have been developed. The general rule gives precision to the rule-of-thumb that the required summation range is of the order $2\pi r$ times $\tfrac12 |f|$ with $r$ and $f$ the values of the (normalized) radial and the focal parameters in image space, irrespective of the degree and order of the radial polynomial involved in the diffraction integral. In the dedicated rule, we have also accounted for the specific way the radial polynomial influences the actual summation range, leading to performances comparable in terms of accuracy and better in terms of computation time than what is offered by the general truncation rule. A salient feature of the double series that manifest itself through the truncation rules is that the computation times stay well within what can be considered practicable, more or less independently of the values of the aperture parameters and the magnitudes of the focal and radial variable. In the case that circle polynomials of very high degree and/or order are involved in the diffraction integrals, the general truncation rule becomes impracticable, and one has to resort to using the dedicated rule. With this full understanding of the double series with regard to truncation matters, it can be said that the advanced ENZ-theory is more or less completed.

\newpage

\section{Additions to arXiv: 1407.6589v1} \label{sec7}
\mbox{} \\[-9mm]

We give in this section two additions to arXiv: 1407.6589v1. The first addition concerns the formulation of truncation rules that are valid for a whole range of radial values $r>0$, rather than a particular $r$. This has the advantage that per focal plane, there is one truncation point that serves all the points $(x,y) = (r\cos{\phi}, r\sin{\phi})$ with r in the specified range. The second addition concerns the integral $I_\textrm{VMML}$, case $|j|=2$, of \cite{ref1}, Sec.~9 and Appendix H that occur for systems with high NA, vector fields, magnification and multi-layered focal region. There are two instances $I^{\pm}_\textrm{VMML}$ of this integral, corresponding to forward propagating waves ($+$-sign) and backward propagating waves ($-$-sign). While $I^{+}_\textrm{VMML}$ behaves to a large extent the same as $I_\textrm{VM}$ with regard to truncation matters, the situation for $I^{-}_\textrm{VMML}$ is drastically different, as evidenced by Figure 6 in \cite{ref1}, cases $m=2,\;3,\;\cdots\;$, with respect to decay of the integrals $I^{\pm}_{mh}(r)$ that replace the Jinc-functions $J_{h+1}(2\pi r)/2\pi r$ in the double series in Eq.~(6). We also use this opportunity to correct some innocent but disturbing errors in \cite{ref1}, Sec.~9, and we remove an apparent singular behavior of $I^{-}_\textrm{VMML}$, $|j|=2$, that would occur in the case that $s_{0,M}-s_{0,h}$, see Subsec.~\ref{subsec7.2} for explanation, is $0$ or very small.

\subsection{Truncation rule valid for a range of radial values}\label{subsec7.1} 
\mbox{} \\[-9mm]

Let $0<\eps<1$, and let $R_{\max} \geq 1/2\pi$ be given. For real $r\geq0$ and real $f$, we let
\beq\label{e89}
R=R(r)=\max{(\tfrac{1}{2\pi},r)},\;\;\; g = g(f)= \max{(1,|f|)}\;. 
\eq
We consider radial ranges of the form $0\leq r\leq R_{\max}$.

\subsubsection{General rule}\label{subsubsec7.1.1}
\mbox{} \\[-9mm]

\begin{figure}[t]
\begin{center}
\plaatje{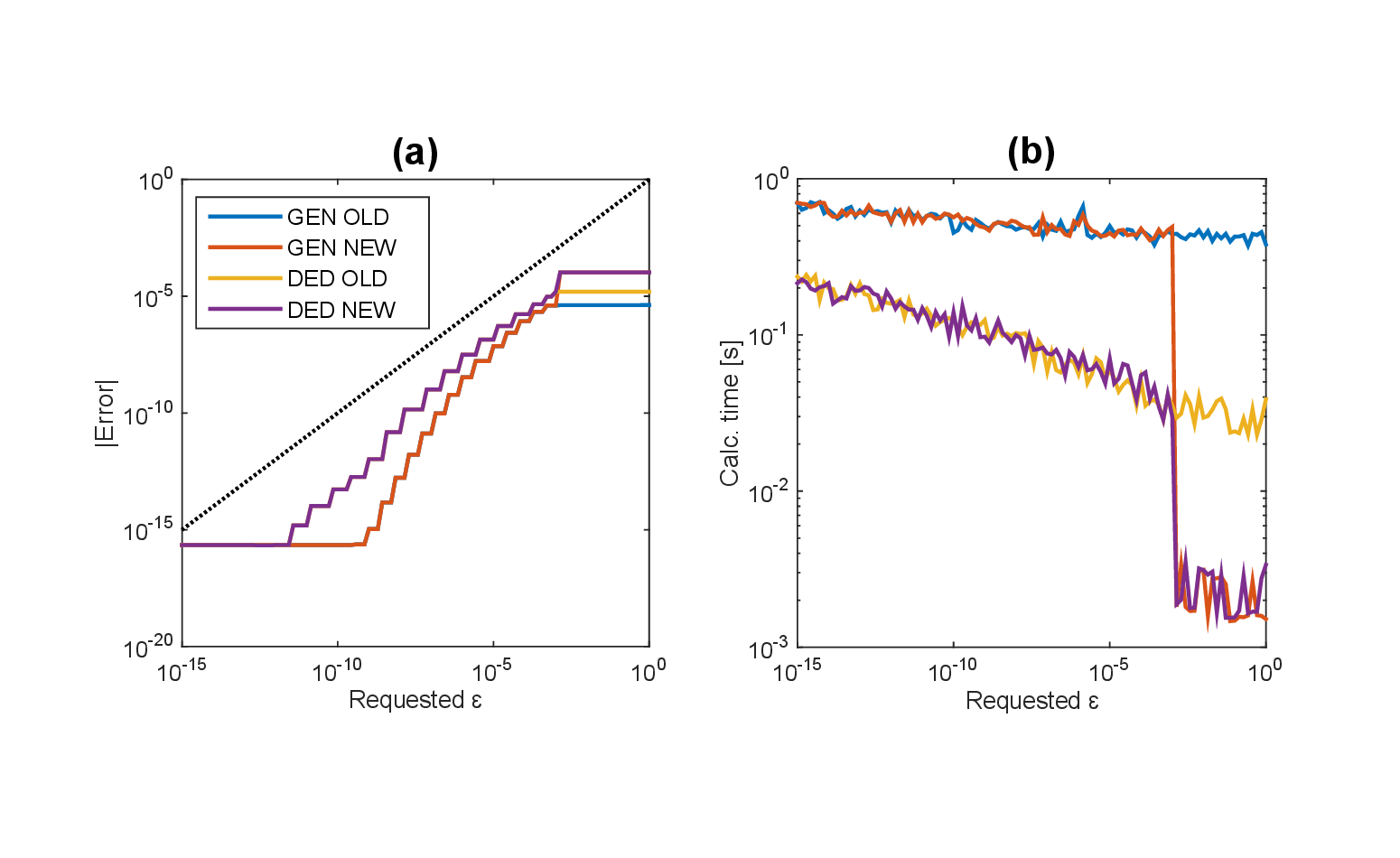}{1.0}
\caption{Absolute accuracy (a) and computation time (b) as a function of requested absolute accuracy $\eps$ using the old general truncation rule (Eq.(\ref{e73})), the new general truncation rule (Eq.(\ref{e93}-\ref{e94})) and the old and new dedicated truncation rules as given in Subsubsec.~\ref{subsubsec4.1.2} starting from the old and new general rules, respectively. The shown curves pertain to the degree $n$ and azimuthal order $m$ of the radial polynomial: $(n,m) = (16,6)$, setting of aperture variables: $s_0=0.8,\;s_{0,M}=0.4$ and setting of focal and radial variable: $f=10,\;r=0.5$. }
\label{fig16}
\end{center}
\end{figure}

For a given $r\geq0$, we have found in Subsec.~\ref{subsec2.3} numbers $H=H(R)$ and $T=T(R)$ such that
\beq \label{e90}
\left| c_t \frac{J_{h+1}(2\pi r)}{2\pi r}\right| < \eps\;\;, \;\;\; t>T\; \textrm{or} \; h+1>H\;.
\eq
For the present purpose, we reformulate the recipe from Subsec.~\ref{subsec2.3} slightly as follows. Let
\beq\label{e91}
\bar{B} = \bar{B}(R) = \ln\left( \frac{2w_0a_0}{\pi^2\eps R\sqrt{R}}\right)\;.
\eq
When $\bar{B}(R)<0$, we set
\beq\label{e92}
H=H(R)=1\;,\;\;T=T(R)=0\;.
\eq
When $\bar{B}(R)\geq 0$, we set
\begin{eqnarray}
&&H=H(R)=\bar{B}(R)+2\pi R \sinh(1)\;,\label{e93}\\
&&T=T(R)=\tfrac{1}{\gamma}\bar{B}(R)+\tfrac{1}{2}g\frac{\sinh(\gamma)}{\gamma}\;,\label{e94}
\end{eqnarray}
with $\gamma$ as in Subsec.~\ref{subsec2.3}. This truncation rule is somewhat more economic than the one in Subsec.~\ref{subsec2.3}, since $T$ and $H$ in Eq.~(\ref{e24}) can be positive in certain cases that $\bar{B}(R)<0$. This is illustrated in Figure \ref{fig16}, where one can observe a sharp decrease in computation time when using the new truncation rules for cases that $\eps$ is relatively large ($B\neq \bar{B}(R)$). 

Requiring Eq.~(\ref{e90}) to hold for all $r$ with $0\leq r\leq R_{\max}$, we define
\begin{eqnarray}
&&H=H^\textrm{gen}_{\max}=\max_{\tfrac{1}{2\pi}\leq R \leq R_{\max}} H(R)\;,\label{e95}\\
&&T=T^\textrm{gen}_{\max}=\max_{\tfrac{1}{2\pi}\leq R \leq R_{\max}} T(R)\;.\label{e96}
\end{eqnarray}
It is evident that $T^\textrm{gen}_{\max} = T(1/2\pi)$, since $\bar{B}(R)$ in Eq.~(\ref{e91}) is a decreasing function of $R\geq 1/2\pi$. To find $H^\textrm{gen}_{\max}$,
we observe that
\beq\label{e97}
H(R) = \left\{ \begin{array}{lll}
0 &,& \;\;R>R_0\;,\\
\ln\left( \frac{2w_0a_0}{\pi^2\eps}\right)-\tfrac{3}{2}\ln R+2\pi R\sinh(1) &,& \;\;\tfrac{1}{2\pi}\leq R \leq R_0\;,
\end{array}\right.
\eq
where
\beq\label{e98}
R_0=\left(\frac{2w_0a_0}{\pi^2\eps}\right)^{2/3}\;.
\eq
On the range $1/2\pi \leq R \leq R_0$, the function $H(R)$ is convex since $H''(R)=3/2R^2 > 0$. Therefore,
\beq\label{e99}
H^\textrm{gen}_{\max} = \max{(H(1/2\pi), \;H(\min{(R_0,R_{\max})}))}\;.
\eq

With this $T=T^\textrm{gen}_{\max}$ and $H = H^\textrm{gen}_{\max}$, we have that Eq.~(\ref{e90}) holds for all $r$ with $0\leq r\leq R_{\max}$. In Subsec.~\ref{subsubsec4.1.1}, Eq.~(\ref{e73}), we just have to replace $H$ and $T$ by $H^\textrm{gen}_{\max}$ and $T^\textrm{gen}_{\max}$, respectively, to achieve that the absolute approximation error is less than $\eps$, simultaneously for all $n$ and $m$ and all $r$ with $0\leq r\leq R_{\max}$.

\subsubsection{Dedicated rule}\label{subsubsec7.1.2}
\mbox{} \\[-9mm]

With a fixed $n$ and $m$ and a given $r\geq 0$, we have shown in Subsec.~\ref{subsec2.4} how to choose $H$ and $T$ such that
\beq \label{e100}
\frac{2w_0a_0}{\pi^2R\sqrt{R}} \exp{(-F(h,t))} < \eps
\eq
for all $(h,2t)\in S^m_n$ with $h+1>H$ or $t>T$. Here $F(h,t)$ is given in Eq.~(\ref{e27}) and involves $R=R(r)$ explicitly. Now we want $H$ and $T$ such that
\beq \label{e101}
\bar{F}(h,t;R):=\varphi(h+1;2\pi R) + \tfrac{3}{2}\ln{R} + \varphi(t;g/2,g/2v_0) > \ln{\left( \frac{2w_0a_0}{\pi^2 \eps}\right)}
\eq
for all $(h,2t)\in S^m_n$ with $h+1>H$ or $t>T$ and all $R$ with $1/2\pi \leq R \leq R_{\max}$.

For a fixed $h,\;t = 0,\;1,\;\cdots\;$, we want to find the minimum of $\bar{F}(h,t;R)$ as a function of $R,\;1/2\pi \leq R \leq R_{\max}$. Noting that $\varphi(t;g/2,g/2v_0)$ is independent of $R$, we can concentrate on minimizing $\varphi(h+1;2\pi R) + (3/2)\ln{R}$. For a fixed $x=h+1=1,\;2,\;\cdots\;$, we consider minimization of 
\beq \label{e102}
\Phi(x;c):=\varphi(x;c)+\tfrac{3}{2}\ln{c} - \tfrac{3}{2}\ln{2\pi}
\eq
over $c,\;1\leq c \leq c_{\max}$ with
\beq \label{e103}
c=2\pi R,\;\;\; c_{\max}=2\pi R_{\max}\;.
\eq
For $x=1$, we have $\varphi(x;c)=0$, and we get
\beq \label{e104}
\min_{1\leq c \leq c_{\max}} \Phi(1;c)=\Phi(1;1)=-\tfrac{3}{2}\ln{2\pi}\;.
\eq
For $x=2,\;3,\;\cdots\;$, we use Eq.~(\ref{a5}) to see that
\beq \label{e105}
\tfrac{\textrm{d}}{\textrm{d}c}[\Phi(x;c)] = -\tfrac{1}{c}\sqrt{x^2-c^2} + \tfrac{3}{2c}\;,\;\; 1 \leq c \leq x\;,
\eq
and this vanishes for $c=\sqrt{x^2-9/4}$. Therefore, $\Phi(x;c)$ decreases in $1\leq c \leq \sqrt{x^2 - 9/4}$ and increases in $\sqrt{x^2 - 9/4} \leq c \leq x$, so that
\beq \label{e106}
\min_{1\leq c \leq c_{\max}} \Phi(x;c)= \left\{ \begin{array}{ll}
-\tfrac{3}{2}\ln{2\pi} &,\;\;x=1\;,\\
\Phi(x;\min(c_{\max},\sqrt{x^2-9/4}))\;&,\;\;x=2,\;3,\;\cdots\;,
\end{array} \right.
\eq
where for the second case in Eq.~(\ref{e106}), we have also used that $\Phi(x;c) = (3/2)\ln{c}-(3/2)\ln{2\pi}$ increases in $c\geq x$.

We conclude that
\begin{eqnarray} \label{e107}
\bar{F}(h,t)&:=& \min_{\tfrac{1}{2\pi}\leq R \leq R_{\max}} \bar{F}(h,t;R)  \\
&=& \varphi(t;g/2,g/2v_0) + \left\{ \begin{array}{ll}
-\tfrac{3}{2}\ln{2\pi}&, \;\;h=0\;,\\
\varphi(h+1;2\pi\hat{R})+\tfrac{3}{2}\ln\hat{R}&,\;\; h=1,\;2,\;\cdots\;,
\end{array}\right. \nonumber
\end{eqnarray}
where
\beq \label{e108}
\hat{R}=\min(R_{\max},\tfrac{1}{2\pi}\sqrt{(h+1)^2-9/4})\;.
\eq
From this point onwards, we can proceed as in Subsec.~\ref{subsec2.4} with $F(h,t)$ of Eq.~(\ref{e27}) replaced by $\bar{F}_{\min}(h,t)$ and $B$ of Eq.~(\ref{e23}) replaced by $\ln(2w_0a_0/\pi^2 \eps)$. Thus, one searches the boundary $\partial S^m_n$, as long as contained in $h+1 \leq H^\textrm{gen}_{\max},\; t\leq T^\textrm{gen}_{\max}$, with $H^\textrm{gen}_{\max}$ and $T^\textrm{gen}_{\max}$ from Subsec.~\ref{subsubsec7.1.1}, for the first and last point $(h,2t)$ where
\beq \label{e109}
\bar{F}_{\min} \leq \ln{\left( \frac{2w_0a_0}{\pi^2 \eps}\right)}\;.
\eq
It is observed that $\bar{F}_{\min}(h,t)$ has the same monotonicity properties as $F(h,t)$ in Eq.~(\ref{e27}). In particular,
\beq \label{e110}
\min_{(h,2t)\in \partial S^m_n} \bar{F}_{\min}(h,t)
\eq
is assumed on edge II of $\partial S^m_n$. When the quantity in Eq.~(\ref{e110}) is larger than $\ln(2w_0a_0/\pi^2\eps)$, we can take $H=1$, $T=0$, and otherwise, we have to carry out the search process described above. With $H^{\textrm{ded},m}_{\max,n}$ and $T^{\textrm{ded},m}_{\max,n}$ found this way, we have that the absolute approximation error in truncating the series in Eq.~(\ref{e70}) at $h+1=H^{\textrm{ded},m}_{\max,n}$ and $t=T^{\textrm{ded},m}_{\max,n}$, is less than $\eps$ for all $r$ with $0\leq r \leq R_{\max}$.

\subsubsection{Illustration of the truncation rules valid for $r$-ranges} \label{subsubsec7.1.3}
\mbox{} \\[-9mm]

\begin{figure}[htp]
\begin{center}
\plaatje{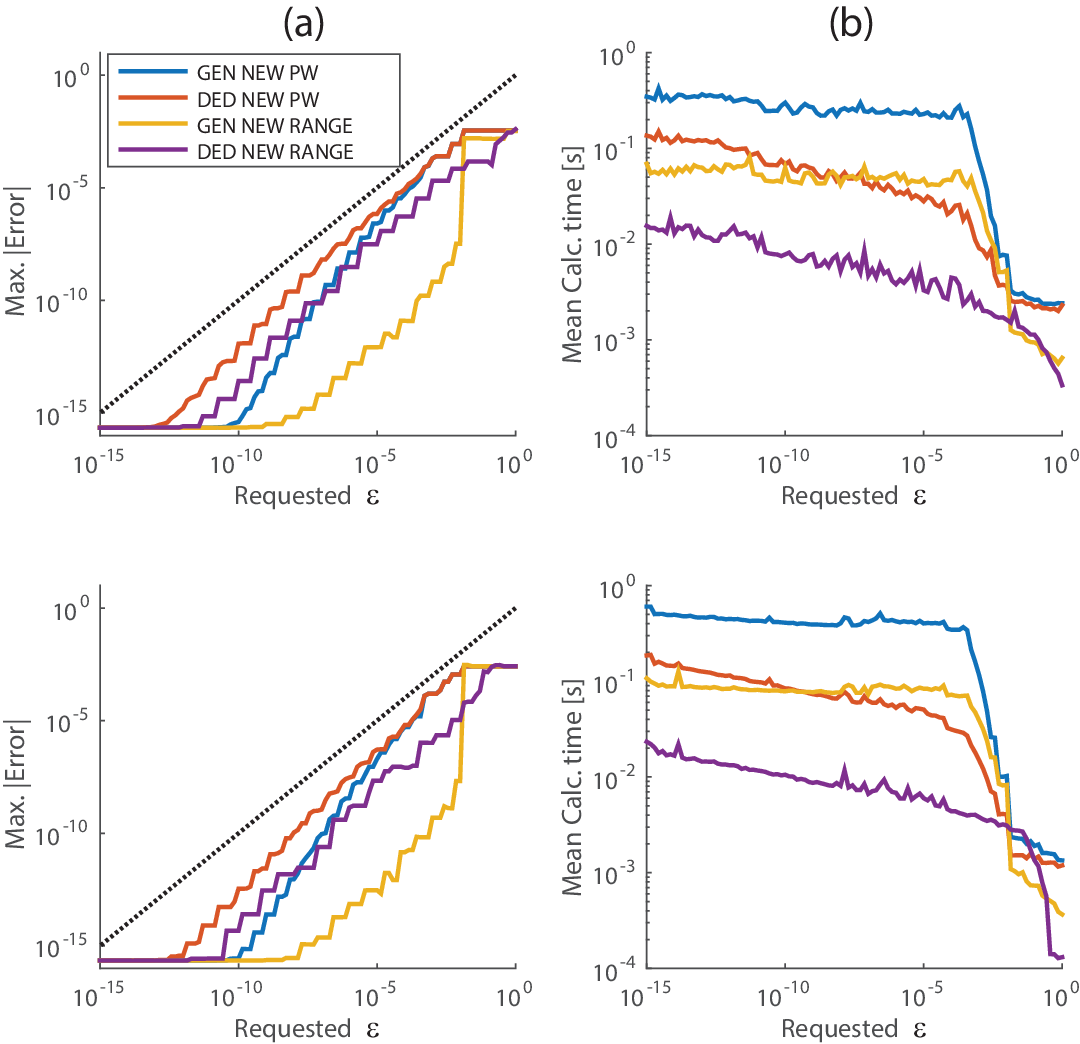}{0.8}
\caption{Absolute accuracy (a) and average computation time (b) as a function of requested absolute accuracy $\eps$ using the new general and dedicated truncation rules valid for a single point (GEN/DED NEW PW) and corresponding truncation rules valid for a given range $0\leq r\leq R_{\max}$ (GEN/DED NEW RANGE) when varying the degree $n$ and azimuthal order $m$ of the radial polynomial from top to bottom according to $(n,m) = (3,1),\;(16,6)$. Setting of aperture variables: $s_0=0.8,\;s_{0,M}=0.4$ and setting of focal and radial variable: $f=10,\;r=R_{\max} \frac{n_r-1}{N_r-1}$ for $n_r = 1,\;2,\cdots,\;N_r$ with $R_{\max}=100$ and $N_r=10$. }
\label{fig17}
\end{center}
\end{figure}

\begin{figure}[htp]
\begin{center}
\plaatje{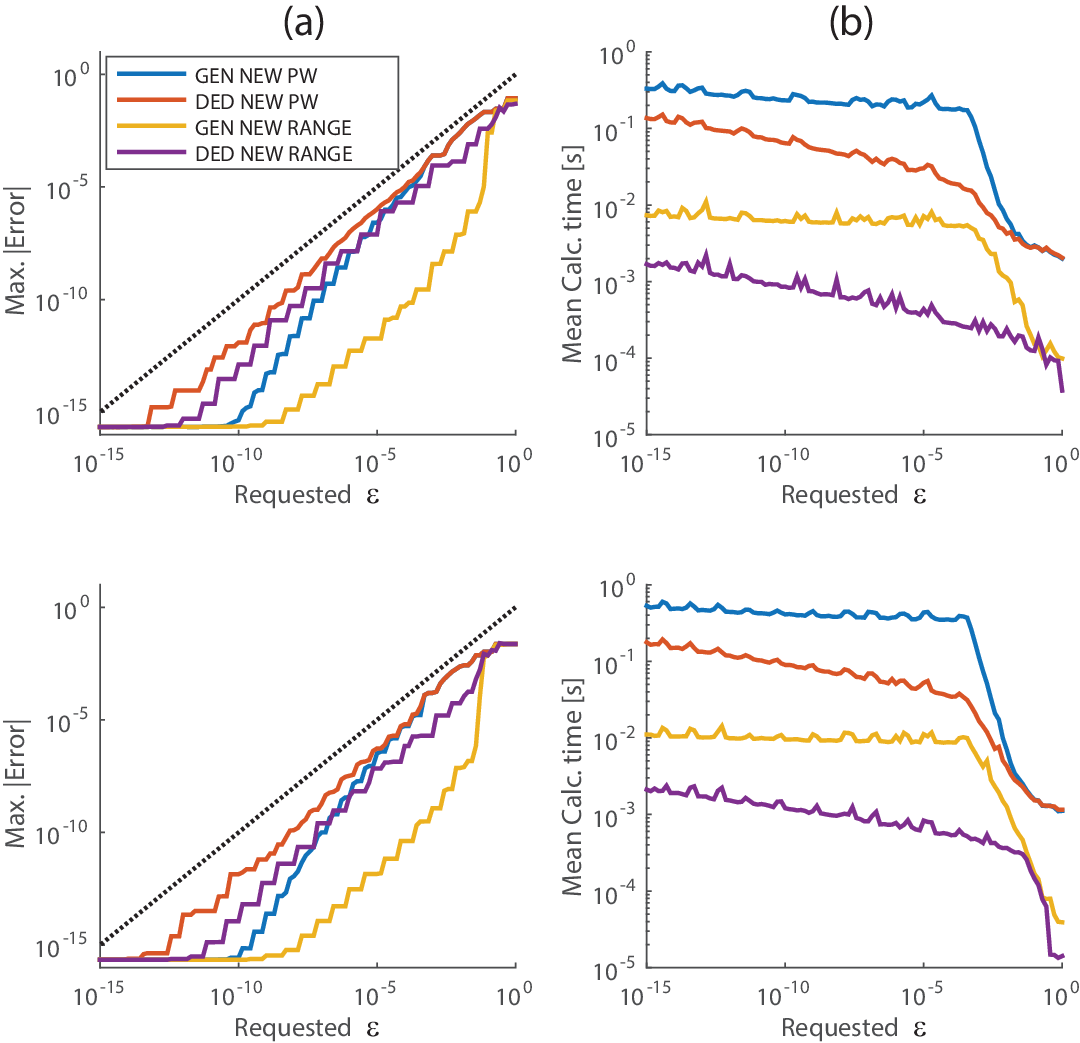}{1.0}
\caption{Same caption as Figure \ref{fig17} but now with $N_r=100$. }
\label{fig18}
\end{center}
\end{figure}

We shall now illustrate the advantage of using the truncation rules valid for an entire range $0\leq r \leq R_{\max}$ over the new general and dedicated truncation rules with pointwise validity. First of all, it is very convenient to have one truncation rule that is valid within a given range of $r$ values, not having to recalculate the summation cut-offs for each point at which the integral $I$ is to be calculated. At first sight, it might seem there is a price to be paid for this convenience in the form of non-optimal summation ranges leading to increased computation times. However, this is not necessarily true since the truncation rules valid for a range $0\leq r \leq R_{\max}$ do enable one to implement an algorithm that computes the integral $I(r)$ for a range of $r$ values simultaneously. By doing so, the overhead of calculating $H$ and $T$ for each value of $r$ is avoided and this appears to compensate by far the increase of computation time due to the non-optimal values of $H$ and $T$ for some $r$ values in the range $0\leq r \leq R_{\max}$. This is illustrated in Figures~\ref{fig17}~and~\ref{fig18} where we have plotted the absolute accuracy (a) and mean calculation times (b) as a function of requested accuracy $\eps$ for both the pointwise and range rules. For the pointwise rules, the mean calculation time is obtained as the average of all single $r$ value computations, while for the range rules the mean calculation time is obtained as the time required to calculate the integral for all $r$ values simultaneously and divide by the number of $r$ values, $N_r$. 

\begin{figure}[b!]
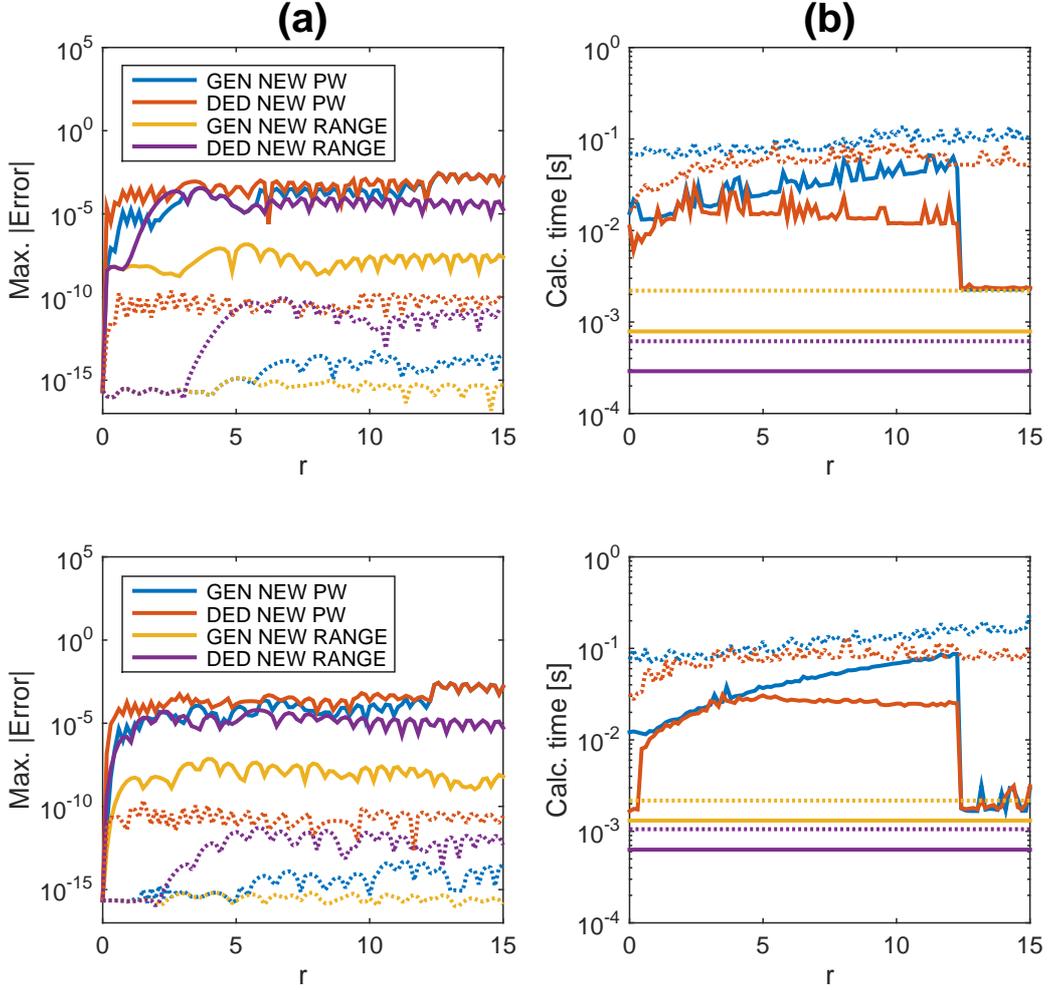

\begin{center}
\plaatje{Fig19}{1.0}
\caption{Absolute accuracy (a) and computation time (b) as a function of the radial variable $r$ using the new general and dedicated truncation rules valid for a single point (GEN/DED NEW PW) and corresponding truncation rules valid for a given range $0\leq r\leq R_{\max}$ (GEN/DED NEW RANGE) where the degree $n$ and azimuthal order $m$ of the radial polynomial is varied from top to bottom according to $(n,m) = (3,1),\;(16,6)$. Solid and dotted lines pertain to a requested accuracy of $\eps = 10^{-2}$, and $\eps= 10^{-8}$, respectively. Setting of aperture variables: $s_0=0.8,\;s_{0,M}=0.4$ and setting of focal and radial variable: $f=10,\;r=R_{\max} \frac{n_r-1}{N_r-1}$ for $n_r = 1,\;2,\cdots,\;N_r$ with $R_{\max}=15$ and $N_r=100$. }
\label{fig19}
\end{center}
\end{figure}

In addition, we show in Figure \ref{fig19} the observed absolute accuracy (a) and calculation time (b) for 100 $r$ values in the range $0\leq r \leq 15$ that are computed with a requested accuracy $\epsilon=10^{-2}$ (solid lines) and $\epsilon=10^{-8}$ (dotted lines) when varying the degree $n$ and azimuthal order $m$ of the radial polynomial from top to bottom according to $(n,m) = (3,1),\;(16,6)$ (see the figure caption for the remaining parameters). In this figure, the calculation time per $r$ value for the range rules is constant as it is obtained as the time required to compute all 100 $r$ values in the range $0\leq r \leq 15$ divided by the number of $r$ values, $N_r=100$. 

The graphs in Figure \ref{fig19}b for the GEN NEW PW and DED NEW PW with $\eps=10^{-2}$ show a sharp drop around $r=12$. This due to the fact that the Jinc-functions have a general $r^{-3/2}$-decay, causing the amplitudes of the decisive terms in the PW truncation rules to drop below the relatively large value $\eps=10^{-2}$ for relatively small values of $r$. In particular, the calculation time for the PW rules is, in general, not always an increasing function of $r$.

\newpage
\subsection{Treatment of $I_\textrm{VMML}$}\label{subsec7.2} 
\mbox{} \\[-9mm]

We recall from \cite{ref1}, Subsec.~3.5 and Sec.~9, that $I_{\rm VMML}$ is given by
\begin{eqnarray} \label{e111}
I_{{\rm VMML}}&=& \il_0^1\,\frac{\{(1-s_{0,h}^2\rho^2)^{1/2}\pm(1-s_{0,M}^2\rho^2)^{1/2}\}^{-|j|+1}} {(1-s_0^2\rho^2)^{1/4}(1-s_{0,M}^2\rho^2)^{3/4}}\,\times\\
&&\exp\Bigl[\frac{if}{u_{0,h}}\,\Bigl(1-\sqrt{1-s_{0,h}^2\rho^2}\,\Bigr)\Bigr] \rho^{|j|}\,R_n^{|m|}(\rho)\,J_{m+j}(2\pi r\rho)\,\rho\,d\rho~\nonumber
\end{eqnarray}
(the $u_0$ in \cite{ref1}, Eq.~(32), should be replaced by $u_{0,h}$). $I_{\rm VMML}$ is the radial part of the diffraction integral corresponding to the Zernike term $Z^m_n$ that occurs when an object at finite distance is imaged by a high-NA optical system in a multi-layered focal region, as it has been given in \cite{ref4}. The basic assumptions in \cite{ref4} are non-absorbing layers with refractive index $n_h$ of layer $h$ satisfying $n_h>n_1s_0$ and absence of non-propagating waves. The $s_{0,M}$ and $s_0$ in Eq.~(\ref{e111}) account for the refractive indices in  object space and homogeneous part of the image space, respectively,  as well as for the magnification due to finite distance of the object to the optical system. The $+$-sign and $-$-sign in Eq.~(\ref{e111}) refer to the forward and backward propagating waves, respectively. Finally, the integer $j$ satisfies $|j|=0,\;1,\;2$.

It is apparent from Eq.~(\ref{e111}) that singular behaviour of $I_{\rm VMML}$, with choice of the $-$-sign and $|j|=2$, occurs when $s_{0,h}^2-s_{0,M}^2$ is zero or very small. This is due to the fact that in \cite{ref1} a factor $s_{0,M}^2-s_{0,h}^2$, that does occur indeed in \cite{ref4}, Eqs.~(33-34) in front of all $V$-functions with $|j|=2$, has been omitted. We thank Prof. J. Braat for observing this to us. Restoring this factor $s_{0,M}^2-s_{0,h}^2$, the singular behavior disappears. In particular, in the cases that $s_{0,M} = s_{0,h}$, the diffraction integral with $+$-sign and $|j|=2$ vanishes, while the one with choice of the $-$-sign and $|j|=2$,  yields
\begin{eqnarray} \label{e112}
&&\int^1_0 \frac{2}{(1-s_0^2\rho^2)^{1/4}(1-s_{0,M}^2\rho^2)^{3/4}}\times \nonumber\\
&&\;\;\;\;\;\;\;\;\;\;\;\;\;\;\;\;\;\;\;
\exp\left[\frac{if}{u_{0,h}}(1-\sqrt{1-s_{0,h}^2\rho^2})\right]
 R^{|m|}_n(\rho)J_{m+j}(2\pi r \rho)\rho\,d\rho~.~\;\;\;\;\;\;\;\;
\end{eqnarray}
Note the mismatch between the azimuthal order $|m|$ of the circle polynomial and the order $m+j,\; |j|=2$ of the Bessel functions in Eq.~(\ref{e112}).

In \cite{ref1}, Sec.~9, the evaluation of the $I_{\rm VMML}$-integral is done separately for the cases $|j|=0,\;1,\;2$, where the cases $|j| = 0,\;1,$ give rise to one or two integrals that behave in all respects the same as the $I_{\rm VM}$-integral. The cases with $|j|=2$ yield a complication, due to the fact that a factor $1/\rho^2$ occurs in front of a difference ($+$-sign) and a sum ($-$-sign) of two algebraic functions. In the case of the $+$-sign, the difference of the algebraic functions vanishes at $\rho^2=0$ so that the factor $1/\rho^2$ cancels, and one can proceed as in the case of the $I_{\rm VM}$-integral. In the case that $|j|=2$ with the $-$-sign the factor $1/\rho^2$ does not cancel, and one ends up with the $I_{\rm VMML}$-integral ($|j|=2,\;-$-sign)
\begin{eqnarray} \label{e113}
I_{{\rm VMML}} &=& \nonumber\\
&&\hspace{-2cm}\il_0^1\,\frac{1}{s_{0,M}^2-s_{0,h}^2}\,\left[\frac{1-s_{0,h}^2\rho^2}
{(1-s_0^2\rho^2)^{\tfrac{1}{4}}(1-s_{0,M}^2 \rho^2)^{\tfrac{3}{4}}}+\frac{(1-s_{0,h}^2\rho^2)^{\tfrac{1}{2}}}{(1-s_0^2\rho^2)^{\tfrac{1}{4}}(1-s_{0,M}^2\rho^2)^{\tfrac{1}{4}}}\right]\,\times \nonumber \\
& & \;\;\;\exp\Bigl[\frac{if}{u_{0,h}}\,\Bigl(1-\sqrt{1-s_{0,h}^2\rho^2}\,\Bigr)\Bigr]\, R_n^{|m|}(\rho)\,J_{m+j}(2\pi r\rho)\,\rho\,d\rho\;.
\end{eqnarray}
This is the $I_{\rm VMML}$-integral of \cite{ref1}, Eq.~(112), where we have carried through a minor correction. Note that the factor $\rho^{|j|}$ that occurs in Eq.~(\ref{e111}) has been canceled, compare with Eq.~(\ref{e112}), and so there is a mismatch between the azimuthal order $m$ of the circle polynomial and the order $m+j$ of the Bessel function. Consequently, in the double series in Eq.~(\ref{e6}), one has now, instead of the Jinc functions $J_{h+1}(2\pi r)/(2 \pi r)$, arising from the basic integral result of the classical Nijboer-Zernike theory, that the integrals
\beq \label{e114}
I^{\pm}_{mh}(r) = \int^1_0 R^{|m|}_n(\rho) J_{m\pm 2}(2\pi r\rho)\rho d\rho
\eq
for the cases $j=\pm2$ appear.

It has been shown in \cite{ref1}, Appendix H, Eq.~(221), that
\beq \label{e115}
I^+_{mh} = \sum^{\infty}_{k=0} D_{mhk} \frac{J_{h+2k+1}(2\pi r)}{2\pi r}\;,
\eq
with bounded coefficients $D$. The decay of the $I^+_{mh}$ in $h\geq1$ is therefore qualitatively the same as the decay of the Jinc functions $J_{h+1}(2\pi r)/(2\pi r)$, compare \cite{ref1}, Figures  5 and 6. Hence, the case $j=2$ does not need separate consideration.

For the case that $j=-2$, it is shown in \cite{ref1}, Appendix H, that one can restrict to $m\geq0$, and that the cases with $m=0,\;1$ lead to $I^+_{0h},\;J_{h+1}(2\pi r)/(2\pi r)$, and these do not need separate consideration. For the cases $m=2,\;3,\;\cdots\;$ there has been given in \cite{ref1}, Appendix H, Eq.~(201), the result
\beq \label{e116}
I^-_{mh} = \sum^{\tfrac{1}{2}(h-m)+1}_{l=0} H_{mhl} \frac{J_{m+2l-1}(2\pi r)}{2\pi r}
\eq
with explicit and well-behaved coefficients $H$ that, however, do not exhibit decay in $h$. The summation range in Eq.~(\ref{e116}) is such that the minimal order of the Bessel functions involved does not tend to $\infty$. Therefore, super exponential decay in $h$ should not be expected. Indeed, it has been shown in \cite{ref1}, Appendix H, Eq.~(203), that
\beq \label{e117}
|I^-_{mh}| \approx \frac{2(m-1)(2\pi r)^{m-2}}{h^m}\;,\;\;h\geq 2\pi r\;.
\eq
Therefore, decay of $I^-_{mh}$ in $h$ is just like $Ch^{-m}$. This slow decay is clearly demonstrated in \cite{ref1}, Figure 6, where larger $m$ give more rapid, but still relatively slow, decay. Interestingly, it can also be observed from \cite{ref1}, Figure 6 that decay in $h \to \infty$ is independent of $r$ when $m=2$, confirming Eq.~(\ref{e117}), the right-hand side of which is independent of $r$ when $m=2$.

It is obvious that the truncation issue for the double series representation of the $I_{\rm VMML}$ as in Eq.~(\ref{e6}), with $I$-functions rather than Jinc functions, for the case $j=-2$ and $-$-sign, cannot be forced into the same framework that worked well for all integrals that can be treated as the $I_{\rm VM}$-integral. Rather than developing a whole truncation strategy for this rare and exceptional case, with clever bounds and convenient arrangements, we just conduct some experiments to show what sort of accuracies can be achieved with a particular amount of computation time. Here we may point out that the formulation of a general truncation rule, in which the product $c_t I^{-}_{mh}$ is bounded by bounding $|c_t|$ and $|I^-_{mh}|$ separately, is impracticable due to the slow decay of $|I^-_{mh}|$. It would be much better to follow the approach that leads to the dedicated rule, in which the product  $c_t I^{-}_{mh}$ is bounded by inspecting the product of upper bounds for $|c_t|$ and $I^-_{mh}$ at the points $(h,2t)$ in $S^m_n$, which is contained in the set $|h-2t|\leq n$. Here advantage can be taken of the facts that $|c_t|$ decays rapidly and that $|I^-_{mh}|$, while not decaying rapidly, is properly bounded as in Eq.~(\ref{e117}) and by $\tfrac{1}{2}(2h+3)^{-\tfrac{1}{2}}$ (using the Cauchy-Schwarz inequality in Eq.~(\ref{e114})).

\begin{figure}[tbp]
\begin{center}
\plaatje{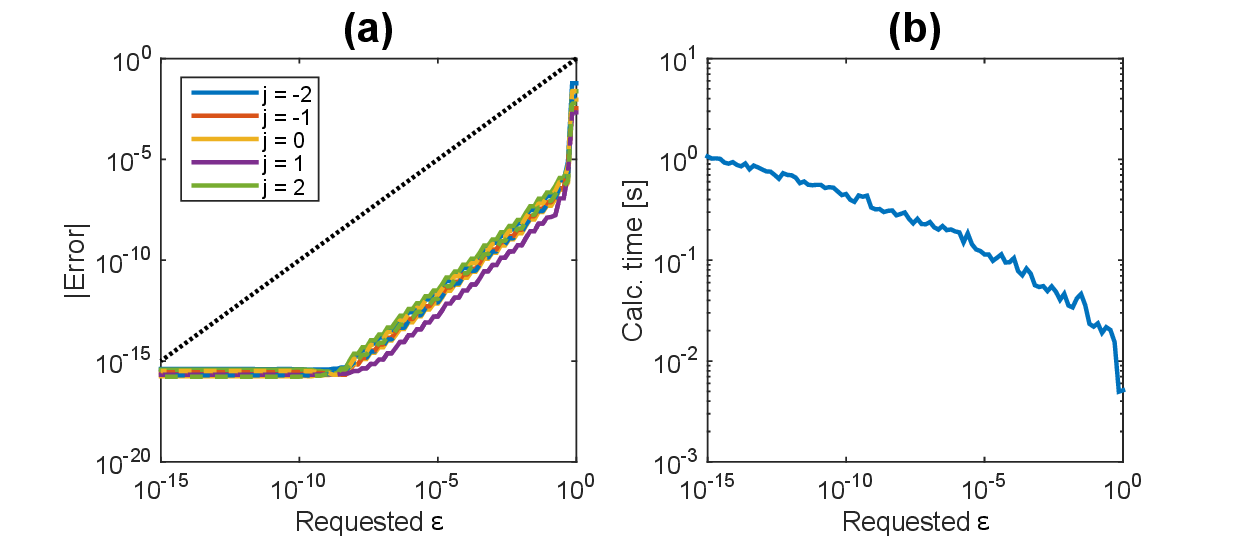}{1.1}
\caption{Absolute accuracy (a) and overall computation time (b) as a function of requested absolute accuracy $\eps$ for the $I_{VMML}$ with parameters $n=m=2$, $j=-2,-1,0,1,2$, $f=10$, $r=1$, $s_0 = 0.95$, $s_{0M} = 0.25$ and $s_{0h} = 0.85$ where the solid and dashed lines pertain to the $-$-sign and $+$-sign cases, respectively. Note that displayed computation times pertain to calculating $I_{VMML}$ for both the $-$-sign and $+$-sign case, and all values of $j$, simultaneously as a function of requested $\eps$. }
\label{fig20}
\end{center}
\end{figure}

In Figure \ref{fig20}, we take a hybrid approach for simplicity. For a given $n$ and $m$, we let $T=T^\textrm{gen}$ and $H=n+2t$, so that we include all $h,\;t$ with $h\leq H$ and $t\leq T$ in the double series, i.e., all $(h,2t)$ in the non-zero range in Figure \ref{fig4} with $t\leq T$. In Figure \ref{fig20}a, one can observe that the given truncation rules for $I_{\rm VMML}$ achieve an error level well within the requested accuracy. At first, it might seem surprising that all curves in Figure \ref{fig20}a show similar behavior while we have extensively discussed in this Chapter that the convergence behavior of $I_{\rm VMML}^-$ with $j=-2$, is a case that needs a special treatment. In this respect we should note that the used implementation, to compute the $I_{\rm VMML}$ integrals, generates values for $I_{\rm VMML}^+$ and $I_{\rm VMML}^-$ and all values of $j=-2,-1,0,1,2$ simultaneously. Consequently, the worst case truncation rules, those pertaining to the $-$-sign and $j=-2$, are applied to all values to be computed and there results only a single curve in Figure \ref{fig20}b which represents the calculation time for all values. Although this approach might seem suboptimal, this is not the case, because it turns out that longer summation ranges due to using the worst case truncation values for $T$ and $H$ are more than compensated by the reduction in overhead achieved by computing all integral values for both signs and all $j$ simultaneously.

\newpage

\setcounter{equation}{0}
\renewcommand{\theequation}{A\arabic{equation}}
\appendix

\section{Results on $\varp$-functions} \label{appA}
\mbox{} \\[-9mm]

In this appendix, we present results on the functions
\beq \label{a1}
\varp(x\,;\,c)=\left\{\ba{llll}
0 & \!\!, & ~~~0\leq x\leq c & \!\!, \\[3mm]
x\,{\rm arccosh}(x/c)-c\,\sqrt{(x/c)^2-1} & \!\!, & ~~~x\geq c & \!\!,
\ea\right.
\eq
and
\beq \label{a2}
\psi(x\,;\,c,d)=\varp(x\,;\,c)-\varp(x\,;\,d)~,~~~~~~x\geq0~,
\eq
where $d>c>0$. In Eq.~(\ref{a1}), we have
\beq \label{a3}
{\rm arccosh}(y)={\rm ln}(y+\sqrt{y^2-1})=\il_1^y\,\frac{dz}{\sqrt{z^2-1}}~,~~~~~~y\geq1~,
\eq
and this is a non-negative, non-decreasing function of $y$. Furthermore, with ``$\,'\,$'' denoting differentiation with respect to $x$,
\beq \label{a4}
\varp'(x\,;\,c)=\left\{\ba{llll}
0 & \!\!, & ~~~0\leq x\leq c & \!\!, \\[3mm]
{\rm arccosh}(x/c) & \!\!, & ~~~x\geq c & \!\!,
\ea\right.
\eq
so that $\varp(x\,;\,c)$ is continuously differentiable in $x\geq0$. From Eqs.~(\ref{a3}--\ref{a4}), it is seen that $\varp(x\,;\,c)$ is non-negative, non-decreasing and convex in $x\geq0$, and strictly so in $x>c$. Also, $\varp(x\,;\,c)$ behaves like $x\,{\rm ln}(2x/ec)$ for large $x>0$, and grows therefore super-linearly.

We next consider $\psi(x\,;\,c,d)$ in Eq.~(\ref{a2}). From
\beq \label{a5}
\frac{\partial\varp}{\partial c}\,(x\,;\,c)=\left\{\ba{llll}
0 & \!\!, & ~~~0\leq x\leq c & \!\!, \\[3mm]
\dfrac{-1}{c}\,\sqrt{x^2-c^2} & \!\!, & ~~~x\geq c & \!\!,
\ea\right.
\eq
we have that $\varp(x\,;\,c)$ is decreasing in $c>0$ for any $x$, and so $\psi(x\,;\,c,d)$ is non-negative. Furthermore,
\beq \label{a6}
\psi'(x\,;\,c,d)=\left\{\ba{llll}
0 & \!\!, & ~~~0\leq x\leq c & \!\!, \\[3mm]
{\rm arccosh}(x/c) & \!\!, & ~~~c\leq x\leq d & \!\!, \\[3mm]
{\rm arccosh}(x/c)-{\rm arccosh}(x/d) & \!\!, & ~~~x\geq d & \!\!,
\ea\right.
\eq
and this shows that $\psi(x\,;\,c,d)$ is non-decreasing in $x\geq0$, and strictly so in $x\geq c$. Moreover, we have for $x>d$
\beq \label{a7}
\psi''(x\,;\,c,d)=\frac{1}{\sqrt{x^2-c^2}}-\frac{1}{\sqrt{x^2-d^2}}<0~,
\eq
and so $\psi(x\,;\,c,d)$ is strictly concave in $x>d$, while $\psi(x\,;\,c,d)$ is strictly convex in $x\in(c,d)$. Finally,
\beq \label{a8}
\psi'(x\,;\,c,d)={\rm ln}\Bigl(\frac{d}{c}\Bigr)+{\rm ln}\Bigl(\frac{x+\sqrt{x^2-c^2}}{x+\sqrt{x^2-d^2}}\Bigr)>{\rm ln}\Bigl(\frac{d}{c}\Bigr)~,~~~~~~ x>d~,
\eq
which shows that $\psi'(x\,;\,c,d)$ decreases to ${\rm ln}(d/c)$ as $x\pr\infty$, and we have directly from Eqs.~(\ref{a1}-\ref{a3})
\beq \label{a9}
\psi(x\,;\,c,d)-x\,{\rm ln}\Bigl(\frac{d}{c}\Bigr)={-}\,\frac{d^2-c^2}{4x}+O\Bigl(\frac{1}{x^3}\Bigr)~,
\eq
for $x>d$, so that $\psi(x\,;\,c,d)-x\,{\rm ln}(d/c)$ increases to 0 as $x\pr\infty$.

In the formulation of the general truncation rule, it has been used that one can find piecewise linear functions bounding $\varp(x\,;\,c)$ and $\psi(x\,;\,c,d)$ from below. Furthermore, in the design of the dedicated truncation rule, it is convenient to have convex functions bounding $\psi(x\,;\,c,d)$ from below (since $\varp(x\,;\,c)$ is itself convex, such an effort does not have to be made for $\varp$).

By convexity of $\varp(x\,;\,c)$, the graph of $\varp$ lies above any tangent line, and so for any $x_0>0$, we have
\beq \label{a10}
\varp(x\,;\,c)\geq\varp(x_0\,;\,c)+(x-x_0)\,\varp'(x_0\,;\,c)~,~~~~~~x\geq0~.
\eq
For a linear lower bound on $\psi(x\,;\,c,d)$, one must choose $x_0\in(c,d)$ such that $\psi'(x_0\,;\,c,d)\leq{\rm ln}(d/c)$, see Eq.~(\ref{a8}), and then
\beq \label{a11}
\psi(x\,;\,c,d)\geq\psi(x_0\,;\,c,d)+(x-x_0)\,\psi'(x_0\,;\,c,d)~,~~~~~~x\geq0~.
\eq
Since $x_0\in(c,d)$ and $\psi(x\,;\,c,d)=\varp(x\,;\,c)$ for $c\leq x\leq d$, we have from Eqs.~(\ref{a1}, \ref{a6}) that
\begin{eqnarray} \label{a12}
\psi(x\,;\,c,d) & \geq & x\,{\rm arccosh}(x_0/c)-c\,\,\sqrt{(x_0/c)^2-1} \nonumber \\[3mm]
& = & \gamma x-c\sinh(\gamma)~,~~~~~~x\geq0~,
\end{eqnarray}
where we have set $\gamma={\rm arccosh}(x_0/c)$. Choosing the largest possible $x_0\in(c,d)$, so that
\beq \label{a13}
\psi'(x_0\,;\,c,d)={\rm ln}(d/c)=:\gamma_0~,
\eq
we have
\beq \label{a14}
x_0=c\cosh(\gamma_0)~,~~~~~~\psi(x_0\,;\,c,d)=\gamma x_0-c\sinh(\gamma_0)~.
\eq
Hence, for any $\gamma\in(0,\gamma_0]$, we have
\beq \label{a15}
\psi(x\,;\,c,d)\geq\gamma x-c\sinh(\gamma)~,~~~~~~x\geq0~.
\eq
Evidently, since $\varp(x\,;\,c)\geq\psi(x\,;\,c,d)$, the latter bound is also valid for $\varp(x\,;\,c)$, without a restriction on $\gamma$. The choice $\gamma=1$ leads to
\beq \label{a16}
\varp(x\,;\,c)\geq x-c\sinh(1)~,~~~~~~x\geq0~.
\eq
The largest convex functon bounding $\psi(x\,;\,c,d)$ from below is given by
\beq \label{a17}
\varp(x\,;\,c,d)=\left\{\ba{llll}
\varp(x\,;\,c) & \!\!, & ~~~0\leq x\leq c\cosh(\gamma_0) & \!\!, \\[3mm]
\gamma_0x-c\sinh(\gamma_0) & \!\!, & ~~~x\geq c\cosh(\gamma_0) & \!\!.
\ea\right.
\eq

We conclude this appendix by showing 3 inequalities. The first one of these reads
\beq \label{a18}
\varp(x\,;\,c)+\tfrac32\,{\rm ln}\,c\geq\varp(x\,;\,1)~,~~~~~~0<c\leq 1~,
\eq
when $x\geq\tfrac12\,\sqrt{13}$, and is required in Appendix~B. We have by Eq.~(\ref{a5}) for $0<c\leq1\leq x$ that
\beq \label{a19}
\frac{d}{dc}\,[\varp(x\,;\,c)+\tfrac32\,{\rm ln}\,c]=\frac1c\,(\tfrac32-\sqrt{x^2-c^2})~,
\eq
and this is negative for all $c\in(0,1]$ when $\sqrt{x^2-1}\geq 3/2$, i.e., when $x\geq\sqrt{13}$. Since there is equality in Eq.~(\ref{a18}) when $c=1$, we get the result.

Next, we show that for $\alpha>0$ and $x\geq c\geq\alpha\geq0$
\beq \label{a20}
\varp(x+\alpha\,;\,c)-\varp(x\,;\,c)-\alpha\,{\rm ln}\Bigl(\frac{x+\alpha}{c}\bigr)\geq0~.
\eq
This is required in Appendix~C with $\alpha=1/2$ and $c\geq1/2$.

To show Eq.~(\ref{a20}), we let $b=\alpha/c$, and we observe from Eq.~(\ref{a4}) that Eq.~(\ref{a20}) holds for $x\geq c$ if and only if \beq \label{a21}
\Phi(w\,;\,b):=\il_w^{w+b}\,{\rm arccosh}(v)\,dv-b\,{\rm ln}(w+b)\geq0
\eq
holds for $w:=x/c\geq1$. Now $\Phi(w\,;\,b=0)=0$, and
\beq \label{a22}
\frac{\partial\Phi}{\partial b}\,(w\,;\,b)={\rm ln}\,\Bigl[ 1+\Bigl(1-\frac{1}{(w+b)^2}\Bigr)^{1/2}\Bigr]-\frac{b}{w+b}
\eq
increases in $w\geq1$ for fixed $b\geq0$. Hence, when $b_0>0$ is such that
\beq \label{a23}
\frac{\partial\Phi}{\partial b}\,(1\,;\,b)\geq0~,~~~~~~0\leq b\leq b_0~,
\eq
we have that
\beq \label{a24}
\frac{\partial\Phi}{\partial b}\,(w\,;\,b)\geq0~,~~~~~~ 0\leq b\leq b_0\,,~~w\geq1~,
\eq
and so, from $\Phi(w\,;\,b=0)=0$, that $\Phi(w\,;\,b)\geq0$ for $w\geq1$ and $0\leq b\leq b_0$. Now with $z=\frac{1}{1+b}\in(0,1]$, \beq \label{a25}
\frac{\partial\Phi}{\partial b}\,(1\,;\,b)={\rm ln}(1+(1-z^2)^{1/2})-1+z
\eq
is a concave function of $z\in(0,1]$, since
\beq \label{a26}
\frac{d}{dz}\,[{\rm ln}(1+(1-z^2)^{1/2})]=\frac{-z}{1-z^2+(1-z^2)^{1/2}}
\eq
decreases from 0 at $z=0$ to $-\infty$ at $z=1$. Furthermore, the right-hand side of Eq.~(\ref{a25}) vanishes at $z=1$, has the value ${\rm ln}\,2-1<0$ at $z=0$, and the value ${\rm ln}(1+\tfrac12\,\sqrt{3})-\tfrac12=0.12\,...>0$ at $z=1/2$. Therefore, the right-hand side of Eq.~(\ref{a25}) is non-negative for $\tfrac12\leq z\leq1$. Hence, with $b=\alpha/c\in[0,1]$, so that $z=(1+b)^{-1}\in[\tfrac12\,,1]$, we have that Eqs.~(\ref{a23}--\ref{a24}) hold with $b_0=1$. It follows that Eq.~(\ref{a21}) holds for $0\leq b\leq1\leq w$, as required.

An inequality converse to Eq.~(\ref{a20}) reads
\beq \label{a26i}
\varp(x+\alpha\,;\,c)-\varp(x\,;\,c)-\alpha\,{\rm ln}\Bigl(\frac{x+\alpha}{c}\bigr)\leq \alpha\,{\rm ln}\,2
\eq
when $\alpha>0$ and $x\geq 0$, $x+\alpha\geq c \geq 0$, and follows easily from Eqs.~(\ref{a3}-\ref{a4}).

In Appendix~C, the inequality in Eq.~(\ref{a20}) is required for all $x\geq0$. We shall comment on this below.

We next show that for $x\geq d\geq c\geq\alpha\geq0$
\begin{eqnarray} \label{a27}
& \mbox{} & \Bigl[\varp(x+\alpha\,;\,c)-\alpha\,{\rm ln}\Bigl(\frac{x+\alpha}{c}\Bigr)\Bigr]-\Bigl[\varp(x+\alpha\,;\,d)-\alpha \,{\rm ln}\Bigl(\frac{x+\alpha}{d}\Bigr)\Bigr] \nonumber \\[3.5mm]
& & \hspace*{2.5cm}\geq~\varp(x\,;\,c)-\varp(x\,;\,d)~.
\end{eqnarray}
This is required in Appendix~C with $\alpha=1/2$ and $c\geq 1/2$.
For $x\geq d$, we have by Eqs.~(\ref{a3}--\ref{a4})
\begin{eqnarray} \label{a28}
& \mbox{} & \Bigl[\varp(x+\alpha\,;\,c)-\varp(x\,;\,c)-\alpha\,{\rm ln}\Bigl(\frac{x+\alpha}{c}\Bigr)\Bigr] \nonumber \\[3.5mm]
& & \hspace*{1.5cm}-~\Bigl[\varp(x+\alpha\,;\,d)-\varp(x\,;\,d) -\alpha\,{\rm ln}\Bigl(\frac{x+\alpha}{d}\Bigr)\Bigr] \nonumber \\[3.5mm]
& & =~\il_x^{x+\alpha}\,\Bigl({\rm arccosh}\Bigl(\frac{y}{c}\Bigr)-{\rm arccosh}\Bigl(\frac{y}{d}\Bigr)\Bigr)\,dy+\alpha\,{\rm ln}\Bigl(\frac{c}{d}\Bigr) \nonumber \\[3.5mm]
& & =~\il_x^{x+\alpha}\,{\rm ln}\Bigl(\frac{y+\sqrt{y^2-c^2}}{y+\sqrt{y^2-d^2}}\Bigr)\,dy\geq0~,
\end{eqnarray}
and this is the required inequality.

The inequalities in Eqs.~(\ref{a20}, \ref{a27}) are required in Appendix~C for all $x\geq0$. Since $\varp(x+\alpha\,;\,c)$ and $\varp(x\,;\,c)$ vanish when $x+\alpha\leq c$, we have that Eq.~(\ref{a20}) holds for all $x\geq0$, except perhaps when $c-\alpha\leq x\leq c$. In this latter case, we have that $\varp(x\,;\,c)=0$, and therefore the left-hand side of Eq.~(\ref{a20}) can be written as
\beq \label{a29}
\varp(x+\alpha\,;\,c)-\alpha\,{\rm ln}\bigl(\frac{x+\alpha}{c}\bigr)= \alpha\,[c'(v\,{\rm arccosh}\,v-\sqrt{v^2-1})-{\rm ln}\,v]~,
\eq
where we have set $c'=c/\alpha$ and $v=(x+\alpha)/c\in[1,1+1/c']$. Now the minimum of
\beq \label{a30}
c'(v\,{\rm arccosh}\,v-\sqrt{v^2-1})-{\rm ln}\,v
\eq
is assumed at $v$ such that $v\,{\rm arccosh}\,v=1/c'$ (this $v$ is indeed in $[1,1+1/c']$), and this minimum increases in $c'$. For the case that $c'=c/\alpha=1$, we find numerically the minimum value $-0.109709667$. Hence, for the case that $\alpha=1/2$, as considered in Appendix~C, we are dealing with a minimum value of the whole left-hand side of Eq.~(\ref{a20}) of the order $-0.05$. This can safely be ignored, and so we declare Eq.~(\ref{a20}) to be valid for all $x\geq0$.

A similar situation arises for the inequality in Eq.~(\ref{a27}) whose validity is ensured for $x\geq d$, $x\leq c-\alpha$ and $c\leq x\leq d-\alpha$ (in the latter case, the second term in $[~]$ in Eq.~(\ref{a28}) is non-positive, while the first term in $[~]$ is non-negative by Eq.~(\ref{a20})). So we only need to consider $c-\alpha\leq x\leq c$ and $d-\alpha\leq x\leq d$ (these two $x$-intervals overlap when $d-\alpha\leq c$). The minimum value of the first term in $[~]$ in Eq.~(\ref{a28}) has been bounded from below by $-0.109709667\alpha$. The second term can be written on $d-\alpha\leq x\leq d$ as
\beq \label{a31}
\alpha\,[d'(v\,{\rm arccosh}\,v-\sqrt{v^2-1})-{\rm ln}\,v]
\eq
with $d'=d/\alpha\geq1$ and $v=(x+\alpha)/d\in[1,1+1/d']$. The function
\beq \label{a32}
f(v)=d'(v\,{\rm arccosh}\,v-\sqrt{v^2-1})-{\rm ln}\,v~,~~~~~~v\geq1~,
\eq
is convex, and so its maximum over $[1,1+1/d']$ occurs at $v=1$, with value $f(1)=0$, or at $v=1+1/d'$, with value
\beq \label{a33}
\frac1u\,[(1+u)\,{\rm arccosh}(1+u)-\sqrt{(1+u)^2-1}]-{\rm ln}(1+u)~,
\eq
where $u=1/d'\in[0,1]$. An elementary analysis of the function in Eq.~(\ref{a33}) shows that it is maximal at $u=0.191487884$, with maximal value $0.2486813544$. Hence, for the case $\alpha=1/2$, as considered in Appendix~C, we are dealing with a maximum value of the second term in $[~]$ in Eq.~(\ref{a28}) that can be bounded by $1/8$. This can be safely ignored, and we thus declare Eq.~(\ref{a27}) to be valid for all $x\geq0$ and $d\geq c\geq\alpha$.

\setcounter{equation}{0}
\renewcommand{\theequation}{B\arabic{equation}}
\section{Bounding Jinc functions} \label{appB}
\mbox{} \\[-9mm]

In this appendix, we bound and estimate Jinc functions $J_{h+1}(2\pi r)/2\pi r$ for $h=0,1,...$ and $r>0$.

We first consider the case that $h+1<2\pi r$. Let $\beta\in(0,\pi/2)$ be fixed, and let $\xi=\nu(\tan\beta-\beta)-\tfrac14\,\pi$. With $\sec\beta=1/\cos\beta>1$, the first term of Debye's asymptotic result \cite{ref2}, 10.19.6, p.~231 as $\nu\pr\infty$ yields the approximation
\beq \label{b1}
J_{\nu}(\nu\sec\beta)+i\,Y_{\nu}(\nu\sec\beta)\approx\Bigl(\frac{2}{\pi\nu \tan\beta}\Bigr)^{1/2}\,e^{i\xi}~,
\eq
where $J_{\nu}$ and $Y_{\nu}$ are the Bessel functions of first and second kind, respectively, and of order $\nu$. With $\nu=h+1$ and $\beta$ such that $h+1=2\pi r\cos\beta$, we have
\beq \label{b2}
\Bigl(\frac{2}{\pi\nu\tan\beta}\Bigr)^{1/2}=\frac{1}{\pi\sqrt{r}}\, \Bigl(1-\Bigl(\frac{h+1}{2\pi r}\Bigr)^2\Bigr)^{-1/4}~.
\eq
The factor $(1-((h+1)/2\pi r)^2)^{-1/4}$ is close to 1 on a large part of the range $0\leq h+1<2\pi r$, and we shall replace it by 1 (this issue is further addressed below). We thus estimate
\beq \label{b3}
\Bigl|\frac{J_{h+1}(2\pi r)}{2\pi r}\bigr|\leq\frac{1}{2\pi^2\,r\,\sqrt{r}}~,~~~~~~0\leq h+1<2\pi r~.
\eq

We next consider the case that $h+1>2\pi r$. With ${\rm sech}\,\alpha=1/\cosh\alpha<1$, the first term of Debye's asymptotic result \cite{ref2}, 10.19.3, p.~231 as $\nu\pr\infty$ yields the approximation
\beq \label{b4}
J_{\nu}(\nu\,{\rm sech}\,\alpha)\approx\frac{\exp(\nu(\tanh\alpha-\alpha))}{(2\pi \nu\tanh\alpha)^{1/2}}~.
\eq
With $\nu=h+1$ and $\alpha$ such that $h+1=2\pi r\,{\cosh}\,\alpha$, we have
\beq \label{b5}
\Bigl(\frac{1}{2\pi\nu\tanh\alpha}\Bigr)^{1/2}=\frac{1}{2\pi\,\sqrt{r}}\, \Bigl(\Bigl(\frac{h+1}{2\pi r}\Bigr)^2-1\Bigr)^{-1/4}~.
\eq
We replace the factor $(((h+1)/2\pi r)^2-1)^{-1/4}$ at the right-hand side of Eq.~(\ref{b5}) by 1 as before, and we observe that
\begin{eqnarray} \label{b6}
\nu(\tanh\alpha-\alpha) & = & 2\pi r\Bigl(\Bigl(\frac{h+1}{2\pi r}\bigr)^2-1\Bigr)^{1/2}-(h+1)\,{\rm arccosh}\Bigl(\frac{h+1}{2\pi r}\Bigr) \nonumber \\[3mm]
& = & {-}\varp(h+1\,;\,2\pi r)~,
\end{eqnarray}
with $\varp$ as in Appendix \ref{appA}.

We thus get on the whole range $h\geq0$ the estimate
\beq \label{b7}
\Bigl|\frac{J_{h+1}(2\pi r)}{2\pi r}\Bigr|\leq\frac{1}{2\pi^2\,r\,\sqrt{r}}\,\exp({-}\varp(h+1\,;\,2\pi r))~.
\eq

In deriving the bound in Eq.~(\ref{b7}), we have set the $(~)^{-1/4}$-factors in Eq.~(\ref{b2}, \ref{b5}) equal to 1. We shall now assess the amount by which the bounding function in Eq.~(\ref{b7}) is off by this simplification. At the point $h+1=2\pi r$ we have $\varp(h+1\,;\,2\pi r)=0$, and we are thus comparing the bound $(2/\pi\nu)^{1/2}$ for $J_{\nu}(\nu)$ by its actual value when $\nu=h+1=2\pi r\pr\infty$. In \cite{ref2}, 10.14.2, p.~227, there is the bound, for $0<x<\nu$,
\beq \label{b8}
0<J_{\nu}(x)<J_{\nu}(\nu)=\frac{2^{1/3}}{3^{2/3}\,\Gamma(2/3)\,\nu^{1/3}}= 0.4473\nu^{-1/3}~.
\eq
The asymptotic value of the maximum of $|J_{\nu}(x)|$ over all $x>0$ is $\approx\:0.6748\nu^{-1/3}$ (assumed near $x=\nu+(\nu/2)^{1/3}$), and this has to be compared with $(2/\pi\nu)^{1/2}$. The ratio of the asymptotic maximum value and $(2/\pi\nu)^{1/2}$ is $\approx\:0.8457\nu^{1/6}$. The quantity $0.8457\nu^{1/6}$ equals 1, 2 and 4 for $\nu=2.73$, 175 and 11194, respectively.

The bound in Eq.~(\ref{b7}) is somewhat awkward to use when $r$ is close to 0. With $R=\max(1/2\pi,r)$, we have
\beq \label{b9}
\Bigl|
\frac{J_{h+1}(2\pi r)}{2\pi r}\Bigr|\leq\frac{1}{2\pi^2\,R\,\sqrt{R}}\,\exp({-}\varp(h+1\,;\,2\pi R))~.
\eq
Indeed, when $r\geq 1/2\pi$, the two right-hand sides of Eqs.~(\ref{b7}, \ref{b9}) are equal. When $0<r<1/2\pi$ and $h\geq1$, the right-hand side of Eq.~(\ref{b7}) is less than the right-hand side of Eq.~(\ref{b9}) which follows from Eq.~(\ref{a18}) with $0<c=r/R<1$ and $x=h+1\geq2>\tfrac12\,\sqrt{13}$. The case $h=0$ needs separate consideration. The inequality to be proved is then
\beq \label{b10}
\Bigl|\frac{J_1(x)}{x}\Bigr|\leq\frac{1}{y}\,\sqrt{\frac{2}{\pi y}}
\eq
when $x>0$ and $y=\max(1,x)$. When $0<x\leq1$, we have $y=1$ and the right-hand side of Eq.~(\ref{b10}) equals $\sqrt{2/\pi}$ which exceeds the maximum value $1/2$ of $|J_1(x)/x|$. When $x\geq1$, the inequality to be shown reads $x\,J_1^2(x)\leq2/\pi$. It follows from \cite{ref3}, \S 13.74 that $x(J_1^2(x)+Y_1^2(x))$ decreases to $2/\pi$ when $x\pr\infty$. The maximum value of $x\,J_1^2(x)$ is just slightly larger than $2/\pi$ (0.6652 near $x=2.00$, compared to $2/\pi=0.6366$). We shall ignore this minor excess.

\setcounter{equation}{0}
\renewcommand{\theequation}{C\arabic{equation}}
\section{Bounding structural quantities} \label{appC}
\mbox{} \\[-9mm]

In this appendix, we bound and estimate the structural quantities $c_t$ required in Eq.~(\ref{e1}). At this point, we are interested in a manageable bound that can be used to formulate transparent truncation rules. To achieve this, we argue somewhat heuristically. We make the observation that the algebraic factor $a(\rho)$ is composed from functions $(1-s^2\rho^2)^{\delta}$ with $|\delta|\leq 3/4$. Any such function can be written as
\begin{eqnarray} \label{c1}
(1-s^2\rho^2)^{\delta} & = & \exp(2\delta\,{\rm ln}(1-(1-(1-s^2\rho^2)^{1/2}))) \nonumber \\[3mm]
& \approx & \exp({-}2\delta(1-(1-s^2\rho^2)^{1/2}))~,
\end{eqnarray}
where the latter function has the appearance of a focal factor with imaginary value of the normalized focal parameter $f/u_0$ of order unity. Moving a factor $\sqrt{1-s_0^2\rho^2}$ from the focal factor to the algebraic factor, see Eqs.~(\ref{e34}--\ref{e35}), we are led to estimate the Zernike coefficients $c_t$ of $a(\rho)\,f(\rho)$ by those of
\beq \label{c2}
\frac{a_0}{\sqrt{1-s_0^2\rho^2}}\,\exp\Bigl(\frac{ig}{u_0}\, (1-\sqrt{1-s_0^2\rho^2})\Bigr)~,
\eq
where $g=\max(1,|f|)$ and $a_0$ is the $R_0^0$-coefficient of $a(\rho)\,\sqrt{1-s_0^2\rho^2}$ as in Eq.~(\ref{e17}). Using the explicit form of the Zernike coefficients $b_t(g)$ of the modified focal factor, see Eqs.~(\ref{e4}, \ref{e35}, \ref{e38}), we thus postulate for $c_t$ the bound
\beq \label{c3}
a_0\,|b_t(g)|=a_0\,\frac{2t+1}{u_0}\,g\,|j_t(g/2)|\,|h_t^{(2)}(g/2v_0)|~.
\eq
Here it has been assumed that $s_0\leq s_{0,M}$. In the case that $s_{0,M}>s_0$, we should replace in the above all $s_0$ by $s_{0,M}$.

We next estimate $j_t$ and $h_t^{(2)}$ using Debye's asymptotic results. We have from Eq.~(\ref{e39}) and Appendix~B
\beq \label{c4}
|j_t(g/2)|=\sqrt{\frac{\pi}{g}}\,|J_{t+1/2}(g/2)|\leq\frac2g~,~~~~~~0\leq t+1/2\leq g/2~,
\eq
where we have replaced a factor $(1-((2t+1)/g)^2)^{-1/4}$ by 1. Similarly, we have from Eq.~(\ref{e40}) and Appendix~B
\beq \label{c5}
|h_t^{(2)}(g/2v_0)|\leq\frac{2v_0}{g}~,~~~~~~0\leq t+1/2\leq g/2v_0~,
\eq
where we have replaced a factor $(1-((2t+1)\,v_0/g)^2)^{-1/4}$ by 1. Hence,
\beq \label{c6}
|b_t(g)|\leq 4\,\frac{v_0}{u_0}~\frac{2t+1}{g}\leq 4\,\frac{v_0}{u_0}~,~~~~~~0\leq t+1/2\leq g/2~.
\eq

On the range $t+1/2\geq g/2$, we need to be more careful since the factor $(2t+1)\,g$ at the right-hand side of Eq.~(\ref{c3}) can become arbitrarily large. We estimate now, in accordance with the equality in Eq.~(\ref{c4}) and Eqs.~(\ref{b4}--\ref{b5}) with $h+1=t+1/2$ and $2\pi r=g/2$
\begin{eqnarray} \label{c7}
& \mbox{} & |j_t(g/2)| \nonumber \\[3.5mm]
& & \leq~ \sqrt{\frac{\pi}{g}}\,\frac{1}{2\pi\,\sqrt{g/4\pi}}\,\Bigl(\Bigl(\frac{t+1/2}{g/2} \Bigr)^2-1\Bigr)^{-1/4} \nonumber \\[3.5mm]
& & \hspace*{6mm}\cdot\:\exp\Bigl({-}\Bigl((t+1/2)\,{\rm arccosh}\Bigl(\frac{t+1/2}{g/2}\Bigr)\Bigr)-\frac{g}{2}\,\Bigl(\Bigl(\frac{t+1/2}{g/2} \Bigr)^2-1\Bigr)^{1/2}\Bigr) \nonumber \\[3.5mm]
& & =~\frac1g\,\Bigl(\Bigl(\frac{t+1/2}{g/2}\Bigr)^2-1\Bigr)^{-1/4}\, \exp({-}\varp(t+1/2\,;\,g/2))~.
\end{eqnarray}
On the range $(t+1/2)\leq\sqrt{2}\,(g/2)$ we replace the factor $(((t+1/2)/(g/2))^2-1)^{-1/4}$ by 1 at the expense of an error whose impact has been assessed in Appendix~B, see around Eq.~(\ref{b8}). For $(t+1/2)\geq\sqrt{2}\,(g/2)$, we have
\beq \label{c8}
\Bigl(\Bigl(\frac{t+1/2}{g/2}\Bigr)^2-1\Bigr)^{-1/4}\leq 2^{1/4}\Bigl(\frac{g/2}{t+1/2}\Bigr)^{1/2}~.
\eq
Hence, we estimate
\beq \label{c9}
|j_t(g/2)|\leq\frac{2^{1/4}}{g}\,\Bigl(\frac{g/2}{t+1/2}\Bigr)^{1/2}\, \exp({-}\varp(t+1/2\,;\,g/2))~,~~~~~~t+1/2\geq g/2~.
\eq
Combining this with the estimate in Eq.~(\ref{c5}), we arrive at
\begin{eqnarray} \label{c10}
& \mbox{} & |b_t(g)|\leq 2^{5/4}\,\frac{v_0}{u_0}\,\Bigl(\frac{t+1/2}{g/2}\Bigr)^{1/2}\, \exp({-}\varp(t+1/2\,;\,g/2))~,\nonumber \\[3.5mm]
& & \hspace*{5.5cm}g/2\leq t+1/2\leq g/2v_0~.
\end{eqnarray}

We proceed in a similar way on the range $t+1/2\geq g/2v_0$ for $h_t^{(2)}(g/2v_0)$, using Debye's asymptotic result, \cite{ref2}, 10.19.3, p.~231
\beq \label{c11}
Y_{\nu}(\nu\,{\rm sech}\,\alpha)\approx \frac{\exp(\nu(\alpha-\tanh\alpha))}{(\tfrac12\,\pi\nu\tanh\alpha)^{1/2}}
\eq
with $\nu=t+1/2$ and $\nu\,{\rm sech}\,\alpha=g/2v_0$. The right-hand side of Eq.~(\ref{c11}) equals
\beq \label{c12}
\Bigl(\frac{4v_0}{\pi g}\Bigr)^{1/2}\,\Bigl(\Bigl(\frac{t+1/2}{g/2v_0}\Bigr)^2-1\Bigr)^{-1/4} \,\exp(\varp(t+1/2\,;\,g/2v_0))~.
\eq
Then from Eq.~(\ref{e40}) and ignoring the relatively small quantity $J_{t+1/2}(g/2v_0)$, we estimate
\begin{eqnarray} \label{c13}
|h_t^{(2)}(g/2v_0)| & \approx & \frac{2v_0}{g}\,\Bigl(\Bigl( \frac{t+1/2}{g/2v_0}\bigr)^2-1\Bigr)^{-1/4}\,\exp(\varp(t+1/2\,;\,g/2v_0)) \nonumber \\[3.5mm]
& \leq & \frac{2^{5/4}v_0}{g}\,\bigl(\frac{g/2v_0}{t+1/2}\bigr)^{1/2}\,\exp(\varp(t+1/2\,;\, g/2v_0))~, \nonumber \\[3.5mm]
& & \hspace*{4.8cm}t+1/2\geq g/2v_0~,
\end{eqnarray}
where the factor $(((t+1/2)/(g/2v_0))^2-1)^{-1/4}$ has been dealt with in the same way as with the corresponding factor in Eq.~(\ref{c7}).

Combining Eqs.~(\ref{c9}, \ref{c13}), we get the estimate
\begin{eqnarray} \label{c14}
|b_t(g)| & \leq & 2^{3/2}\,\frac{v_0}{u_0}\,\Bigl(\frac{t+1/2}{g/2}\Bigr)^{1/2} \,\exp({-}\varp(t+1/2\,;\,g/2)) \nonumber \\[3.5mm]
& & \cdot\:\Bigl(\frac{g/2v_0}{t+1/2}\Bigr)^{1/2}\,\exp(\varp(t+1/2\,;\, g/2v_0))~,~~~~~~t+1/2\geq g/2v_0~. \nonumber \\
\mbox{}
\end{eqnarray}

We have established now the estimates in Eqs.~(\ref{c6}, \ref{c10}, \ref{c14}) on $|b_t(g)|$ on the ranges $0\leq t+1/2\leq g/2$, $g/2\leq t+1/2\leq g/2v_0$ and $t+1/2\geq g/2v_0$, respectively. According to Appendix~A, Eqs.~(\ref{a20}, \ref{a27}), extended to all $x\geq0$ at the expense of a negligible error when $\alpha=1/2$, see end of Appendix~A, we thus have \\[3mm]
$
\ba{ll}
|b_t(g)|\leq 4\,\dfrac{v_0}{u_0}~, & 0\leq t+1/2\leq g/2~, \\
\mc{2}{r}{\mbox{(C15)}} \\[3mm]
|b_t(g)|\leq 2^{5/4}\,\dfrac{v_0}{u_0}\,\exp({-}\varp(t\,;\,g/2))~, & g/2\leq t+1/2\leq g/2v_0~, \\
\mc{2}{r}{\mbox{(C16)}} \\[3mm]
|b_t(g)|\leq 2^{3/2}\,\dfrac{v_0}{u_0}\,\exp({-}\varp(t\,;\,g/2)+\varp(t\,;\, g/2v_0))~,~ & t+1/2\geq g/2v_0~.~\mbox{(C17)}
\ea
$ \\[3mm]
Since $\varp(t\,;\,g/2)=0$ for $t\leq g/2$ and $\varp(t\,;\,g/2v_0)=0$ for $t\leq g/2v_0$, the three estimates in Eqs.~(C15--C17) can be combined into a single one, viz.
\setcounter{equation}{17}
\beq \label{c18}
|b_t(g)|\leq 4\,\frac{v_0}{u_0}\,\exp({-}\varp(t\,;\,g/2)+\varp(t\,;\,g/2v_0))~,~~~~~~t\geq0~.
\eq
Using this in Eq.~(\ref{c3}), we see that $|c_t|$ is estimated by
\beq \label{c19}
4a_0\,w_0\,\exp({-}\varp(t\,;\,g/2)+\varp(t\,;\,g/2v_0))~,~~~~~~t\geq0~,
\eq
where
\beq \label{c20}
w_0=\frac{v_0}{u_0}=\frac{1}{1+\sqrt{1-s_0^2}}~.
\eq

The validity of Eq.~(\ref{c19}) as a bound for $|c_t|$ should be subjected to the same side comment as validity of Eq.~(\ref{b7}) for the Jinc function $J_{h+1}(2\pi r)/2\pi r$. There are now two relatively small regions, around $t+1/2=g/2$ and around $t+1/2=g/2v_0$, where the bound in Eq.~(\ref{c19}) is too low by a factor that increases very slowly as $g\pr\infty$. Fortunately, we consider values of $s_0\leq0.99$, which implies that $v_0\leq 0.75$, so that the exceptional regions do not overlap as $g\pr\infty$.

For the sake of computation of the quantities $b_k$ in Eq.~(\ref{e38}), involving the products of spherical Bessel and Hankel functions, with a specified accuracy, we note the bounds for $k\geq0$
\beq\label{c21}
|j_k(g/2)| \leq \frac2g\;, \;\; |h_k(g/2v_0)|\leq \frac{2^{7/4}v_0}{g}\exp{(\varphi(k;g/2v_0))}\;.
\eq
The first bound follows from Eqs.~(\ref{c4}),~(\ref{c9}) and (\ref{a20}) with $\alpha=1/2$ and $c=g/2$, and the second bound follows from Eqs.~(\ref{c5}),~(\ref{c13}) and (\ref{a26i}) with $\alpha=1/2$ and $c=g/2v_0$. 
Since $|j_k(f/2)|\leq 1$, we may replace the argument $g/2$ in the first inequality in Eq.~(\ref{c21}) by $f/2$. In the second inequality, we can replace the argument $g/2v_0$ by $f/2v_0$ only when $|f/v_0|\geq 1$.

\setcounter{equation}{0}
\renewcommand{\theequation}{D\arabic{equation}}
\section{Proof of validity of truncation rules} \label{appD}
\mbox{} \\[-9mm]

In this appendix, we give the proofs for the results in Subsecs.~\ref{subsec2.3}--\ref{subsec2.4} on truncation rules. We first show that the quantity in Eq.~(\ref{e11}) is less than $\eps\in(0,1)$ when $H$ and $T$ are chosen according to Eq.~(\ref{e24}) with $B$ given in Eq.~(\ref{e23}).

From Appendix~A, we have for $d\geq c>0$ that
\beq \label{d1}
\varp(x\,;\,c)\geq\varp(x\,;\,c)-\varp(x\,;\,d)\geq\gamma x-c\sinh(\gamma)~,~~~~~~x\geq0~,
\eq
where $\gamma\leq{\rm ln}(d/c)$. Taking $x=h+1$, $c=2\pi R$, $d=ec$ (so that $\gamma\leq1$), we get by taking $\gamma=1$
\beq \label{d2}
\varp(h+1\,;\,2\pi R)\geq h+1-2\pi R\sinh(1)~.
\eq
The right-hand side of Eq.~(\ref{d2}) exceeds $B$ of Eq.~(\ref{e23}) when $h+1\geq H$, and then from Eq.~(\ref{e14})
\beq \label{d3}
\Bigl|\frac{J_{h+1}(2\pi r)}{2\pi r}\Bigr|\leq\frac{\eps}{4w_0a_0}~,~~~~~~h+1\geq H~.
\eq

Next, for $x=t$, $c=g/2$, $d=g/2v_0$ in Eq.~(\ref{d1}) (so that $\gamma\leq{\rm ln}(1/v_0))$, we get by taking $\gamma=\min(1,{\rm ln}(1/v_0))$
\beq \label{d4}
\varp(t\,;\,g/2)-\varp(t\,;\,g/2v_0)\geq\gamma t-\tfrac12\,g\sinh(\gamma)~.
\eq
The right-hand side of Eq.~(\ref{d4}) exceeds $B$ of Eq.~(\ref{e23}) when $t\geq T$, and then from Eq.~(\ref{e16})
\beq \label{d5}
|c_t|\leq 2\eps\,\pi^2\,R\,\sqrt{R}~,~~~~~~t\geq T~.
\eq
Since for all $h\geq0$, $t\geq0$ by Eqs.~(\ref{e14}, \ref{e16})
\beq \label{d6}
\Bigl|\frac{J_{h+1}(2\pi r)}{2\pi r}\Bigr|\leq\frac{1}{2\pi^2\,R\,\sqrt{R}}~,~~~~~~|c_t|\leq 4w_0a_0~,
\eq
we find that
\beq \label{d7}
|c_t|\,\Bigl|\frac{J_{h+1}(2\pi r)}{2\pi r}\Bigr|<\eps
\eq
when $h+1\geq H$ and/or $t\geq T$. This means that the quantity in Eq.~(\ref{e11}) is less than $\eps$.

As to the dedicated truncation rule, we use continuity, monotonicity and convexity of $F(h,t)$ as a function of both $h$ and $t$, see Eqs.~(\ref{e27}--\ref{e28}). It thus follows easily that the right-hand side of Eq.~(\ref{e30}) is less than $\eps$ when $h+1>H$ or $t>T$ when $H$ and $T$ are chosen as $H=H_n^m=\max(h_1,h_2)+1$, $T=T_n^m=\max(t_1,t_2)$ (for the case that $M$ in Eq.~(\ref{e31}) $\leq\:B$; otherwise we simply have $H=1,\;T=0$). Here the points $(h_1,t_1)$ and $(h_2,t_2)$ are found as the first and the last point $(h,2t)$ on $\partial S_n^m$ with $F(h,t)>B$ when inspecting the 4 line segments of the boundary $\partial S_n^m$ in counterclockwise manner through integer $h$ and $t$ with $h$ same parity as $n$. This means that with this choice of $H$ and $T$ the quantity in Eq.~(\ref{e10}) is less than $\eps$.

It also follows that $F(h,t)$ increases along both edge~I and edge IV in Fig.~1 when $(h,2t)\pr\infty$. Also $F(h,t)$ increases along edge III when $t$ increases and $h$ is kept fixed at $|m|$. Therefore, the minimum $M$ in Eq.~(\ref{e31}) is to be found on edge~II. On this edge~II, it follows from convexity of $F$ that the minimum is attained on a set of points $(n-2t,2t)$ with $t$ in a closed interval contained in $[0,\tfrac12\,(n-|m|)]$ (which reduces to a single point $t$ when $F$ is strictly convex on edge~II).

\setcounter{equation}{0}
\renewcommand{\theequation}{E\arabic{equation}}
\section{Asymptotics, bounds and truncation issues for coefficients of algebraic functions} \label{appE}
\mbox{} \\[-9mm]

We consider in this appendix the (computation of the) Zernike coefficients of the modified algebraic function
\begin{eqnarray} \label{E1}
& \mbox{} & \hspace*{-8mm}A(\rho)=a(\rho)\,\sqrt{1-s_0^2\rho^2}=\sum_{l=0}^{\infty}\,a_l\,R_{2l}^0(\rho) \nonumber \\[3mm]
& & \hspace*{-8mm}=~ (1-s_0^2\rho^2)^{3/4}\,(1-s_{0,M}^2\rho^2)^{-3/4}+(1-s_0^2\rho^2)^{1/4}\,(1-s_{0,M}^2\rho^2) ^{-1/4}~,
\end{eqnarray}
see Eqs.~(\ref{e3}, \ref{e41}). This $A$ is the sum of two functions
\beq \label{E2}
a_{\alpha\beta}(\rho)=(1-s_{\alpha}^2\rho^2)^{\alpha}\,(1-s_{\beta}^2\rho^2)^{\beta}=\sum_{l=0}^{\infty}\, a_{l,\alpha\beta}\,R_{2l}^0(\rho)~.
\eq
We let for such an $a_{\alpha\beta}$
\beq \label{E3}
S=\max(s_{\alpha},s_{\beta})~,~~~~~~s=\min(s_{\alpha},s_{\beta})~,
\eq
\beq \label{E4}
\Delta=\arg(S)~,~~~~~~\delta=\arg(s)~,
\eq
so that
\beq \label{E5}
a_{\alpha\beta}(\rho)=(1-s^2\rho^2)^{\delta}\,(1-S^2\rho^2)^{\Delta}~.
\eq
Observe that in the cases in Eq.~(\ref{E1}) we have $\Delta+\delta=0$. 

We consider the power series coefficients $r_{N,\alpha\beta}$ of $a_{\alpha\beta}(\rho)$, and the computation of the $a_{l,\alpha\beta}$ according to
\beq \label{E6}
a_{l,\alpha\beta}=\sum_{N=l}^{\infty}\,b_N(l)\,r_{N,\alpha\beta}~;~~~~~~b_N(l)=\frac{2l+1}{l+1} \,\Bigl(\!\ba{c} N \\ l\ea\!\Bigr)\Bigl/\Bigl(\!\ba{c} N+l+1 \\ N \ea\!\Bigr)~.
\eq
It will be shown below that for $\delta\in({-}1,1)$ and $N=1,2,...$
\beq \label{E7}
r_{N,\delta,{-}\delta}=\frac{1}{\pi}\sin(\pi\delta)\,\il_{1/S^2}^{1/s^2}\,\Bigl(\frac{1-s^2x}{S^2x-1} \Bigl)^{\delta}\,\frac{dx}{x^{N+1}}~.
\eq
Hence, $r_{N,\delta,{-}\delta}$ has the sign of $\delta$, and it will also be shown that for $\delta\in(0,1)$ and $N=0,1,...$
\beq \label{E8}
r_{N,\delta,{-}\delta}\geq|r_{N,{-}\delta,\delta}|~.
\eq

It follows easily from Eq.~(\ref{E7}) that $r_{N,\delta,{-}\delta}$ decreases as a function of $s\in(0,S]$ when $\delta>0$. Hence, for $\delta\in(0,1)$,
\beq \label{E9}
r_{N,\delta,{-}\delta}\leq\lim_{s\downarrow 0}\,r_{N,\delta,{-}\delta}= C_{\rho^{2N}}[(1-S^2\rho^2)^{-\delta}]~.
\eq
Since the $b_N(l)$ in Eq.~(\ref{E6}) are all non-negative, it follows from Eqs.~(\ref{E8}, \ref{E9}) that for $\delta\in(0,1)$
\beq \label{E10}
|a_{l,{-}\delta,\delta}|\leq a_{l,\delta,{-}\delta}\leq ZC_l\,[(1-S^2\rho^2)^{-\delta}]~,
\eq
where $ZC_l$ abbreviates ``the $l^{{\rm th}}$ Zernike coefficient of the function in $[~]$''.

We shall show below that for $\Delta\in({-}1,1)$, the asymptotic behavior of the Zernike coefficients of $(1-S^2\rho^2)^{\Delta}$ is given by
\beq \label{E11}
ZC_l\,[(1-S^2\rho^2)^{\Delta}]\sim \frac{2\sqrt{\pi}}{\Gamma({-}\Delta)}~\frac{(1-S^2)^{\frac12\Delta+\frac14}}
{1+\sqrt{1-S^2}}~\frac{V^l}{(l+1)^{\Delta+\frac12}}
\eq 
as $l\pr\infty$, where
\beq \label{E12}
V=\frac{1-\sqrt{1-S^2}}{1+\sqrt{1-S^2}}~.
\eq
For $\Delta=0$, we have $\Gamma({-}\Delta)=\infty$, and the right-hand side of Eq.~(\ref{E11}) vanishes. For $\Delta={-}1/2$, the right-hand side of Eq.~(\ref{E11}) is exactly equal to $ZC_l\,[(1-S^2\rho^2)^{-1/2}]$, see \cite{ref1}, Eq.~(134), and also for the case that $\Delta=1/2$, there is good agreement between $ZC_l\,[(1-S^2\rho^2)^{1/2}]$, given by [1], Eq.~(135), and the right-hand side of Eq.~(\ref{E11}).

The maximum modulus of the right-hand side of Eq.~(\ref{e11}) occurs at $l=0$ and decreases in $l=0,1,...$ unless $\Delta<{-}1/2$ and $S$ is extremely close to 1. In the relevant case that $\Delta={-}3/4$, monotonicity of the modulus is guaranteed as long as $V\leq 2^{-1/4}$, i.e., $S\leq 2(2^{-1/8}+2^{1/8})^{-1}=0.9963$. 

In Sec.~\ref{sec3}, Eqs.~(\ref{e50}--\ref{e52}), it is required to find for a given $\eta>0$, $E>0$, and $\Delta\in(-1,0)$, $V\in(0,1)$ an $L>0$ such that
\beq \label{E13}
l\geq L\Rightarrow\frac{E\,V^l}{(l+1)^{\Delta+1/2}}<\eta~.
\eq
Under the monotonicity assumption, an approximation of the required $L$ is found by rewriting the equation $E\,V^L(L+1)^{-\Delta-1/2}=\eta$ for $L$ as
\beq \label{E14}
L=\frac{{\rm ln}(E/\eta)-(\Delta+1/2)\,{\rm ln}(L+1)}{{\rm ln}(1/V)}~,
\eq
and to iterate this equation twice, starting with $L=0$. This yields the quantity at the right-hand side of Eq.~(\ref{e52}), with $\Delta={-}\delta$.

We next address the truncation issue when computing $a_{l,\alpha\beta}$ according to Eq.~(\ref{E6}). It is sufficient to consider this for the function $(1-S^2\rho^2)^{\Delta}$ with $\Delta\in({-}1,0)$, see Eqs.~(\ref{E9}, \ref{E10}). We have
\beq \label{E15}
C_{\rho^{2N}}\,[(1-S^2\rho^2)^{\Delta}]=\frac{\Gamma(N-\Delta)\,S^{2N}} {\Gamma({-}\Delta)\,\Gamma(N+1)}\sim\frac{S^{2N}}{\Gamma({-}\Delta)\,N^{\Delta+1}}~.
\eq
Thus, the terms in the series in Eq.~(\ref{E6}) are approximated as
\beq \label{E16}
	\frac{(2l+1)\Gamma^2(N+1)}{\Gamma(N+1+l)\Gamma(N+1-l)} \frac{S^{2N}}{\Gamma(-\Delta)N^{\Delta+1}(N+l+1)}\;.
\eq
For a given $N$, the maximum of 
\beq \label{E17}
	\frac{(2l+1)\Gamma^2(N+1)}{\Gamma(N+1+l)\Gamma(N+1-l)}\;,\;\;\; l = 0,\;1,\;\cdots,\;N\;,
\eq
is approximately $\sqrt{2N/e}$ and occurs at $l$ near $\sqrt{N/2}$. Thus the truncation errors $\sum_{N=N_L}^{\infty} b_n(l)r_{N,\alpha\beta}$ for the series in Eq.~(\ref{E6}) are all bounded by
\beq \label{E18}
	\sqrt{\frac{2}{e}}\, \frac{1}{\Gamma(-\Delta)} \sum\limits^\infty_{N=N_L} \frac{S^{2N}}{N^{\Delta+3/2}}\;.
\eq
Now, by partial integration and $\Delta+3/2>0$,
\begin{eqnarray} \label{E19}
	\sum\limits^\infty_{N=N_L} \frac{S^{2N}}{N^{\Delta+3/2}} &\approx& \int\limits^\infty_{N_L} \frac{e^{-x\ln{(S^{-2})}}}{x^{\Delta+3/2}}\, \textrm{d}x	
	< \frac{e^{-N_L\ln{(S^{-2})}}}{N_L^{\Delta+3/2}\ln(S^{-2})}\nonumber \\ &=& \frac{S^{2N_L}}{N_L^{\Delta+3/2}\ln{S^{-2}}} < \frac{S^{2N_L}}{N_L^{\Delta+3/2}(1-S^{2})}\;,
\end{eqnarray}
and so the quantity in Eq.~(\ref{E18}) is realistically bound by
\beq\label{E20}
	\sqrt{\frac{2}{e}}\,\frac{S^{2N_L}}{\Gamma(-\Delta)N_L^{\Delta+3/2}(1-S^{2})}\;.
\eq
We recall that $a_{l,\alpha\beta}$ are required for all $l\leq L$, where $L$ satisfies $V^L = \tfrac{\eta}{E}(L+1)^{\Delta+1/2}$, see Eq.~(\ref{E13}), with
\beq \label{E21}
	E = \frac{2\sqrt{\pi}}{\Gamma(-\Delta)} \frac{(1-S^2)^{\tfrac12 \Delta+\tfrac14}}{1+\sqrt{1-S^2}}\;.
\eq
We now propose to take $N_L = 2L/\sqrt{1-S^2}$. Then 
\beq \label{E22}
	S^{2N_L} = \exp{\left(\frac{2L\ln(S^2)}{\sqrt{1-S^2}}\right)} < V^L = \frac{\eta}{E}(L+1)^{\Delta+1/2}\;,
\eq
where the inequality in Eq.~(\ref{E22}) follows from 
\beq \label{E23}
	\frac{2}{y}\ln(1-y^2)<\ln\left( \frac{1-y}{1+y}\right)\;,\;\; 0 < y < 1\;,
\eq
with $y=\sqrt{1-S^2}$. Thus, all truncation errors are bounded by
\begin{eqnarray} \label{E24}
	\sqrt{\frac{2}{e}} \frac{(L+1)^{\Delta+1/2}\eta}{E\Gamma(-\Delta)N_L^{\Delta+3/2}(1-S^2)} \approx \frac{\eta\sqrt{2/e}}{2\sqrt{\pi}\, 2^{\Delta+3/2}} \frac{1+\sqrt{1-S^2}}{L\sqrt{1-S^2}}\;,
\end{eqnarray}
where we have used the definitions of $E$ and $N_L$. This quantity (\ref{E24}) is well below $\eta/2$ for somewhat larger values of $L$. In fact, from Eq.~(\ref{E22}) and in the relevant case $\Delta=-3/4$ ( so that $(L+1)^{\Delta+1/2}\leq1$)
\begin{eqnarray} \label{E25}
	\ln{\left( \frac{\eta}{E}\right)} > \ln{V^L} &=& L \ln{\left( \frac{1-\sqrt{1-S^2}}{1+\sqrt{1-S^2}}\right)} \nonumber \\ 
	&=& L\ln{\left( 1-\frac{2\sqrt{1-S^2}}{1+\sqrt{1-S^2}}\right)} \approx -\frac{2L\sqrt{1-S^2}}{1+\sqrt{1-S^2}}\;,\;\;\;
\end{eqnarray}
and so the quantity in Eq.~(\ref{E24}) is realistically estimated at $(\Delta=-3/4)$
\beq \label{E26}
	\frac{\eta\sqrt{2/e}}{\sqrt{\pi}\,2^{\Delta+3/2}\ln{\left( E/\eta \right)}} = \frac{0.2878\,\eta}{\ln(E/\eta)}\;.
\eq

We still owe the reader a proof of the results in Eq.~(\ref{E7}, \ref{e11}). As to Eq.~(\ref{E7}), we consider the general case in Eq.~(\ref{E5}). Setting $x=\rho^2$, we have by Cauchy's formula
\beq \label{E27}
r_{N,\alpha\beta}=C_{x^N}\,[(1-s^2x)^{\delta}(1-S^2x)^{\Delta}]= \frac{1}{2\pi i}\,\oint\,\frac{(1-s^2z)^{\delta}(1-S^2z)^{\Delta}}{z^{N+1}}\,dz~,
\eq
with integration contour a circle of radius $<\:1/S^2$ in positive sense. We choose principal values of the roots $(1-s^2z)^{\delta,\Delta}$, and we deform the contour so that the positive real axis from the first branch point $z=1/S^2$ onwards, passing along the second branch point $z=1/s^2$, to $z=\infty$ is enclosed. When $N=1,2,...$ and $\delta,\Delta>{-}1$, $\delta+\Delta<1$, this can be done without problems. Since
\beq \label{E28}
(1-s^2(x\pm io))^{\delta}=(s^2x-1)^{\delta}\,e^{\mp\pi i\delta}~,~~~~x>1/s^2~,
\eq
\beq \label{E29}
(1-S^2(x\pm io))^{\Delta}=(S^2x-1)^{\Delta}\,e^{\mp\pi i\Delta}~,\hspace*{5mm}x>1/S^2~,
\eq
it follows that
\begin{eqnarray} \label{E30}
r_{N,\alpha\beta} & = & \frac{1}{2\pi i}\,\il_{1/S^2}^{1/s^2}\,(1-s^2x)^{\delta}\,(S^2x-1)^{\Delta}\, (e^{-\pi i\Delta}-e^{\pi i\Delta})\,\frac{dx}{x^{N+1}} \nonumber \\[3.5mm]
& & +~\frac{1}{2\pi i}\,\il_{1/s^2}^{\infty}\,(s^2x-1)^{\delta}\,(S^2x-1)^{\Delta} \,(e^{-\pi i(\delta+\Delta)}-e^{\pi i(\delta+\Delta)})\,\frac{dx}{x^{N+1}} \nonumber \\[3.5mm]
& = & \frac{-\sin\pi\Delta}{\pi}\,\il_{1/S^2}^{1/s^2}\, \frac{(1-s^2x)^{\delta}\,(S^2x-1)^{\Delta}}{x^{N+1}}\,dx \nonumber \\[3.5mm]
& & -~ \frac{\sin(\delta+\Delta)}{\pi}\,\il_{1/s^2}^{\infty}\,\frac{(s^2x-1)^{\delta}\, (S^2x-1)^{\Delta}}{x^{N+1}}\,dx~.
\end{eqnarray}
When $\delta+\Delta=0$, the second integral in Eq.~(\ref{E30}) is canceled, and we get Eq.~(\ref{E7}).

We now show Eq.~(\ref{E8}). We have for $\delta\in(0,1)$ 
\beq \label{E31}
r_{0,\delta,{-}\delta}=r_{0,{-}\delta,\delta}=1~,~~~~~~r_{1,\delta,{-}\delta} ={-}r_{1,{-}\delta,\delta}=(S^2-s^2)\,\delta
\eq
as readily follows from Eq.~(\ref{E5}). From Eq.~(\ref{E7}) we have
\begin{eqnarray} \label{E32}
& \mbox{} & r_{N,\delta,{-}\delta}+r_{N,{-}\delta,\delta} \nonumber \\[3.5mm]
& & =~\frac{\sin\pi\delta}{\pi}\,\il_{1/S^2}^{1/s^2}\,\Bigl[\Bigl( \frac{1-s^2x}{S^2x-1}\Bigr)^{\delta}-\Bigl(\frac{1-s^2x}{S^2x-1}\Bigr)^ {-\delta}\Bigr]\,\frac{dx}{x^{N+1}}~,
\end{eqnarray}
and this vanishes when $N=1$. The function $g(x)$ in $[~]$ in the integral in Eq.~(\ref{E32}) decreases in $x\in[1/S^2,1/s^2]$ since $\delta>0$, and has there a single zero, at $x=2/(s^2+S^2)=:x_0$. Then for $N>1$, we have
\beq \label{E33}
\il_{1/S^2}^{1/s^2}\,\frac{g(x)}{x^{N+1}}\,dx=\il_{1/S^2}^{1/s^2}\, \frac{g(x)}{x^2}\,\Bigl(\frac{1}{x^{N-1}}-\frac{1}{x_0^{N-1}}\Bigr)\,dx~,
\eq
and this is positive since the integrand of the second integral is positive for all $x\neq x_0$. Since $r_{N,\delta,{-}\delta}$ is positive and $r_{N,{-}\delta,\delta}$ is negative, see Eq.~(\ref{E7}), we get Eq.~(\ref{E8}).

We finally show the asymptotic result in Eq.~(\ref{E11}). We have from $R_{2l}^0(\rho)=P_l(2\rho^2-1)$, where $P_l$ is the Legendre polynomial of degree $l$, the substitutions
\beq \label{E34}
x=2\rho^2-1\in[{-}1,1]~,~~~~~~a=1-\tfrac12\,S^2\,,~~b=\tfrac12\,S^2~,
\eq
Rodriguez' formula
\beq \label{E35}
P_l(x)=\frac{({-}1)^l}{2^l\,l!}\,\Bigl(\frac{d}{dx}\Bigr)^l\,[(1-x^2)^l]~,
\eq
and $l$ partial integrations, that
\begin{eqnarray} \label{E36}
ZC_l\,[(1-S^2\rho^2)^{\Delta}] & = & 2(2l+1)\,\il_0^1\,(1-S^2\rho^2)^{\Delta}\,R_{2l}^0(\rho)\,\rho\,d\rho \nonumber \\[3.5mm]
& = & \frac{(l+1/2)\,\Gamma(l-\Delta)}{l!\,\Gamma({-}\Delta)}\,\il_{-1}^1\,(a-bx)^{\Delta}\, \Bigl(\frac{b}{2}~\frac{1-x^2}{a-bx}\Bigr)^l\,dx~. \nonumber \\
\mbox{}
\end{eqnarray}
The remaining integral in Eq.~(\ref{E36}) can be approximated by using Laplace's method. The stationary point of the integrand is found by setting $((1-x^2)/(a-bx))'=0$, and this yields $x=V$ when we restore the parameter $S$, see Eqs.~(\ref{E34}, \ref{e12}). We have furthermore
\beq \label{E37}
a-bx|_{x=V}=\sqrt{1-S^2}~,~~~~~~\frac{b}{2}~\frac{1-x^2}{a-bx}\Bigl|_{x=V}=V~,
\eq
and
\beq \label{E38}
\Bigl({\rm ln}\Bigl(\frac{1-x^2}{a-bx}\Bigr)\Bigr)''\Bigl|_{x=V}
={-}\,\frac{(1+\sqrt{1-S^2})^2}{2\,\sqrt{1-S^2}}~.
\eq
This then yields
\begin{eqnarray} \label{E39}
ZC_l\,[(1-S^2\rho^2)^{\Delta}] & \approx & \frac{(l+1/2)\,\Gamma(l-\Delta)}{\Gamma(l+1)\,\Gamma({-}\Delta)}\,(\sqrt{1-S^2}) ^{\Delta}\,V^L \nonumber \\[3.5mm]
& & \cdot\:\il_{-\infty}^{\infty}\,\exp\Bigl({-}l\,\frac{(1+\sqrt{1-S^2})^2} {4\,\sqrt{1-S^2}}\,(x-V)^2\Bigr)\,dx \nonumber \\[3.5mm]
& = & \frac{2\sqrt{\pi}}{\Gamma({-}\Delta)}~\frac{(1-S^2)^{\frac12\Delta+\frac14}} {1+\sqrt{1-S^2}}~\frac{(l+1/2)\,\Gamma(l-\Delta)}{\Gamma(l+1)\,l^{1/2}} \,V^l \nonumber \\[3.5mm]
& \approx & \frac{2\sqrt{\pi}}{\Gamma({-}\Delta)}~\frac{(1-S^2)^{\frac12\Delta+\frac14}} {1+\sqrt{1-S^2}}~\frac{V^l}{(l+1)^{\Delta+\frac12}}~,
\end{eqnarray}
as required.

\end{document}